\documentclass[a4paper,11pt]{article}
\usepackage{pos}

\usepackage{mdframed}
\usepackage{simplewick}
\usepackage{slashed}
\usepackage{bm}
\renewcommand{\vec}[1]{\bm{#1}}
\newcommand{\GY}{\color{gray}}
\definecolor{green}{rgb}{0.,0.7,0.}
\def\uR{\color{red}u}
\def\dR{\color{red}d}
\def\sR{\color{red}s}
\def\cR{\color{red}c}
\def\tR{\color{red}t}
\def\bR{\color{red}b}
\def\uG{\color{green}u}
\def\dG{\color{green}d}
\def\sG{\color{green}s}
\def\cG{\color{green}c}
\def\tG{\color{green}t}
\def\bG{\color{green}b}
\def\uB{\color{blue}u}
\def\dB{\color{blue}d}
\def\sB{\color{blue}s}
\def\cB{\color{blue}c}
\def\tB{\color{blue}t}
\def\bB{\color{blue}b}
\def\bV{\mathbf{V}}
\def\bM{\mathbf{M}}
\def\bU{\mathbf{U}}
\def\cU{{\cal U}}
\def\bS{\mathbf{V}}
\def\bH{\mathbf{H}}
\def\bmd{\mathbf{d}}
\def\bmu{\mathbf{u}}
\def\bme{\mathbf{e}}
\def\bmn{\boldsymbol{\nu}}
\def\bmf{\mathbf{f}}
\newcommand{\eq}[1]{\begin{align}#1\end{align}}
\newcommand{\pmat}[1]{\begin{pmatrix}#1\end{pmatrix}}
\newcommand{\ba}{\begin{array}}
\newcommand{\ea}{\end{array}}
\newcommand{\bma}{\begin{pmatrix}}
\newcommand{\ema}{\end{pmatrix}}
\newcommand{\dis}{\displaystyle}
\renewcommand{\H}{{\cal H}}
\def\lag{{\cal L}}
\def\d{\partial}
\def\dd{{\rm d}}
\def\D{{\cal D}}
\def\ii{{\rm i}}
\def\e{{\rm e}}
\def\braket#1#2{\left\langle #1\vphantom{#2}
  \right. \kern-2.5pt\left| #2\vphantom{#1}\right\rangle }
\def\ketbra#1#2{\left| #1\vphantom{#2}
  \right\rangle \kern-2.5pt\left\langle #2\vphantom{#1}\right| }
\def\bra#1{\left\langle #1\right| }
\def\ket#1{\left| #1\right\rangle }

\def\BraKet#1#2{\left\langle #1\vphantom{#2}
  \right. \kern-2.5pt\left\| #2\vphantom{#1}\right\rangle }
\newcommand{\Raw}{\Rightarrow}
\newcommand{\nn}{\nonumber}
 
\numberwithin{equation}{section}
	\DeclareSymbolFont{CMletters}{OML}{cmm}{m}{it}
	\DeclareMathSymbol{v}{\mathord}{CMletters}{`v}
	\DeclareSymbolFont{epsilon}{OML}{ntxmi}{m}{it}
	\DeclareMathSymbol{\epsilon}{\mathord}{epsilon}{"0F}

\title{Quantum field theory and the structure of the Standard Model}
\ShortTitle{Quantum field theory and the structure of the Standard Model}

\author*[a]{Jos\'e Ignacio Illana}
\author[a,b]{Alejandro Jim\'enez Cano}

\affiliation[a]{Departamento de F{\'\i}sica Te\'orica y del Cosmos, Universidad de Granada,\\
   	E-18071 Granada, Spain}

\affiliation[b]{Laboratory of Theoretical Physics, Institute of Physics, University of Tartu,\\
	50411 Tartu, Estonia}

\emailAdd{jillana@ugr.es}
\emailAdd{alejandro.jimenez.cano@ut.ee}

\abstract{%
The Standard Model of the electroweak and strong interactions of particle physics is a quantum field theory. Elementary particles are not indivisible `pieces' of matter but energy bundles of fields, whose properties and interactions are a consequence of the principles of symmetry. These lecture notes provide a brief introduction to the construction of the Standard Model from its basic ingredients: Poincar\'e symmetry, gauge invariance and spontaneous symmetry breaking. The full Lagrangian is derived in detail and the most relevant aspects of the electroweak phenomenology are discussed with special emphasis on the determination of the input parameters and the consistency checks of the model. Some exercises are proposed to fix the main ideas.
}

\FullConference{%
  Corfu Summer Institute 2021 "School and Workshops on Elementary Particle Physics and Gravity"\\
  29 August - 9 October 2021\\
  Corfu, Greece
}

\tableofcontents

\newcommand\boxexercise[2]{\begin{mdframed} \noindent{\bf Exercise #1.} #2 \end{mdframed}}
\newcommand\subsubsubsection[1]{\subsubsection*{#1}}
\allowdisplaybreaks

\begin{document}
\maketitle

\section{Particles, fields and symmetries}

Relativistic quantum field theory (QFT) reconciles quantum mechanics and special relativity \cite{Weinberg,Schwartz,Maggiore,Lahiri,Peskin,Pokorski}. Wave equations of quantum mechanics, whether relativistic or not, cannot account for systems in which the number of particles changes, and their relativistic versions suffer from a number of pathologies: negative probability densities, negative-energy solutions and violation of causality, among others. The formulation of quantum fields: 
{\it i)} provides a natural framework to describe multiparticle states with varying occupation,
{\it ii)} makes sense of negative-energy solutions (they are interpreted as antiparticles),
{\it iii)} solves the causality problem (outside the lightcone particle and antiparticle contributions to the field commutator cancel each other), and
{\it iv)} justifies the spin-statistics connection, that is no longer a postulate but a theorem.

\subsection{Basics: Poincar\'e symmetry}

\subsubsection{Guided by symmetry}

Relativistic fields are irreducible representations (irreps) of the Poincar\'e group, including Lorentz transformations (space rotations and Lorentz boosts) and spacetime translations. Examples of field irreps are the scalar $\phi(x)$, the four-vector $V_\mu(x)$ and the symmetric tensor $h_{\mu\nu}(x)$, where $\mu,\nu\in\{0,1,2,3\}$ are Lorentz indices.\footnote{%
A general tensor $T^{\mu\nu\rho\dots}$ with $N$ Lorentz indices contains representations of spin $j=0,1,\dots,N$ under rotations. The scalar field $\phi$ (no indices) has spin $j=0$. The vector field $V^\mu=(V^0,\vec{V})$ contains spin $j=0$ ($V^0$) and spin $j=1$ ($\vec{V}$).
}
The Lorentz subgroup is locally isomorphic to SU(2)$\otimes$SU(2) whose irreps are labeled by $(j_-,j_+)$ so it also admits spinorial representations where $j_\pm$ can be half-integer, as for example the Weyl bispinor fields $\psi_L(x)\sim(\frac{1}{2},0)$ and $\psi_R(x)\sim(0,\frac{1}{2})$ that are two non-equivalent (conjugated) representations of spin $j=\tfrac{1}{2}$ under rotations. The Dirac four-spinor field $\psi(x)=\psi_L(x)\oplus\psi_R(x)$ is a reducible representation of the Lorentz group that contains left and right-handed chiralities, exchanged by a parity transformation.

In order to describe the field dynamics, one introduces the action
\begin{equation}
  S[\phi_i] = \int \dd^4x \ \lag(\phi_i(x),\d_\mu\phi_i(x)) \,,
\label{eq:action}
\end{equation}
where the Lagrangian (density) $\lag(x) = \lag(\phi_i,\d_\mu\phi_i)$ is a local function of the fields and their derivatives, and $\phi_i$ stands for any type of fields ($\phi$, $\psi$, $V_\mu$, etc.). The action must be invariant under Poincar\'e transformations according to the covariance principle of special relativity.
As an example, the Lagrangian of a free Dirac field $\psi(x)$ is
\eq{
\lag_0 = \overline\psi(\ii\slashed{\d}-m)\psi \,,
\label{eq:lagD}
}
where $\gamma^\mu$ are the Dirac or `gamma' matrices, $\slashed{\d} \equiv \gamma^\mu \d_\mu$ (slash notation), $\overline\psi \equiv \psi^\dagger\gamma^0$ is the Dirac adjoint 
and the constant $m$ is the Dirac mass. 

The Lagrangian contains all the information about the particular theory under study. 
Following Noether's (first) theorem, any continuous global symmetry of the action is in correspondence with a conservation law. In particular, the conservation of energy, linear momentum and angular momentum are the consequence of the invariance under time translations, space translations and space rotations, respectively, all of which are Poincar\'e symmetries.

Given a Lagrangian, one derives the equations of motion (Euler-Lagrange equations), which describe the classical evolution of the fields. They are obtained from the principle of least action: the field configuration must be a stationary point of the action, {\it i.e.} $\delta S=0$. Then
\begin{align}
\delta S &= \int \dd^4x \sum_i \left( \frac{\d\lag}{\d\phi_i}\delta\phi_i + \frac{\d\lag}{\d(\d_\mu\phi_i)}\delta(\d_\mu\phi_i) \right) 
         = \int \dd^4x \sum_i \left( \frac{\d\lag}{\d\phi_i}-\d_\mu\frac{\d\lag}{\d(\d_\mu\phi_i)} \right)\delta\phi_i = 0\,, \label{eq:varS}
\end{align}
where we have used integration by parts and we have dropped the boundary term under the assumption that field variations vanish at infinity. Since \eqref{eq:varS} must hold for any variation $\delta \phi_i$, we get for each field the corresponding equation of motion (EoM):
\begin{equation}
  \frac{\d\lag}{\d\phi_i}-\d_\mu\frac{\d\lag}{\d(\d_\mu\phi_i)}=0\,.
\end{equation}
For instance, in the case of a free Dirac field, the EoM is the well-known Dirac equation,
\begin{equation}
(\ii\slashed{\d}-m)\psi(x)=0\,,
\end{equation}
whose general solution is
\begin{equation}
\psi(x)=\int\frac{\dd^3 p}{(2\pi)^3\sqrt{2E_{\vec{p}}}}\sum_{s=1,2}
\left( a_{\vec{p},s}u^{(s)}(p)\e^{-\ii p\cdot x} + b^*_{\vec{p},s}v^{(s)}(p)\e^{\ii p\cdot x} \right)
\label{eq:freefield}
\end{equation}
with $p^2=E_{\vec{p}}^2-|\vec{p}|^2=m^2$, where $u^{(s)}(p)$ and $v^{(s)}(p)$ are constant four-spinors of positive and negative energy, respectively ($\pm E_{\vec{p}}$ with $E_{\vec{p}}\equiv +\sqrt{|\vec{p}|^2+m^2}$) verifying 
\eq{
(\slashed{p}-m)u^{(s)}(p)=0\,,\quad (\slashed{p}+m)v^{(s)}(p)=0\,,\quad s=1,2,
}
and $a_{\vec{p},s}$, $b_{\vec{p},s}$ are for the moment just complex coefficients with convenient normalizations.\footnote{
Natural units $\hbar=c=1$ are used throughout this course.} 

\subsubsection{Quantization\label{sec:quantization}}

The approach we have seen so far is purely classical. Now we will address the problem of quantizing the theory. It is quite remarkable that there is not a unique way of making a classical theory quantum. We will mostly focus on the so-called canonical quantization. In this method the notion of particle emerges in a quite transparent way. However, functional methods are very useful and will be applied to quantize gauge theories.

To quantize a classical theory \`a la canonical, we need to have a well-defined Hamiltonian formulation of the theory. Let us briefly recall how this works. First we compute the conjugate momenta of every field, $\Pi_i(x)=\d\lag/\d(\d_0\phi_i)$, then we perform the Legendre transform of the Lagrangian with respect to the velocities $\dot{\phi_i}=\d_0\phi_i$ and finally we invert the definition of canonical momenta to express the velocities in terms of them, {\it i.e.} $\dot\phi = \dot \phi(\phi, \Pi)$.
The resulting object is the Hamiltonian (density),
\begin{equation}
  \mathcal{H}(\phi,\Pi) 
                 = \sum_i \Pi_i\dot\phi_i -\lag(\phi,\dot\phi)\,.
\end{equation}
Once the theory is written in terms of fields and conjugate momenta, the canonical quantization proceeds as follows:
\begin{enumerate}
\item
Promote the fields and canonical momenta to {\em operators} acting on a certain Hilbert space.

\item 
Impose the canonical {\em quantization rules}. These are either commutation or anticommutation relations between the fields and their conjugate momenta at equal times. For example,
\eq{
[\phi(t,\vec{x}),\Pi_\phi(t,\vec{y})] &= \ii\delta^3(\vec{x}-\vec{y})\,,
\quad
[\phi(t,\vec{x}),\phi(t,\vec{y})] = [\Pi_\phi(t,\vec{x}),\Pi_\phi(t,\vec{y})] = 0
\\
\{\psi(t,\vec{x}),\Pi_\psi(t,\vec{y})\} &= \ii\delta^3(\vec{x}-\vec{y})\,,
\quad
\{\psi(t,\vec{x}),\psi(t,\vec{y})\} = \{\Pi_\psi(t,\vec{x}),\Pi_\psi(t,\vec{y})\} = 0
\,.
}
Which rules must be imposed depends on what is the type of fields (bosonic or fermionic), and correspond to those leading to a Hamiltonian bounded from below, so for a consistent theory one cannot freely choose. For instance, in the case of the Dirac fermion field we have to use anticommutators for the fields \eqref{eq:freefield}, and then
\eq{
\{a_{\vec{p},r},a^\dagger_{\vec{k},s}\}=\{b_{\vec{p},r},b^\dagger_{\vec{k},s}\}=(2\pi)^3\delta^3(\vec{p}-\vec{k})\delta_{rs}\ , \quad
{}\{a_{\vec{p},r},a_{\vec{k},s}\}=\dots=0.
\label{eq:anticom}
}
As a consequence, $a_{\vec{p},s}$, $b_{\vec{p},s}$ (and their Hermitian adjoints) become operators that annihilate (create) fermionic modes of well-defined momentum $\vec{p}$, mass $m$ and spin component $s$ on the Fock space of multiparticle states, that we call particles and antiparticles, respectively. The vacuum $\ket{0}$ is defined by $a_{\vec{p},s}\ket{0}=b_{\vec{p},s}\ket{0}=0$; the states with one particle or antiparticle (conveniently normalized) are given by
\eq{
\mbox{one particle} \equiv \sqrt{2E_{\vec{p}}}\, a^\dagger_{\vec{p},s}\ket{0}\,,\quad
\mbox{one antiparticle} \equiv \sqrt{2E_{\vec{p}}}\, b^\dagger_{\vec{p},s}\ket{0}\,,
}
and general fermionic multiparticle states are proportional to $a^\dagger_{\vec{p_1},s_1}a^\dagger_{\vec{p_2},s_2}\cdots
b^\dagger_{\vec{q_1},r_1}b^\dagger_{\vec{q_2},r_2}\cdots\ket{0}$. Hence they are antisymmetric under the exchange of any pair (or symmetric if they were bosons) enforced by the quantization rules. This way we obtain the spin-statistics connection.

\item 
Apply {\em normal ordering} to the Hamiltonian (and any other observable made of fields): move all creation operators to the left of annihilation operators, adding a minus sign each time you exchange the position of any annihilation or creation operator if they are fermionic. This is a prescription that subtracts the infinite contribution of the vacuum to the expectation value of the energy of the system (renormalization). In fact, after some algebra we find that the Hamiltonian of Dirac fields would be
\eq{
\int\dd^3 x\, \H(x)
  =\int\frac{\dd^3 p}{(2\pi)^3} E_{\vec{p}} \sum_{s=1,2} (a^\dagger_{\vec{p},s}a_{\vec{p},s} - b_{\vec{p},s}b^\dagger_{\vec{p},s})
}
and from \eqref{eq:anticom} we may write $-b_{\vec{p},s}b^\dagger_{\vec{p},s} = b^\dagger_{\vec{p},s}b_{\vec{p},s}+V$ where $V=\dis\lim_{\vec{k}\to\vec{p}}(2\pi)^3\delta^3(\vec{p}-\vec{k})$ is the infinite volume of the system, that gives not only an infinite vacuum (zero-point) energy but also an infinite vacuum energy density. After normal ordering ($:\!\!b_{\vec{p},s}b^\dagger_{\vec{k},r}\!\!:\,= -b^\dagger_{\vec{k},r}b_{\vec{p},s}$) the Hamiltonian reads
\begin{equation}
H = \int\dd^3 x :\H(x):\;
  =\int\frac{\dd^3 p}{(2\pi)^3} E_{\vec{p}} \sum_{s=1,2} (a^\dagger_{\vec{p},s}a_{\vec{p},s} + b^\dagger_{\vec{p},s}b_{\vec{p},s})\,.
\end{equation}
We see that both particles and antiparticles contribute {\em positively} to the energy:
\eq{
H\; a^\dagger_{\vec{k},r}\ket{0} = E_{\vec{k}} \;a^\dagger_{\vec{k},r}\ket{0}\,,\quad
H\; b^\dagger_{\vec{k},r}\ket{0} = E_{\vec{k}} \;b^\dagger_{\vec{k},r}\ket{0}\,.
}

\end{enumerate}

\subsubsection{One-particle representations}

A particle can be understood as an equivalence class of states connected by Poincar\'e transformations (see e.g. \cite{Schwartz}). In other words, the representation space of the Poincar\'e group contains all the possible states of one particle of the considered type. This is called a {\em one-particle representation} of the Poincar\'e group.
This irreducible representation cannot be an arbitrary one. It must be unitary, so that scalar products between states remain invariant if we choose a different (inertial) observer,
\begin{equation}
\braket{\psi_1}{\psi_2} = \bra{\psi_1} {\cal P}^\dagger{\cal P}\ket{\psi_2}.
\end{equation}
Therefore, Poincar\'e tranformations ${\cal P}$ are represented by unitary operators in this space, and their generators $J^i$ (rotations), $K^i$ (boosts) and $P^\mu$ (translations) by Hermitian operators.

This is actually a bit tricky, because the representation under which our fields (scalars, spinors, vectors) transform is not necessarily unitary. So it is important not to confuse the representation of the fields with the representations of the corresponding one-particle Hilbert spaces.

Rotations form a compact subgroup, hence their finite dimensional irreps are unitary. However the Lorentz group and the Poincar\'e group are non-compact. Therefore, the unitary representations of the Poincar\'e group are infinite-dimensional, or, in other words, the Hilbert space of one-particle states has infinite dimension.

\subsubsubsection{Wigner's classification}

In order to identify the unitary representations of the Poincar\'e group we first need to find the Casimir operators of the group, {\it i.e.} those that commute with all generators. One can check that the two Casimir operators of the Poincar\'e group are 
\begin{equation}
m^2\equiv P_\mu P^\mu\quad \text{and}\quad W_\mu W^\mu\,,
\end{equation}
where 
\begin{equation}
  W^\mu=-\frac{1}{2}\epsilon^{\mu\nu\rho\sigma}J_{\nu\rho}P_\sigma
\end{equation}
is the Pauli-Lubanski vector, which depends on both the translational generators $P_\mu$ and the Lorentz generators $J_{\nu\rho}$.\footnote{
The antisymmetric tensor $J^{\mu\nu}$, where $J^{ij}=\epsilon^{ijk}J^k$ and $J^{0i}=K^i$, includes the generators of rotations and boosts.}
The eigenvalues of $P_\mu P^\mu$ and $W_\mu W^\mu$ label the irreps. Notice that they are Lorentz invariant, so one can choose a convenient frame to explore the features of the different representations. As a result, the unitary irreps of the Poincar\'e group with $P^0>0$ are characterized by just two numbers, the mass $m$ and the spin $j$, leading to Wigner's classification:
\begin{itemize}

\item 
Case $m>0$. Then it is convenient to choose the rest frame, in which $P^\mu=(m,0,0,0)$ and
\begin{equation}
  W_\mu W^\mu = -m^2 j(j+1).
\end{equation}
This corresponds to a {\em massive particle of spin $j$} with $2j+1$ degrees of freedom ($j_3=-j,-j+1,\dots,j$) because after performing a boost that brings any $P^\mu$ to the rest frame, the group of rotations ${\rm SU}(2)$ is the `little group' of transformations that leave invariant this choice of $P_\mu$.

\item 
Case $m=0$. Now there is no rest frame. We are forced to choose a frame in a given direction, say the $\hat{z}$ axis, $P^\mu=(\omega,0,0,\omega)$. Then we get 
\begin{equation}
  W_\mu W^\mu = -\omega^2[(J^1+K^2)^2+(J^2-K^1)^2]\,.
\end{equation}
This corresponds to a {\em massless particle of spin $j$} with just 2 degrees of freedom, independently of the value of $j$, the helicities $h=\pm j$ (projection of the spin in the direction of motion). This is, again a consequence of the residual symmetry we have: in this case we can perform rotations in the plane perpendicular to $P^\mu$ so the little group contains ${\rm SO}(2)$. Actually when $m=0$, the states with $h=+j$ and $h=-j$ belong to two different irreps of the Poincar\'e group, although we say they are the same particle with opposite helicities, {\it e.g.} the left-handed and right-handed photon.
 
\end{itemize}

The embedding of unitary representations of the Poincar\'e group (particles) in a field theory is non trivial. For example the vector field $V_\mu$ describes both spin 0 and spin 1. In order to construct a unitary field theory for massive spin 1 one has to choose carefully the Lagrangian so that the physical theory never excites the spin-0 component. And for massless spin 1 one has also to choose a Lagrangian that only propagates the states with transverse polarization (left and right-handed helicities), which can be done introducing the gauge invariance, a symmetry that is connected to the charge conservation and also to the fundamental interactions as we will discover soon. 

\subsection{Particle physics with quantum fields}

\subsubsection{S-matrix elements and the LSZ formula}

The purpose of physics is to make predictions of quatities that are measurable. The observables in particle physics, essentially cross-sections and decay widths, are written in terms of S-matrix elements for scattering processes, which express the probability amplitude that they occur. 
Taking just one type of scalar fields for simplicity, the S-matrix element
\begin{equation}
{}_{\rm out}\braket{\vec{p}_1\vec{p}_2\cdots\vec{p}_{n_f}}{\vec{k}_1\vec{k}_2\cdots\vec{k}_{n_i}}_{\rm in}\label{Smat}
\end{equation}
describes a scattering process involving identical spin-0 particles, where $n_i$ of them 
are incoming with momenta $\{\vec{k}_i\}$ and $n_f$ are outgoing with momenta $\{\vec{p}_i\}$.

In the following, we will make contact with the quantum field formulation and show the connection between particle scattering amplitudes like \eqref{Smat} and expectation values of interacting fields in the vacuum. Note that only {\em free} fields are related to particles and antiparticles through creation and annihilation operators, as in \eqref{eq:freefield}. However, particles are no longer free during scattering. Nonetheless, it is physically reasonable to assume that fields are {\em asymptotically} free long time before and after the interaction, so one expects
\begin{equation}
\phi(x) \xrightarrow[t\to-\infty]{} Z_\phi^{1/2}\,\phi_{\rm in}(x)\,, \quad
\phi(x) \xrightarrow[t\to+\infty]{} Z_\phi^{1/2}\,\phi_{\rm out}(x)\,,
\end{equation}
where $\phi(x)$ represents an interacting field and $\phi_{\rm in}(x)$ and $\phi_{\rm out}(x)$ are asymptotic incoming and outcoming free fields, as in \eqref{eq:freefield}. The constant $Z_\phi$ that we have introduced (wave function renormalization) will acquire a meaning when we interpret the output of our calculations at the quantum level.

Perhaps one of the most important results in QFT is the Lehmann-Symanzik-Zimmermann (LSZ) reduction formula, that relates S-matrix elements with the (Fourier transform of) vacuum expectation values of time-ordered field products (correlators). For scalar fields, the LSZ formula reads
\begin{align}
\int\!\left(\prod_{i=1}^{n_i}\dd^4x_i\ \e^{-\ii k_i\cdot x_i}\right) 
\int\!\left(\prod_{j=1}^{n_f}\dd^4y_j\ \e^{+\ii p_j\cdot y_j}\right) &\;
\bra{0}T\{\phi(x_1)\cdots\phi(x_{n_i})\phi(y_1)\cdots\phi(y_{n_f})\}\ket{0}
\nn\\
=
\left(\prod_{i=1}^{n_i} \frac{\ii\sqrt{Z_\phi}}{k_i^2-m^2}\right) 
\left(\prod_{j=1}^{n_f}\frac{\ii\sqrt{Z_\phi}}{p_j^2-m^2}\right)
&\;
{}_{\rm out}\braket{\vec{p}_1\vec{p}_2\cdots\vec{p}_{n_f}}{\vec{k}_1\vec{k}_2\cdots\vec{k}_{n_i}}_{\rm in} + \dots
\label{eq:LSZ}
\end{align}
where we have introduced the time-ordered product, defined as
\eq{
T\{\phi(x)\phi(y)\}=\left\{\ba{l}
\phi(x)\phi(y)\,,\quad x^0>y^0 \\
\phi(y)\phi(x)\,,\quad x^0<y^0
\ea\right.\,,
}
and the dots stand for terms less singular in the limit of on-shell momenta. Physical particles, {\it i.e.} asymptotic states, are on-shell ($p^2-m^2=0$). Observe that 
the S-matrix element can be read from the residues of the correlators in \eqref{eq:LSZ}.

In the fermion case, a minus sign must be added each time the position of two adjacent fermionic fields is exchanged. The correlators are in fact Green's functions of certain operators, as we will show in section~\ref{sec:propagators} for some particular cases. We will refer to them as $(n_i+n_f)$-point functions $G(\vec{p}_1\cdots\vec{p}_{n_f};\vec{k}_1\cdots\vec{k}_{n_i})$ in momentum space.

\subsubsection{Perturbation theory}

\subsubsubsection{Feynman diagrams}

The next question is how to compute the correlators of interacting fields. The problem is solved in the interaction picture, where the correlators of interacting fields, 	$\phi(x)$, can be expressed in terms of fields in the interaction picture, $\phi_I(x)$, that evolve as free fields. Then
\begin{equation}
\bra{0}T\{\phi(x_1)\cdots\phi(x_n)\}\ket{0} =
\frac{\bra{0}T\left\{\phi_I(x_1)\cdots\phi_I(x_n)\exp\left[-\ii\dis\int\dd^4x\ {\cal H}_I(x)\right]\right\}\ket{0}}{\bra{0}T\left\{\exp\left[-\ii\dis\int\dd^4x\ {\cal H}_I(x)\right]\right\}\ket{0}}.
\end{equation}
Here ${\cal H}_I$ is the interaction Hamiltonian, ${\cal H}_{\rm int}={\cal H}-{\cal H}_0$, in the interaction picture, whose functional dependence on $\phi_I$ is the same as that of ${\cal H}_{\rm int}$ on $\phi$.
The expression above admits a perturbative treatment by Taylor expanding the exponential. In practice, we can compute every correlator by summing over products of all possible `contractions' of two fields (Wick's theorem):
\begin{equation}
\text{contraction} \equiv
\contraction{}{\phi}{(x_i)}{\phi}\phi_I(x)\phi_I(y) = 
D_F(x-y) =
\bra{0}T\{\phi_I(x)\phi_I(y)\}\ket{0} 
\equiv \text{Feynman propagator}.
\end{equation}
The use of Feynman diagrams and the implementation of Feynman rules provide a systematic method to organize and compute the Green's functions perturbatively in terms of propagators and interaction vertices.

At this point it is important to remark that the functional quantization based on path integrals provides an alternative method to compute the correlators. In the case of scalar fields, one has
\begin{align}
\bra{0}T\{\hat\phi(x_1)\cdots\hat\phi(x_n)\}\ket{0} =
\frac{\int{\cal D}\phi\;\phi(x_1)\cdots\phi(x_n)\;{\rm e}^{\ii S[\phi]}}{\int{\cal D}\phi\;{\rm e}^{\ii S[\phi]}}\,.
\end{align}
where $S[\phi]$ is the action \eqref{eq:action}. On the left-hand side we have written $\hat\phi$ with a hat to emphasize that these are operators. The right-hand side 
is now a functional integral that can be computed perturbatively or using numerical methods and does not involve operators but just functions $\phi(x)$, which makes more obscure the particle interpretation. This expression is particularly useful for gauge theories and is also the starting point for the computation of non-perturbative effects in lattice field theory.

\subsubsubsection{Propagators and causality}

One of the central notions in theoretical physics is the principle of causality. In a relativistic theory that preserves causality, two arbitrary events  
separated by a spacelike interval cannot influence each other. In particular, information cannot travel outside the lightcone, namely faster than light. In a QFT with a scalar field, causality requires that the commutator $[\phi(x),\phi^\dagger(y)]=0$ if $(x-y)^2<0$. Next we show that, for this to be possible, both particles and antiparticles are needed.

A free scalar field is a combination of positive and negative energy waves:
\begin{equation}
\phi(x) =
\int \frac{\dd^3 p}{(2\pi)^3\sqrt{E_{\vec{p}}}} \left( 
  a_{\vec{p}}{\rm e}^{-\ii p\cdot x}
+ b^\dagger_{\vec{p}}{\rm e}^{\ii p\cdot x} \right)\,.
\end{equation}
From the commutation relations of creation and annihilation operators in the quantum field one gets
\begin{align}
[\phi(x),\phi^\dagger(y)]
&=\int \frac{\dd^3 p}{(2\pi)^3\sqrt{E_{\vec{p}}}}
 \int \frac{\dd^3\vec{q}}{(2\pi)^3\sqrt{E_{\vec{q}}}}
 \big(
 {\rm e}^{-\ii(p\cdot x-q\cdot y)}[a_{\vec{p}},a^\dagger_{\vec{q}}] +
 {\rm e}^{\ii(p\cdot x-q\cdot y)}[b^\dagger_{\vec{p}},b_{\vec{q}}]
 \big)
\nonumber\\
&= \Delta(x-y)-\Delta(y-x)\,,
\end{align}
where the first contribution comes from particles and the other from antiparticles, and
\begin{equation}
\Delta(x-y)= \int \frac{\dd^3 p}{(2\pi)^3 E_{\vec{p}}}{\rm e}^{-\ii p\cdot(x-y)}.
\end{equation}
When the interval is spacelike, $(x-y)^2<0$, it is always possible to choose a frame where $x-y \equiv (0,\vec{r})$ and then
\begin{equation}
\Delta(x-y)=\Delta(y-x) \propto \frac{m}{r}{\rm e}^{-mr} \ne 0 \quad
\mbox{for } mr\gg 1.
\end{equation}
Therefore, if there were only particles $[\phi(x),\phi^\dagger(y)] = \Delta(x-y) \ne 0$ and causality would be violated. However, thanks to the existence of {\em both} particles and antiparticles, causality is preserved because then $[\phi(x),\phi^\dagger(y)] = \Delta(x-y) - \Delta(y-x) = 0$ outside the lightcone.

In fact one can see that the Feynman propagator of a complex field is made of two pieces,
\begin{equation}
D_F(x-y)=\bra{0}T\{\phi(x)\phi^\dagger(y)\}\ket{0} =
\theta(x^0-y^0)\Delta(x-y) + \theta(y^0-x^0)\Delta(y-x)
\label{eq:FP}
\end{equation}
that have a clear physical interpretation. The first term contributes only if $x^0>y^0$ and is the probability amplitude that a particle created in $y$ propagates to $x$ where it is annihilated. The second one contributes only if $y^0>x^0$ and gives the probability amplitude for the propagation of an antiparticle from $x$ to $y$. It is useful to express the Feynman propagator in a compact form, that reveals its Fourier transform, as 
\begin{equation}
D_F(x-y)=\int\frac{\dd^4 p}{(2\pi)^4} \frac{\ii}{p^2-m^2+\ii\varepsilon}{\rm e}^{-\ii p\cdot(x-y)}\,,
\end{equation}
where the (Feynman) prescription $\varepsilon\to0^+$ (often omitted) slightly displaces the poles from the real $p^0$ axis to $p^0=\pm E_{\vec{p}}(1-\ii\varepsilon/2E^2_{\vec{p}})$ so that the integral on the complex $p^0$ plane reproduces the causal time ordering of \eqref{eq:FP}. The Feynman propagator may also be derived using the path integral formulation in which the above prescription ensures the convergence of the integrals.

\subsubsubsection{Rewriting the LSZ formula}

Interactions introduce quantum corrections to the propagator that can be resummed in momentum space as follows:
\begin{align}
\raisebox{2mm}{\includegraphics[scale=0.5]{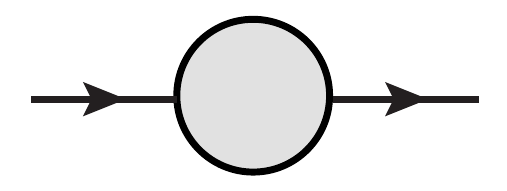}}&
\raisebox{5.5mm}{\boldmath$=$}
\raisebox{5mm}{\includegraphics[scale=0.5]{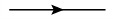}}
\raisebox{5.5mm}{\boldmath$+$}
\raisebox{1.5mm}{\includegraphics[scale=0.5]{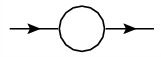}\put(-48,11){\mbox{1PI}}}
\raisebox{5.5mm}{\boldmath$+$}
\raisebox{1.5mm}{\includegraphics[scale=0.5]{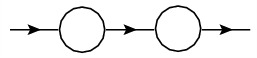}\put(-48,11){\mbox{1PI}}\put(-96,11){\mbox{1PI}}}
\raisebox{5.5mm}{\boldmath$+\ \cdots$}
\nonumber\\[-1ex]
&=\frac{\ii}{p^2-m_0^2}+\frac{\ii}{p^2-m_0^2}[-\ii M^2(p^2)]\frac{\ii}{p^2-m_0^2}+\dots
\nonumber\\
&=\frac{\ii}{p^2-m_0^2-M^2(p^2)}\,,
\end{align}
where 1PI stands for one-particle irreducible diagrams (those that cannot be split into two by cutting a single internal line), whose contribution is encoded in the `self-energy' $M^2(p^2)$, and $m_0$ is here the mass parameter in the Lagrangian ({\em bare} mass). We can now perform a Taylor expansion about $p^2=m^2$, where $m$ is the {\it physical} pole mass,
\eq{
p^2-m_0^2-M^2(p^2)
=(p^2-m^2)\left(1-\left.\frac{\dd M^2}{\dd p^2}\right|_{p^2=m^2}\right)\quad
}
that yields
\eq{
\raisebox{-3.5mm}{\includegraphics[scale=0.5]{FIGS/Dsum.pdf}}=
\frac{\ii Z_\phi}{p^2-m^2}+\mbox{regular near }p^2=m^2,
}
with
\begin{equation}
 m^2=m_0^2+M^2(m^2)\ , \qquad
Z_\phi=\left(1-\left.\frac{\dd M^2}{\dd p^2}\right|_{p^2=m^2}\right)^{-1}.
\end{equation}
Notice that $Z_\phi$ is the residue of the propagator that accounts for the field or wave function renormalization due corrections induced by the interactions.

The previous result allows to factor out external legs from `amputated' diagrams, obtained by cutting the external propagators to the connected diagrams (only connected diagrams contribute to S-matrix elements),
\begin{align}
G(\vec{p}_1\cdots\vec{p}_{n_f};\vec{k}_1\cdots\vec{k}_{n_i}) = 
\raisebox{-.45\height}{
\includegraphics[width=35mm]{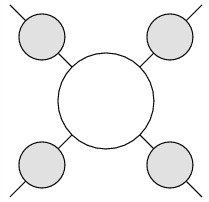}
\put(-64,45){amptd}
\put(-16,55){$\vdots$}
\put(-84,55){$\vdots$}
\put(-16,32){$\vdots$}
\put(-84,32){$\vdots$}
\put(0,90){$p_1$}
\put(0,1){$p_{n_f}$}
\put(-110,90){$k_1$}
\put(-110,1){$k_{n_i}$}
}
\end{align}
This way one can explicitly cancel the poles in both sides of equation~\eqref{eq:LSZ} and rewrite the LSZ formula in a simpler form:
\begin{align}
{}_{\rm out}\braket{\vec{p}_1\vec{p}_2\cdots\vec{p}_{n_f}}{\vec{k}_1\vec{k}_2\cdots\vec{k}_{n_i}}_{\rm in} = \left(\sqrt{Z}\right)^{n_i+n_f}\raisebox{18mm}{\includegraphics[width=35mm,angle=-90]{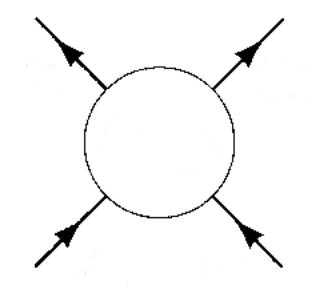}
\put(-57,-50){amptd}
\put(-10,-30){$\vdots$}
\put(-80,-30){$\vdots$}
\put(-10,-78){$\vdots$}
\put(-80,-78){$\vdots$}
\put(0,-10){$p_1$}
\put(0,-94){$p_{n_f}$}
\put(-96,-10){$k_1$}
\put(-96,-94){$k_{n_i}$}
}
\equiv (2\pi)^4\delta^4\left(\sum_{i=1}^{n_i} k_i - \sum_{j=1}^{n_f} p_j\right)\ii{\cal M}
\,.
\end{align}
In the last equality the usual Lorentz-invariant amplitude ${\cal M}$ has been introduced, extracting a prefactor that expresses the 4-momentum conservation in the process. The external momenta are on-shell.

\subsubsubsection{Renormalization}

We already have all the ingredients needed to compute scattering amplitudes, but there is still one problem to overcome: Feynman rules instruct us to integrate over loop momenta, which results very often in ultraviolet divergent expressions.\footnote{The perturbative expansion in powers of a coupling constant in $\lag_{\rm int}$ is at the same time an expansion in powers of $\hbar$ (quantum corrections) and an expansion in the number of loops, given by the number of momenta of internal lines of a diagram that are not fixed by momentum conservation at every vertex.} In order to make sense of these divergences we must `renormalize' the theory. The idea is the following. One assumes that fields and parameters in the original (classical) Lagrangian ({\it bare}) must be replaced by new ones ({\it renormalized}) in such a way that predictions for physical observables  are finite at any given order in perturbation theory when expressed in terms of renormalized fields and couplings. In fact these redefinitions are needed even if the corrections are finite. In order to isolate and manipulate the divergences one has first to `regularize' the integrals using a parameter (regulator) that in a certain limit leads to the divergent result, like a cut-off to the modulus of the loop momentum $\Lambda\to\infty$, or the number of dimensions $d=4-\epsilon$ with $\epsilon\to0$ in dimensional regularization. Keep in mind that fields and parameters (couplings and masses) are not observables but ancillary objects in terms of which the actual observables (cross-sections, decay widths) are expressed.

\begin{figure}
\centering
\includegraphics[scale=0.6]{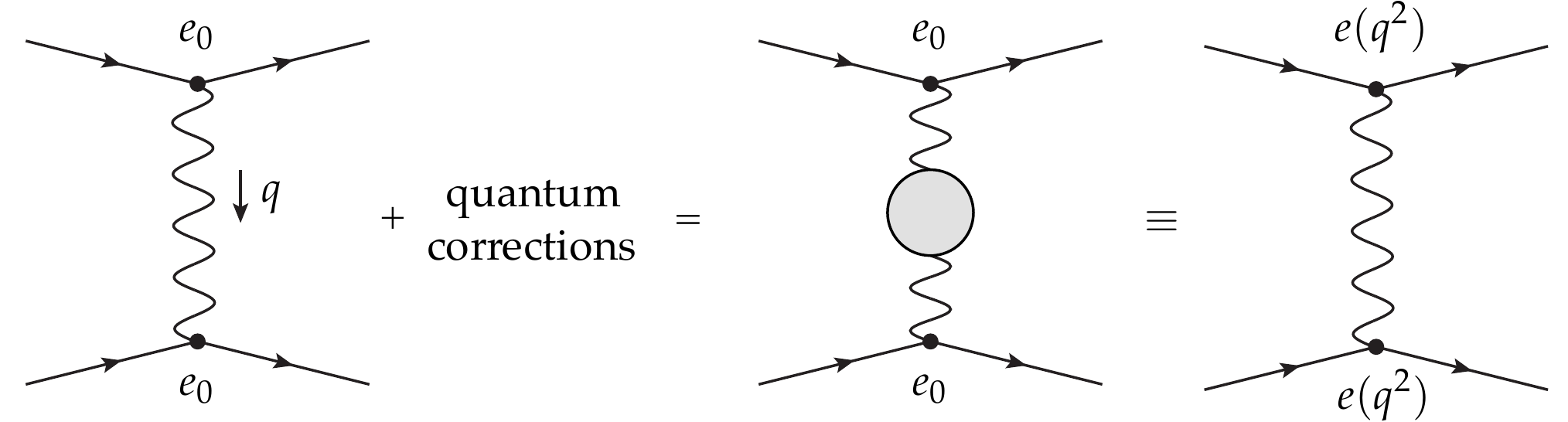}
\caption{The running of the electromagnetic coupling constant from quantum corrections.\label{fig:chargerenorm}}
\end{figure}

An important consequence of the renormalization program is the {\em running} of the renormalized coupling `constants': the value of the couplings acquires a dependence with the scale at which they are extracted from experiment. Let us illustrate this with an example. Consider the determination of the electromagnetic coupling from the scattering of two charged particles. If we include loop corrections (fig.~\ref{fig:chargerenorm}) the amplitude is proportional to 
\eq{
  \frac{e_0^2}{1-\Pi(q^2)} 
= \frac{e_R^2}{1-[\Pi(q^2)-\Pi(0)]} 
\equiv e^2(q^2)
}
where $e_0$ is the bare coupling, $\Pi(q^2)$ is a (divergent) quantum correction to the photon propagator, higher orders are neglected and $q^2$ is the momentum transfer in the process. To make sense of this result we have introduced the renormalized coupling $e_R$ with $e_0\equiv Z_A ^{-1/2} e_R$. The renormalization constant $Z_A=[1-\Pi(0)]^{-1}$ is the residue of the $q^2=0$ pole of the photon propagator.\footnote{Actually the renormalization of the coupling is
$e_0\equiv Z_e e_R$. However the preservation of the gauge symmetry at the quantum level implies $Z_e Z_\psi Z_A^{1/2} = Z_\psi \Raw Z_e = Z_A^{-1/2}$ (Ward-Takahashi identity). Therefore the redefinition of the coupling is universal, independent of the type of fermions.} Comparing with the experiment we extract the effective coupling $e^2(q^2)$, a running coupling, that must be finite but depends on the `renormalization scale' $q^2$ where it has been measured. In fact, the renormalized coupling $e_R= e(0)$.

\subsection{Global symmetries and gauge invariance}

\subsubsection{Internal symmetries and the gauge principle}

\subsubsubsection{The free Lagrangian}

The free Lagrangian for the Dirac field \eqref{eq:lagD} is invariant under spacetime (Poincar\'e) symmetries {\em and} also under `internal' symmetries, acting only on the fields, not changing the spacetime coordinates. They consist of {\em global} ${\rm U}(1)$ phase transformations,
\begin{equation}
\psi(x) \mapsto \e^{-\ii Q\theta}\psi(x)
\end{equation}
where $Q$ and $\theta$ are real constants. Then, as a consequence of Noether's theorem, there must exist a divergentless current and a conserved charge associated to this continuos symmetry,
\begin{equation}
\d_\mu {\cal J}^\mu = 0\,,\quad \d_t {\cal Q} = 0\quad\mbox{with}\quad{\cal Q} = \int\dd^3 x\ {\cal J}^0\,.
\end{equation}
The Noether's current corresponding to the U(1) invariance of the Lagrangian is
\begin{equation}
{\cal J}^\mu = Q\ \overline\psi\gamma^\mu\psi\,.
\label{eq:U1current}
\end{equation}
After field quantization, the conserved charge becomes an operator on Fock space,
\eq{
{\cal Q}\ = Q\int\dd^3 x\ :\overline\psi\gamma^0\psi: 
   \ = Q\int\frac{\dd^3 p}{(2\pi)^3}\ \sum_{s=1,2} \left(
        a^\dagger_{\vec{p},s}a_{\vec{p},s}
       -b^\dagger_{\vec{p},s}b_{\vec{p},s}\right)\,,
}
where we have applied the normal ordering prescription for fermionic operators.
From this we can easily check that particles and antiparticles carry opposite charges $\pm Q$:
\begin{equation}
{\cal Q}\; a^\dagger_{\vec{k},r}\ket{0} = +Q\; a^\dagger_{\vec{k},s}\ket{0}
\mbox{ (particle)}\ , \quad
{\cal Q}\; b^\dagger_{\vec{k},s}\ket{0} = -Q\; b^\dagger_{\vec{k},s}\ket{0}
\mbox{ (antiparticle)}.
\end{equation}

\subsubsubsection{Gauge invariance dictates interactions}

It is evident that the free Lagrangian is not invariant under phase transformations where  $\theta=\theta(x)$, different for every spacetime point,
\begin{equation}
\psi(x) \mapsto \e^{-\ii Q\theta(x)}\psi(x)\,.
\label{eq:gaugeU1}
\end{equation}
In order to impose that physics is invariant under these {\em local} or {\em gauge} U(1) transformations it is enough to perform the minimal substitution
\eq{
\d_\mu \to D_\mu = \d_\mu + \ii e Q A_\mu
\label{eq:U1covder}
}
that introduces a {\em gauge field} $A_\mu(x)$ transforming as
\begin{equation}
A_\mu(x) \mapsto A_\mu(x) + \frac{1}{e}\d_\mu\theta(x)\,.\label{eq:gaugeU12}
\end{equation}
This basically ensures that $D_\mu\psi \mapsto \e^{-\ii Q\theta(x)} D_\mu\psi$, so it transforms the same as $\psi$, hence the name of {\em covariant derivative}.
 
\boxexercise{1 (a)}{%
Show that the $\overline\psi\slashed{D}\psi$ is invariant under gauge transformations \eqref{eq:gaugeU1} -- \eqref{eq:gaugeU12}.
}

The outcome of this replacement is the generation of an interaction between $\psi$ and $A_\mu$ given by the scalar product of the conserved current \eqref{eq:U1current} and the gauge field,
\begin{equation}
\lag_{\rm int} 
= -e\ Q\ \overline\psi\gamma^\mu\psi A_\mu
= -e {\cal J}^\mu A_\mu\,,
\end{equation}
which is proportional to the coupling constant $e$ (a property of the gauge field) and the conserved charge $Q$ (a property of the fermion field). 

Finally we can provide dynamics for the vector field we have introduced without spoiling gauge invariance. This is achieved by adding the following kinetic term,
\begin{equation}
\lag_{G} = -\dis\frac{1}{4} F_{\mu\nu}F^{\mu\nu}, 
\end{equation}
where $F_{\mu\nu}=\d_\mu A_\nu-\d_\nu A_\mu$ is a gauge invariant antisymmetric tensor, that has the same form as the electromagnetic tensor. In fact, applying the Euler-Lagrange equations for $A_\mu$ to the full invariant Lagrangian ($\lag_0 + \lag_{\rm int} + \lag_{G}$),
\begin{equation}
\lag_{\text{inv}}  = \overline\psi(\ii\slashed{D}-m)\psi
                 -\dis\frac{1}{4}F_{\mu\nu}F^{\mu\nu}\,,
\end{equation}
one obtains precisely the Maxwell's equations, $\d_\mu F^{\mu\nu} = e{\cal J}^\mu$, where $F^{\mu\nu}$ is the electromagnetic field strength and the four-current ${\cal J}^\mu=(\rho,\vec{j})$ includes the electric charge-density and charge-current in units of $e$.

Similarly, one gets gauge interactions for a complex scalar field $\phi$, whose free Lagrangian is invariant under global U(1) phase transformations,
\eq{
\phi(x) \mapsto \e^{-\ii Q\theta(x)}\phi(x)\,,
}
by substituting the covariant derivative \eqref{eq:U1covder}, which results in the gauge invariant Langrangian:
\begin{equation}
\lag_{\text{inv}} = (D_\mu\phi)^\dagger D^\mu\phi-m^2\phi^\dagger\phi
                -\lambda(\phi^\dagger\phi)^2
                -\dis\frac{1}{4}F_{\mu\nu}F^{\mu\nu}\,.
\end{equation}

We have seen that making local a global symmetry requires the existence of interactions, of a type that is determined by the symmetry. This way of introducing the interactions is known as the {\em gauge principle}. 

A very important comment is here in order. We often call {\em gauge symmetry} a
local transformation of the fields with $\theta=\theta(x)$ that leaves invariant the Lagrangian. Although this gauge invariance implies the existence of a global symmetry, which can be properly called a `symmetry', with physical consequences like charge conservation, a local transformation is {\em not a symmetry} of our system (states are not transformed), but a {\em redundancy} of our description of physics: we can redefine fields at every point of spacetime with no physical consequences. The gauge invariance is necessary to have a local description of massless spin-1 particles (two degrees of freedom) with four-vector fields, which are Lorentz invariant objects with too many polarizations (two of them are spurious). The gauge symmetry is more a gauge {\em freedom}.

\subsubsubsection{The gauge principle in non-Abelian gauge theories}

Next we will apply the gauge principle to a more general symmetry group than just U(1).
A general gauge symmetry group $G$ is a compact $N$-dimensional Lie group, whose elements \begin{equation}
{\rm g} \in G\,, \quad
{\rm g}(\vec{\theta}) = \e^{-\ii T_a\theta^a}\,, \quad
a=1,\dots,N\,,
\end{equation}
are given by a set of real and continuous parameters $\{\theta^a\}$ in terms of $N$ generators $\{T_a\}$ that form the basis of the Lie algebra of the group. The generators $T_a$ of gauge groups are Hermitian if the transformation is unitary, and in general the Lie algebra structure is totally determined by the commutators between the elements of the basis,
\begin{equation}
[T_a,T_b]=\ii f_{abc} T_c\,,
\end{equation}
with $f_{abc}$ the structure constants characterizing the group. The structure constants vanish if and only if the group is Abelian (recall the Baker-Campbell-Hausdorff formula).

The finite-dimensional irreducible representations of a compact Lie group are unitary: the $g(\vec{\theta})$ are represented by unitary $d\times d$ matrices $U(\vec{\theta})$ that are expressed in terms of the corresponding Lie algebra representation of $\{T_a\}$. These matrices act on some $d$-dimensional vector space whose elements are called $d$-multiplets,
\begin{equation}
\Psi(x) \mapsto U(\vec{\theta})\Psi(x)\ , \qquad
\Psi=\pmat{\psi_1\\\vdots\\\psi_d}\,.
\end{equation}
In our context, the multiplet components are fields.

Examples of Lie groups that often appear in quantum field theories are U(1) (Abelian, with $N=1$ generator) and SU($n$), which is the group of $n\times n$ unitary matrices of unit determinant (non-Abelian, with $N=n^2-1$ generators). The unitary irreps of Abelian groups, like U(1), are one-dimensional. Prominent irreps of SU($n$) groups are the `fundamental representation' ($d=n$) and the `adjoint representation' ($d=N$). The elements of the $N$-dimensional matrices representing the generators in the adjoint representation are $(T_a)_{bc} = -\ii f_{abc}$, totally antisymmetric for SU($n$). Let us briefly list the main properties of these groups.
 
\begin{itemize}

\item ${\rm U}(1)$. The only generator is representated by a real number ($Q$), that labels each one-dimensional representation.

\item ${\rm SU}(2)$. It has 3 generators. The structure constants are $f_{abc} = \epsilon_{abc}$ (Levi-Civita symbol). The generators in the fundamental representation ($d=2$) can be chosen $T_a=\tfrac{1}{2}\sigma_a$, the 3 Pauli matrices. The adjoint representation has dimension 3.

\item
${\rm SU}(3)$. It has 8 generators. The totally antisymmetric structure constants are given by $f_{123}=1$, $f_{458}=f_{678}=\frac{\sqrt{3}}{2}$, $f_{147}=f_{156}=f_{246}=f_{247}=f_{345}=-f_{367}=\frac{1}{2}$ and the others not related to these by permuting indices are zero. The generators in the fundamental representation ($d=3$) can be chosen $T_a=\tfrac{1}{2}\lambda_a$, the 8 Gell-Mann matrices. The adjoint representation has dimension 8.

\end{itemize} 

Consider now the free Lagrangian for a fermion field multiplet, 
\eq{
\lag_0 = \overline\Psi(\ii\slashed{\d}-m)\Psi\,,
}
invariant under an $N$-dimensional Lie group $G$ of global transformations,
\eq{
\Psi(x) \mapsto U(\vec{\theta})\Psi(x)\,.
\label{eq:gaugeSU}
}
We can get a Lagrangian invariant under local (gauge) transformations $\vec{\theta} = \vec{\theta}(x)$ by substituting the covariant derivative
\begin{equation}
\d_\mu \to D_\mu = \d_\mu - \ii g \widetilde W_\mu\,, \quad
\widetilde W_\mu \equiv T_a W^a_\mu\,,
\end{equation}
where one gauge field $W^a_\mu(x)$ per generator $T_a$ has been introduced, transforming as
\begin{equation}
\widetilde W_\mu(x) \mapsto 	
U\widetilde W_\mu(x)U^\dagger - \frac{\ii}{g}(\d_\mu U)U^\dagger\,.
\label{eq:gaugeSU2}
\end{equation}
The first term implements the global transformation of a multiplet of $N$ vector fields in the adjoint representation and the second accounts for the local dependence with the spacetime point $x$. Then $D_\mu\Psi \mapsto U D_\mu\Psi$, transforming the same as $\Psi$, just as we need. The choice of sign for the coupling $g$ in the covariant derivative is conventional.

\boxexercise{1 (b)}{
Show that $\overline\Psi\slashed{D}\Psi$ is invariant under gauge transformations \eqref{eq:gaugeSU} -- \eqref{eq:gaugeSU2}.
}

The new Lagrangian contains interactions of fermions in $\Psi$ with every $W^a_\mu$, 
\begin{equation}
\lag_{\rm int} = g\, \overline\Psi\gamma^\mu T_a\Psi\, W^a_\mu
= g\, {\cal J}^\mu_a W^a_\mu\,,
\end{equation}
where each ${\cal J}^\mu_a$ is the Noether's current associated to the invariance of the Lagrangian under the symmetry generated by $T_a$. The strength of the interaction of gauge field $W^a_\mu$ to two fermion fields $\psi_i$ and $\psi_j$ of the $d$-multiplet is proportional to the coupling $g$ and is given by the element $(T_a)_{ij}$ of the corresponding generator in that representation. The fermion charges under group $G$ are eigenvalues of the generators in the given representation. Fermion singlets belong to the trivial one-dimensional representation with $T_a=0$ and hence do not couple to gauge fields.

The next step is to add kinetic terms for the gauge fields respecting the gauge invariance. Interestingly, this cannot be done without introducing at the same time interactions among the gauge fields when the symmetry is non-Abelian. The minimal choice is the Yang-Mills Lagrangian,
\begin{equation}
\lag_{\rm YM}=-\dis\frac{1}{2}{\rm Tr}\left\{\widetilde W_{\mu\nu}\widetilde W^{\mu\nu}\right\} 
= -\frac{1}{4}W_{\mu\nu}^a W^{a\,\mu\nu}\,,
\label{eq:YM0}
\end{equation}
where
\begin{align}
\widetilde W_{\mu\nu} \equiv T_a W^a_{\mu\nu} &\equiv
D_\mu\widetilde W_\nu - D_\nu\widetilde W_\mu
= \d_\mu\widetilde W_\nu - \d_\nu\widetilde W_\mu 
- \ii g [\widetilde W_\mu, \widetilde W_\nu]\,,
\end{align}
from which one derives the field strengths:
\begin{equation}
W^a_{\mu\nu} = \d_\mu W_\nu^a - \d_\nu W_\mu^a + g f_{abc} W_\mu^b W_\nu^c\,.
\label{ex21}
\end{equation}
These are a generalization of $F_{\mu\nu}$ to the non-Abelian case. They transform in the adjoint representation of the gauge group,
\begin{equation}
 \widetilde W_{\mu\nu} \mapsto U \widetilde W_{\mu\nu} U^\dagger\,.
\end{equation}
Note that, besides the kinetic terms, $\lag_{\rm YM}$ contains cubic and quartic self-interactions of gauge fields completely determined by the gauge group properties:
\begin{align}
\lag_{\rm kin} &= 
-\frac{1}{4}(\d_\mu W_\nu^a - \d_\nu W_\mu^a)
            (\d^\mu W^{a\,\nu} - \d^\nu W^{a\,\mu}),\\
\lag_{\rm cubic} &= -\frac{1}{2} g f_{abc}\ (\d_\mu W_\nu^a-\d_\nu W_\mu^a)
W^{b\,\mu} W^{c\,\nu},
\\
\lag_{\rm quartic} &=-\frac{1}{4} g^2 f_{abe} f_{cde}\ W^a_\mu W^b_\nu W^{c\,\mu} W^{d\,\nu}.\label{ex22}
\end{align}
The self-couplings of gauge fields in non-Abelian theories have profound consequences. For instance, in quantum chromodynamics where gluons interact with each other, it is the main reason for confinement (see section \ref{sec:QCD}). 

\boxexercise{2}{Reproduce expressions \eqref{ex21}-\eqref{ex22}.}

\subsubsection{Quantization of gauge theories\label{sec:propagators}}

So far we have discussed only classical gauge theories. The quantization of gauge fields involves a number of subtleties, related to the fact that the quanta (gauge bosons) are massless spin~1 particles with just two degrees of freedom embedded in vector field that has four. In fact, the propagator of the gauge field does not exist! Remember that the Feynman propagator is the basic correlator of the quantum field theory. It is a Green's function for the free equation of motion.
For a scalar field,
\begin{equation}
D_F(x-y)=\bra{0}T\{\phi(x)\phi^\dagger(y)\}\ket{0} =
\int\frac{\dd^4p}{(2\pi)^4}\frac{\ii}{p^2-m^2+\ii\varepsilon}\e^{-\ii p\cdot(x-y)}
\end{equation}
is indeed a Green's function (the analogue to the inverse) of the Klein-Gordon operator
\begin{equation}
(\Box_x+m^2)D_F(x-y)=-\ii\delta^4(x-y) \quad\Leftrightarrow\quad
\widetilde D_F(p) = \frac{\ii}{p^2-m^2+\ii\varepsilon}.
\end{equation}
Similarly, the propagator of a fermion field, 
\begin{equation}
S_F(x-y)=\bra{0}T\{\psi(x)\overline\psi(y)\}\ket{0} =
(\ii\slashed{\d}_x+m)
\int\frac{\dd^4p}{(2\pi)^4}\frac{\ii}{p^2-m^2+\ii\varepsilon}\e^{-\ii p\cdot(x-y)} 
\end{equation}
is a Green's function of the Dirac operator,
\begin{equation}
(\ii\slashed{\d}_x-m)S_F(x-y)=\ii\delta^4(x-y) \quad\Leftrightarrow\quad
\widetilde S_F(p) = \frac{\ii}{\slashed{p}-m+\ii\varepsilon}.
\end{equation}
However, the propagator of a gauge field cannot be defined. For the simpler Abelian case (Maxwell's Lagrangian) the equation of motion is 
\eq{
\frac{\d\lag}{\d A_\nu}-\d_\mu\frac{\d\lag}{\d(\d_\mu A_\nu)} = 0 
\quad\Raw\quad
\d_\mu F^{\mu\nu} = [g^{\mu\nu}\Box-\d^\mu\d^\nu]A_\mu = 0.
}
The Green's function should be the inverse of the differential operator in brackets, but this operator is not invertible because $-k^2 g^{\mu\nu}+k^\mu k^\nu$ is singular (it has a zero eigenvalue, with eigenvector $k_\mu$). The origin of this problem is gauge invariance. The usual solution consists of modifying the Lagrangian adding a gauge-fixing term, $\lag = \lag_{\rm G} + \lag_{\rm GF}$, where in the so called $R_\xi$ gauges
\begin{equation}
\lag_{\rm GF} = -\dis\frac{1}{2\xi}(\d^\mu A_\mu)^2\,.
\label{eq:lagGF}
\end{equation}
The Euler-Lagrange equation of the modified Lagrangian is then
\begin{equation}
\left[g^{\mu\nu}\Box-\left(1- \frac{1}{\xi}\right)\d^\mu\d^\nu\right]A_\mu = 0.
\end{equation}
The propagator can now be computed and in momentum space is given by
\begin{equation}
-k^2g^{\mu\nu}+\left(1-\frac{1}{\xi}\right)k^\mu k^\nu
\quad\xrightarrow{\text{inverse}}\quad
\widetilde D_{\mu\nu}(k)=\frac{\ii}{k^2+\ii\varepsilon}\left[- g_{\mu\nu}+(1-\xi)\frac{k_\mu k_\nu}{k^2}\right]\,,
\end{equation}
where we have introduced the Feynman $\varepsilon$-prescription.

Let us justify this procedure, which is more transparent in functional quantization. The gauge invariance of Maxwell's Lagrangian under field transformations,
\eq{
A_\mu(x)\mapsto A_\mu(x) + \d_\mu\Omega(x)\,,
\label{eq:gaugetrans}
}
implies that $A_\mu$ provides a redundant description of the electromagnetic field, because any four-vector in the same `gauge orbit' leads to the same physics. As a consequence, we would be overcounting (infinite) equivalent configurations in the path integral, that leads to divergent Green's functions. To prevent this we impose a gauge condition,
\eq{
F[A_\mu]=0\,,
} 
and integrate only over one representative of each equivalence class by replacing
\eq{
\int \D A_\mu \to
\int \D A_\mu\, \delta(F[A_\mu])\det M\,,
\label{eq:FP1}
}
where we have introduced a Dirac delta function and the corresponding Jacobian with
\eq{
M(x,y) = \left.\frac{\delta F[A_\mu(x)]}{\delta\Omega(y)}\right|_{F[A_\mu]=0}\,.
}
Consider now a class of gauge conditions of the form
\eq{
F[A_\mu]-C(x) = 0\,,
\label{eq:FP2}
}
where $C(x)$ is an arbitrary function independent of $A_\mu$. Note that $\det M$ is independent of $C(x)$. We can use this to replace the delta function in \eqref{eq:FP1} by a functional of $F$ obtained by averaging over $C$ with an arbitrary functional $G[C]$ as follows,
\eq{
\int\D A_\mu\,\det M\int\D C\,\delta[F[A_\mu]-C(x)]G[C] 
=\int\D A_\mu\, G[F[A_\mu]]\, \det M\,,
\label{eq:FP3}
}
where an irrelevant normalization factor has been ignored. It is convenient to take
\eq{
G[C] = \exp\left\{-\frac{\ii}{2\xi}\int\dd^4x\, C^2(x)\right\}\ .
\label{eq:FP4}
}
Note that up to a normalization factor one has
\eq{
G[C] \xrightarrow[\xi\to0]{} \delta[C]\ .
}
Next comes a mathematical trick (see {\em e.g.} \cite{Pokorski}): we may turn the determinant into a functional integral over Grassmann variables $\eta(x)$ and $\bar\eta(x)$,
\eq{
\det M =& \int\D\bar\eta\,\D\eta\,\exp\left\{-\ii\int\dd^4x\,\bar\eta M\eta\right\}
\,.
\label{eq:FP5}
}
Putting together \eqref{eq:FP3}, \eqref{eq:FP4} and \eqref{eq:FP5} we see that the implementation of the gauge condition \eqref{eq:FP1} is equivalent to the replacement
\eq{
\int\D A_\mu\,\exp\left\{\ii\int\dd^4x\,\lag_{\rm G}\right\} \to
\int\D A_\mu\D\bar\eta\,\D\eta\,\exp\left\{\ii\int\dd^4x\,\lag\right\} 
}
where $\lag =\lag_{\rm G} + \lag_{\rm GF} + \lag_{\rm FP}$ includes a generic gauge-fixing term
\eq{
\lag_{\rm GF} = -\frac{1}{2\xi}F^2[A_\mu]
\label{eq:lagGF2}
}
and the so-called Faddeev-Popov term
\eq{
\lag_{\rm FP} = -\bar\eta M\eta.
}
The (auxiliary) Faddeev-Popov fields $\eta(x)$ and $\bar\eta(x)$ are anticommuting scalar fields, without spinor indices. They are unphysical, they never appear as external fields but only in internal lines of Feynman diagrams.\footnote{Because of their anticommuting character, a close loop of Faddeev-Popov ghosts adds a minus sign to the diagram, the same as closed fermion loops.}

In an Abelian gauge theory, like quantum electrodynamics, the FP ghosts can be ignored because they do not couple to gauge fields under the usual gauge condition. Their contribution only changes the normalization of the path integral. In this case it is enough to add the gauge-fixing term \eqref{eq:lagGF2}, that comes from \eqref{eq:FP4} using the Lorenz condition 
\eq{
F[A_\mu]=\d^\mu A_\mu\,. 
}
This leads to the $R_\xi$ gauges, a generalization of the Lorenz gauge. 

However, in a non-Abelian theory, like the electroweak standard model or quantum chromodynamics, FP ghosts are usually needed and will contribute in loops. An exception is the axial gauge, that in turn would lead to more complicated gauge boson propagators. To see this, consider the Yang-Mills Lagrangian \eqref{eq:YM0}. From \eqref{eq:gaugeSU2}, the $N$ gauge fields $\{W_\mu^a\}$ transform under $U=\exp\{-\ii T_a\theta^a\}$ as
\eq{
W_\mu^a &\mapsto W_\mu^a - f_{abc} W_\mu^b \theta^c - \frac{1}{g}\d_\mu\theta^a
}
which coincides with the Abelian case (\ref{eq:gaugetrans}) when $f_{abc}=0$, $W_\mu^a\to W_\mu$ and $\theta^a\to -g\Omega$. The gauge condition
\eq{
F[W_\mu^a] = 0\ , \quad\forall a=1,\dots,N
}
in an $R_\xi$ gauge has the form
\eq{
F[W_\mu^a] = \d^\mu W_\mu^a \quad\Raw\quad \lag_{\rm GF} = -\sum_a\frac{1}{2\xi_a}(\d^\mu W_\mu^a)^2\,.
\label{eq:lagGF3}
}
Now the Faddeev-Popov determinant does not vanish because
\eq{
F[W_\mu^a] \mapsto F[W_\mu^a] - f_{abc}\d^\mu(W_\mu^b\theta^c) - \frac{1}{g}\Box\theta^a\,,
}
that implies
\eq{
M_{ab} = \left.\frac{\delta F[W_\mu^a]}{\delta\theta^b}\right|_{F[W_\mu^a] = 0}
= -\frac{1}{g}\left(
\delta_{ab}\Box  - g f_{abc} W_\mu^c\d^\mu
\right)\,.
}
Finally, introducing FP ghosts $\eta=\{\eta^a\}$ and $\bar\eta=\{\bar\eta^a\}$, that absorb the factor $-g^{-1}$, we may write
\eq{
\det M 
&=\int\D\bar\eta\,\D\eta\, \exp\left\{-\ii\int\dd^4x\, \bar\eta^a(\delta_{ab}\Box - g f_{abc} W_\mu^c\d^\mu)\eta^b\right\} \nn\\
&=\int\D\bar\eta\,\D\eta\, \exp\left\{\ii\int\dd^4x\, \lag_{\rm FP} \right\}
}
with
\eq{
\lag_{\rm FP} 
= (\d^\mu\bar\eta^a)(\d_\mu\eta^a-gf_{abc}\eta^bW_\mu^c)
= (\d^\mu\bar\eta^a)(D_\mu^{\rm adj})_{ab}\eta^b
\label{eq:lagFP}
}
where we have introduced the covariant derivative in the adjoint representation,
\eq{
D_\mu^{\rm adj} = \d_\mu - \ii g T^{\rm adj}_c W^c_\mu\ , \quad
(T^{\rm adj}_c)_{ab}=-\ii f_{abc}\ .
}

As a conclusion to this section, the quantization of gauge theories requires the introduction of a gauge-fixing, like \eqref{eq:lagGF} or \eqref{eq:lagGF3}, that allows to define the gauge field propagators. In the $R_\xi$ gauges, the propagators read
\eq{
\widetilde D^{ab}_{\mu\nu}(k) = \frac{\ii\delta_{ab}}{k^2+\ii\varepsilon}\left[- g_{\mu\nu}+(1-\xi_a)\frac{k_\mu k_\nu}{k^2}\right]\,.
}
Propagators are not physical observables, so the gauge dependence with the parameter $\xi_a$ is not worrisome; it will cancel out in physical amplitudes.
Particular cases of interest are the Landau gauge ($\xi_a=0$) and the Feynman-'t Hooft gauge ($\xi_a=1$). The latter has a simpler form, very helpful for loop calculations,
\eq{
\widetilde D^{ab}_{\mu\nu}(k) = -\frac{\ii\delta_{ab}g^{\mu\nu}}{k^2+\ii\varepsilon}.
}
If the gauge theory is non-Abelian one also needs to introduce interactions with FP ghosts \eqref{eq:lagFP}. These are anticommuting scalar fields that only appear in internal lines, never as external legs. They are produced in pairs and are needed in order to preserve the gauge symmetry at the quantum level: they cancel unphysical degrees of freedom of virtual gauge bosons in loops. This procedure ensures that we do not count field configurations of $W_\mu^a$ which are pure gauge, nor count separately fields which differ only by a gauge transformation.

\boxexercise{3}{Obtain the Feynman rules for cubic and quartic self-interactions among gauge fields in a general non-Abelian gauge theory, as well as those for the interactions of Faddeev-Popov ghosts with gauge fields.}

\boxexercise{4}{Consider the 1-loop self-energy diagrams for non-Abelian gauge theories in the figure. Calculate the diagrams in the Feynman-'t Hooft gauge and show that the sum does not have the tensor structure $g_{\mu\nu}k^2-k_\mu k_\nu$ required by the gauge invariance of the theory unless diagram (c) involving ghost fields is included.

\centerline{\includegraphics[width=0.5\linewidth]{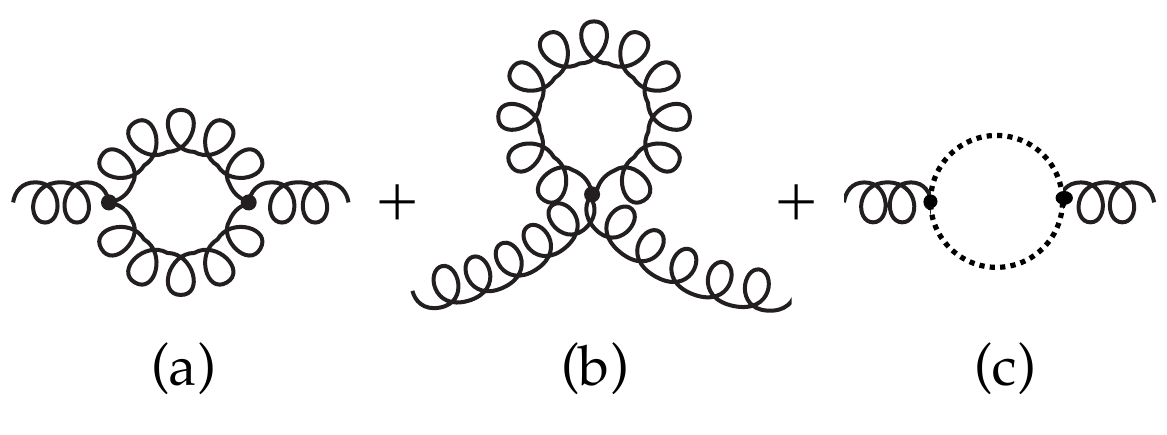}}

\noindent
{\em Hint}: Take Feynman rules from previous excercise and use dimensional regularization. It is convenient to use the Passarino-Veltman tensor decomposition of loop integrals:
\begin{align}
\frac{\ii}{16\pi^2}\{B_0,\,B_\mu,\,B_{\mu\nu}\}&=\mu^\epsilon
\int\frac{\dd^D q}{(2\pi)^D}\frac{\{1,\,q_\mu,\,q_\mu q_\nu\}}{q^2(q+k)^2}
\nn\\
\mbox{where}\quad B_0&=\Delta_\epsilon + \mbox{finite} \nn\\
B_\mu&=k_\mu B_1\ , \quad B_1=-\frac{\Delta_\epsilon}{2}+\mbox{finite} \nn\\
B_{\mu\nu}&=g_{\mu\nu}B_{00}+k_\mu k_\nu B_{11}\ , \quad
B_{00} = -\frac{k^2}{12}\Delta_\epsilon+\mbox{finite}\ , \quad
B_{11} = \frac{\Delta_\epsilon}{3}+\mbox{finite} \nn
\end{align}
with $\Delta_\epsilon=2/\epsilon-\gamma+\ln4\pi$ and $D=4-\epsilon$. 
You may check that the ultraviolet divergent part has the expected structure or find the final result in terms of scalar integrals, that for 
massless fields
read:
\begin{align}
B_1 = -\frac{1}{2}B_0\ , \quad
B_{00} = -\frac{k^2}{4(D-1)}B_0\ , \quad
B_{11} = \frac{D}{4(D-1)}B_0\ .\nn
\end{align}
Do not forget a symmetry factor (1/2) in front of (a) and (b), and a factor $(-1)$ in (c).
}

We finally know how to build the Lagrangian of a quantum gauge field theory. Provided a gauge symmetry group and matter fields transforming in given group representations, the covariant derivatives specify the form of the interactions mediated by the gauge fields encoded in a gauge invariant piece $\lag_{\text{inv}}$, that has to be supplemented by gauge-fixing terms and, if necessary, by interactions with unphysical Faddeev-Popov ghosts,
\eq{
\lag_{\text{inv}} + \lag_{\rm GF} + \lag_{\rm FP}.
}

However, mass terms for the gauge fields break explicitly the gauge invariance. In fact the Proca Lagrangian
\eq{
\lag = -\frac{1}{4}F_{\mu\nu}F^{\mu\nu} + \frac{1}{2}M^2 A_\mu A^\mu
}
is not invariant under U(1) gauge transformations if $M\ne0$, which on the other hand allows to define the propagator,
\eq{
\widetilde D_{\mu\nu}(k) = \frac{\ii}{k^2-M^2+\ii\varepsilon}
\left(- g_{\mu\nu}+\frac{k^\mu k^\nu}{M^2}\right)\,.
\label{eq:Procaprop}
}
This is a serious issue if we wish to describe the fundamental interactions inspired by the gauge principle, since in particular weak interactions are mediated by massive gauge bosons. Fortunately, there is a way to cope with massive gauge mediators without spoiling the nice properties of the gauge symmetry, as we will see in next section.

\boxexercise{5}{Derive the propagator of the massive vector field of the Proca Lagrangian \eqref{eq:Procaprop} and check that the EoM imply $\d^\mu A_\mu=0$ (it is not a gauge condition!) consistently with the description of massive spin-1 particles with 3 degrees of freedom.}

\subsection{Spontaneous symmetry breaking}

\subsubsection{Discrete symmetry}

\begin{figure}
\centering
\begin{tabular}{cc}
\includegraphics[scale=0.3]{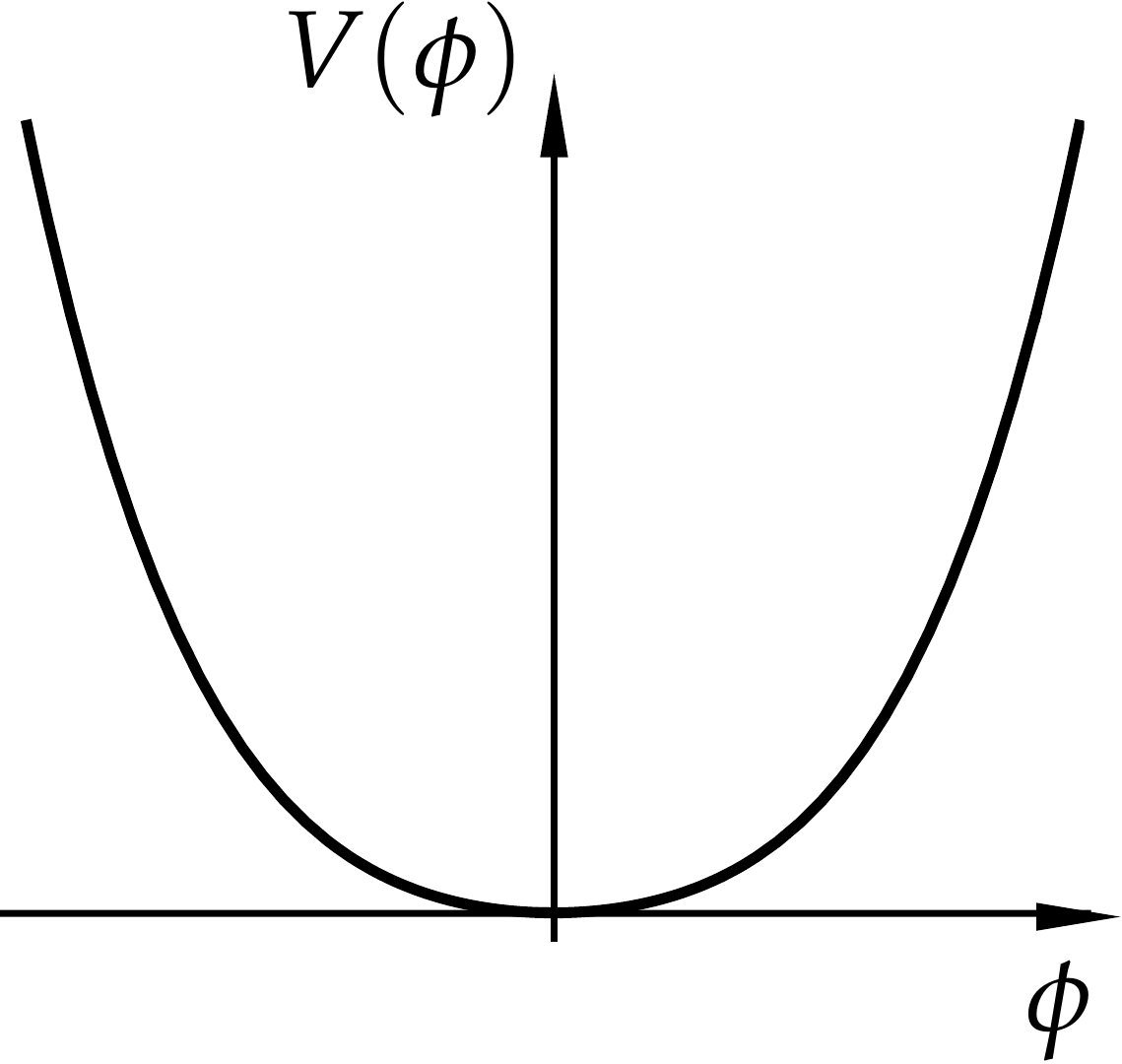} &\quad
\includegraphics[scale=0.3]{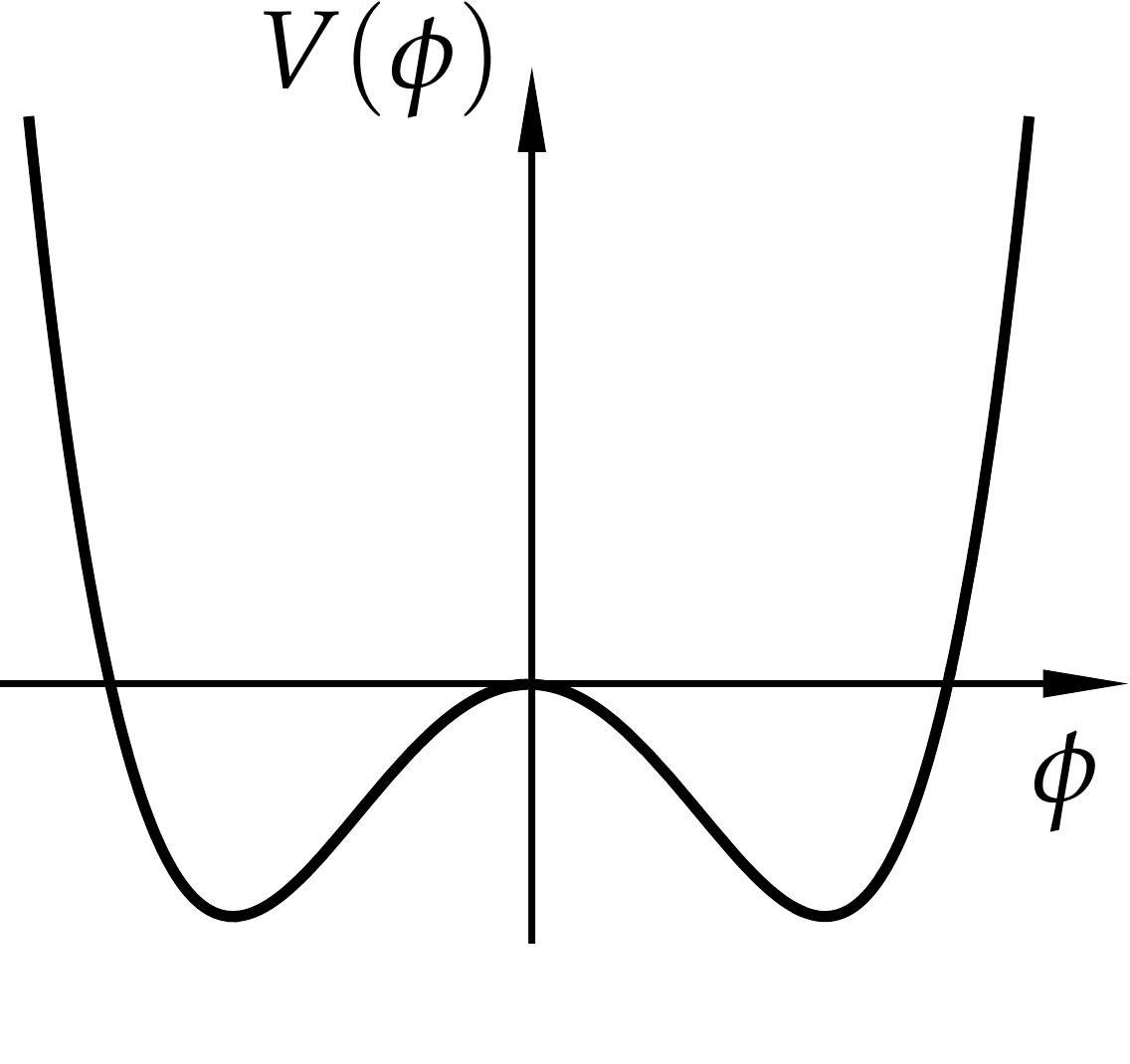} 
\end{tabular}
\caption{Potential in \eqref{eq:Vreal} for $\mu^2>0$ (left) and $\mu^2<0$ (right), symmetric under $\phi\mapsto-\phi$.
\label{fig:discrete}}
\end{figure}

In order to understand the basic ideas behind the spontaneous symmetry breaking, let us first consider a real scalar field $\phi(x)$ with Lagrangian
\begin{equation}
\lag = \frac{1}{2}(\d_\mu\phi)(\d^\mu\phi) - V(\phi)\,,\quad
V(\phi) = \frac{1}{2}\mu^2\phi^2 + \frac{\lambda}{4}\phi^4.
\label{eq:Vreal}
\end{equation}
This Lagrangian is invariant under a discrete $\mathbb{Z}_2$ symmetry $\phi \mapsto -\phi$. The Hamiltonian is given by
\begin{equation}
\quad{\cal H} = \frac{1}{2}(\dot\phi^2+(\nabla\phi)^2)+V(\phi)\,,
\end{equation}
where the constants $\mu^2$ and $\lambda$ are real so that the Hamiltonian is real/Hermitian, and $\lambda>0$ to ensure there exists a ground state.
We distinguish two cases depending on the sign of $\mu^2$ (fig.~\ref{fig:discrete}).
The interesting case is $\mu^2<0$ for which the minimum is not zero and degenerate, $\phi = v \equiv \pm\sqrt{-\mu^2/\lambda}$.

For a quantum field the configuration of minimum energy must be interpreted as the expectation value (VEV) of the field in the ground state, the vacuum. But if $\bra{0}\phi\ket{0}=v\ne0$ we have a problem, because $\ket{0}$ must be annihilated by any annihilation operator $a_{\vec{p}}$ in $\phi$, a requirement for the construction of the Fock space of its multiparticle states. Therefore, we must perform a redefinition
\eq{
\phi(x) \equiv v + \eta(x)
} 
with $\eta(x)$ the field describing the quantum fluctuations, $\bra{0}\eta\ket{0} = 0$. Then, at the quantum level, the same system is described by the following Lagrangian in terms of $\eta(x)$:
\begin{align}
\lag &= \frac{1}{2}(\d_\mu\eta)(\d^\mu\eta)- \lambda v^2 \eta^2 
 -\lambda v \eta^3 -\frac{\lambda}{4}\eta^4
 +\frac{1}{4}\lambda v^4\,,
\end{align}
where $\eta$ has a mass $\sqrt{2\lambda v^2}$.
Note that the $\mathbb{Z}_2$ symmetry of the original Lagrangian is broken, or hidden to be more precise. We say that the symmetry is `spontaneously' broken because it is due to a non-invariant vacuum, not to an external agent. One may think that $\lag(\eta)$ exhibits an `explicit' breaking of the symmetry. However this is not the case: the fact that the coefficients of terms $\eta^2$, $\eta^3$ and $\eta^4$ are not independent (they are determined by just two parameters, $\lambda$ and $v$) is a remnant of the original symmetry. The last constant term can be omitted as it has no effect on the field dynamics.

\subsubsection{Continuous global symmetry}

\begin{figure}
\centering
\includegraphics[scale=0.3]{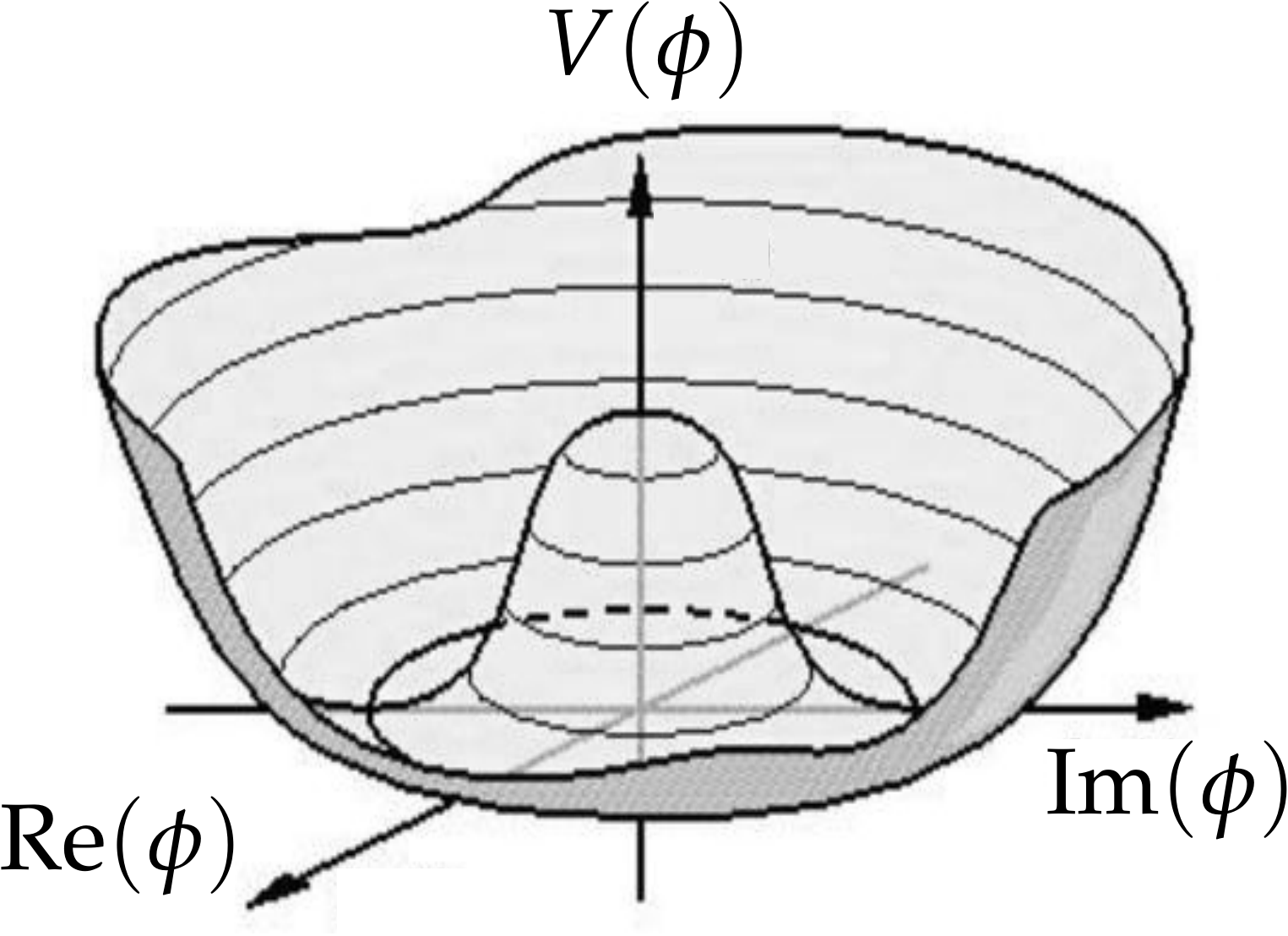}
\caption{Mexican hat potential.\label{fig:Mexican}}
\end{figure}

Consider now a complex scalar field $\phi(x)$ with Lagrangian
\begin{equation}
\lag = (\d_\mu\phi)^\dagger(\d^\mu\phi) - V(\phi)\,,\quad
V(\phi) = \mu^2\phi^\dagger\phi + \lambda(\phi^\dagger\phi)^2\,,
\end{equation}
which is  invariant under global ${\rm U}(1)$ transformations $\phi \mapsto \e^{-\ii Q\theta}\phi$.
For $\lambda>0,\ \mu^2<0$ (fig.~\ref{fig:Mexican}) the potential has a Mexican hat shape with a degenerate minimum,
\begin{equation}
\bra{0}\phi\ket{0} \equiv \frac{v}{\sqrt{2}}\ , \quad
|v|=\sqrt{\frac{-\mu^2}{\lambda}}.
\end{equation}
We may choose $v\in\mathbb{R}^+$ without loss of generality.\footnote{If we take any  complex value $|v|\e^{\ii\alpha}$ the conclusions will be the same for redefined fields, $\eta\to(\eta\cos\alpha-\chi\sin\alpha)$ and $\chi\to(\eta\sin\alpha+\chi\cos\alpha)$.} In terms of the quantum fluctuations,
\begin{equation}
\phi(x)\equiv \frac{1}{\sqrt{2}}[v+\eta(x)+\ii\chi(x)], \quad
\bra{0}\eta\ket{0}=\bra{0}\chi\ket{0}=0\,,
\end{equation}
the Lagrangian reads
\begin{equation}
\lag = \frac{1}{2}(\d_\mu\eta\d^\mu\eta) 
      +\frac{1}{2}(\d_\mu\chi\d^\mu\chi)
      -\lambda v^2\eta^2
      -\lambda v \eta(\eta^2+\chi^2)
      -\frac{\lambda}{4}(\eta^2+\chi^2)^2
      +\ \frac{1}{4}\lambda v^4,
\end{equation}
Observe that the quantum Lagrangian $\lag(\eta,\chi)$ is no longer invariant under ${\rm U}(1)$. The spontaneous breaking of this symmetry leaves one massless scalar field, $\chi$, whereas $\eta$ has a mass proportional to the VEV, $m_\eta = \sqrt{2\lambda}\,v$. 

In order to understand what are the consequences of the spontaneous breaking we will explore next the case of a group with more symmetries. Take an ${\rm SO}(3)$ triplet of real scalar fields, $\Phi(x)$, whose self-interactions are given by a similar Mexican hat potential,
\begin{equation}
\lag = \frac{1}{2}(\d_\mu\Phi)^{\sf T}(\d^\mu\Phi)
      -\frac{1}{2}\mu^2\Phi^{\sf T}\Phi - \frac{\lambda}{4}(\Phi^{\sf T}\Phi)^2.
\end{equation}
This theory is invariant under global ${\rm SO}(3)$ transformations $\Phi \mapsto \e^{-\ii T_a\theta^a}\Phi$· For $\lambda>0,\ \mu^2<0$ the triplet acquires a VEV 
\begin{equation}
\bra{0}\Phi^{\sf T}\Phi\ket{0} = v^2 = -\mu^2/\lambda\,.
\end{equation}
We express the quantum field as $\Phi(x) \equiv \big(\varphi_1(x),\, \varphi_2(x),\, v + \varphi_3(x)\big)^{\sf T}$ and define the complex combination $\varphi \equiv \dis\frac{1}{\sqrt{2}}(\varphi_1+\ii\varphi_2)$. Then, the Lagrangian can be rewritten as
\begin{align}
\lag &= (\d_\mu\varphi)^\dagger(\d^\mu\varphi)
     +\frac{1}{2}(\d_\mu\varphi_3)(\d^\mu\varphi_3)\nonumber\\
      & \quad-\lambda v^2\varphi_3^2
      -\lambda v (2\varphi^\dagger\varphi + \varphi_3^2)\varphi_3
      -\frac{\lambda}{4}(2\varphi^\dagger\varphi + \varphi_3^2)^2
      +\frac{1}{4}\lambda v^4,
\end{align}
which is not symmetric under ${\rm SO}(3)$ but is invariant under the ${\rm U}(1)$ transformation
\begin{equation}
\varphi \mapsto \e^{-\ii Q\theta}\varphi\quad (Q\ \text{is arbitrary}),\qquad\qquad
\varphi_3 \mapsto \varphi_3\quad (Q=0).
\end{equation}
In other words, the group ${\rm SO}(3)$ has broken spontaneously into a ${\rm U}(1)$ subgroup. Since there are $3-1=2$ broken generators, 2 real scalar fields (or, equivalently, one complex scalar $\varphi$) remain massless, while the other scalar gets a mass proportional to the VEV:
\begin{equation}
 m_{\varphi_1}=m_{\varphi_2}=0\quad (m_{\varphi}=0)\,,\quad m_{\varphi_3}=\sqrt{2\lambda v^2}\,.
\end{equation}

The two examples we have just analyzed illustrate the \emph{Goldstone theorem} \cite{Nambu:1960tm, Goldstone:1961eq}: the number of massless particles ({\em Nambu-Goldstone bosons}) is equal to the number of spontaneously broken generators of the symmetry. It is not difficult to understand what is behind this result. By definition of a symmetry, if the Hamiltonian is invariant under the group $G$ with generators $T_a$, we have
\begin{equation}
[T_a,H] = 0\ , \quad a=1,\dots,N,
\end{equation}
where the generator is denoted with the same symbol as its representation in the Fock space (Noether charge operator). And by definition of the vacuum state, 
\begin{equation}
H\ket{0} = 0 \quad\Raw\quad 
H(T_a\ket{0})=T_a H\ket{0}=0.
\end{equation}
Therefore: 
\begin{itemize}

\item
If $\ket{0}$ is such that $T_a\ket{0}=0$ for all generators, there is a non-degenerate minimum: {\em the} vacuum, that will remain invariant. 

\item
But if $\ket{0}$ is such that $T_{a'}\ket{0}\ne 0$ for some (broken) generators $T_{a'}$, there is a degenerate minimum: for any choice ({\em true} vacuum) we will have $\e^{-\ii T_{a'}\theta^{a'}}\ket{0} \ne \ket{0}$, so it will not remain invariant. In this case there are excitations from $\ket{0}$ to $\e^{-\ii T_{a'}\theta^{a'}}\ket{0}$ (flat directions of the potential) that cost no energy, so they correspond to massless particles (the Goldstone bosons). 

\end{itemize}

\subsubsection{Gauge symmetry}

Take the simplest ${\rm U}(1)$ gauge invariant Lagrangian for a complex scalar field $\phi(x)$:
\begin{equation}
\lag = -\frac{1}{4}F_{\mu\nu}F^{\mu\nu} + (D_\mu\phi)^\dagger(D^\mu\phi)
      -\mu^2\phi^\dagger\phi - \lambda(\phi^\dagger\phi)^2 \ , \quad
      D_\mu \equiv \d_\mu + \ii e Q A_\mu,
\end{equation}
which is invariant under the tansformations
\begin{equation}
\phi(x) \mapsto \e^{-\ii Q\theta(x)}\phi(x) \,, \quad
A_\mu(x) \mapsto A_\mu(x) + \dis\frac{1}{e}\d_\mu\theta(x).
\end{equation}
If $\lambda>0$ and $\mu^2<0$ the potential has a Mexican hat shape with a minimum at $\bra{0}\phi\ket{0}=v/\sqrt{2}$ where $|v|=\sqrt{-\mu^2/\lambda}$. We will choose $v\in\mathbb{R}^+$ as before. Then we write
\begin{equation}
\phi(x)\equiv \frac{1}{\sqrt{2}}[v+\eta(x)+\ii\chi(x)]
\end{equation}
where $\eta$ and $\chi$ are two real fields with null VEVs that describe particle excitations. In terms of these quantum fields the Lagrangian reads 
\begin{align}
\lag=
      & -\frac{1}{4}F_{\mu\nu}F^{\mu\nu} +\frac{1}{2}(\d_\mu\eta)(\d^\mu\eta)  +\frac{1}{2}(\d_\mu\chi)(\d^\mu\chi) \nonumber\\
      &-\lambda v^2\eta^2 -\lambda v \eta(\eta^2+\chi^2) -\frac{\lambda}{4}(\eta^2+\chi^2)^2 \ +\ \frac{1}{4}\lambda v^4 \nonumber\\     
      &+\ e Q v A_\mu\d^\mu\chi  +e Q A_\mu (\eta\d^\mu\chi-\chi\d^\mu\eta) \nonumber\\
      &+\ \dis\frac{1}{2}(eQv)^2 A_\mu A^\mu +\frac{1}{2}(eQ)^2 A_\mu A^\mu (\eta^2+2v\eta+\chi^2).
\end{align}
Several comments are in order at this point:
\begin{itemize}

\item
At first sight, one of the scalar fields, $\chi$, seems massless (the Goldstone boson field) and the other one has a mass $m_{\eta}=\sqrt{2\lambda}\,v$. The global symmetry has broken spontaneously. We cannot say that the gauge symmetry has broken, because it is not really a symmetry, as we have discussed before.

\item
The gauge field $A_\mu$ acquires a mass $M_A = |eQv|$, proportional to the VEV of $\phi$. 

\item
There is a cross term $A_\mu\d^\mu\chi$ that mixes $A_\mu$ and $\chi$, producing kinetic terms that are neither diagonal nor invertible. Therefore, it is premature to infer the masses of $A_\mu$ and $\chi$ until we have made sense of this term.

\item
We still have to add a gauge-fixing term $\lag_{\rm GF}$.

\end{itemize}
The cross term can be removed and the gauge fixed at the same time by introducing the following gauge-fixing Lagrangian:
\begin{equation}
\lag_{\rm GF} = -\frac{1}{2\xi}(\d_\mu A^\mu - \xi M_A\chi)^2\,,
\end{equation}
which in particular adds a term to the kinetic mixing above yielding an irrelevant total derivative, $M_A\d_\mu(A^\mu \chi)$, that can be ignored. Therefore
\begin{align}
\quad \lag + \lag_{\rm GF} = & -\frac{1}{4}F_{\mu\nu}F^{\mu\nu}
+\frac{1}{2}M^2_A A_\mu A^\mu-\frac{1}{2\xi}(\d_\mu A^\mu)^2 \nonumber \\
  & + \frac{1}{2}(\d_\mu\chi)(\d^\mu\chi) -\frac{1}{2}\xi M_A^2\chi^2 + \mbox{interactions}. 
\end{align}
The resulting propagators of $A_\mu$ and $\chi$ are, respectively:
\begin{align}
\widetilde D_{\mu\nu}(k) &= \frac{\ii}{k^2-M^2_A+\ii\varepsilon}\left[- g_{\mu\nu}+(1-\xi)\frac{k_\mu k_\nu}{k^2-\xi M^2_A}\right], \label{ex:prop1}\\
\widetilde D(k) &=\frac{\ii}{k^2- \xi M^2_A+\ii\varepsilon}.\label{ex:prop2}
\end{align}
This confirms that the interaction of $A_\mu$ with $\phi$ has provided the gauge boson with a mass proportional to $\bra{0}\phi\ket{0}$. Notice also that $\chi$ has a gauge-dependent mass, an indication that it is not `physical'.

\boxexercise{6}{Prove the expressions \eqref{ex:prop1} and \eqref{ex:prop2} for the propagators of $A_\mu$ and $\chi$. In addition, show that the propagator of $\eta$ is
\eq{
\widetilde D_\eta(k) =\frac{\ii}{k^2- M^2_\eta+\ii\varepsilon},\quad\text{with}\quad M^2_\eta=-2\mu^2=2\lambda v^2.
}
}

We can better understand the consequences of the spontaneous breaking of the symmetry in the context of a gauge theory if we use a more transparent parametrization of the quantum fluctuations of $\phi$. Let us now define 
\begin{equation}
\phi(x) \equiv \e^{\ii Q\zeta(x)/v}\frac{1}{\sqrt{2}}[v+\eta(x)] \ , \qquad
\bra{0}\eta\ket{0}=\bra{0}\zeta\ket{0}=0.
\end{equation}
Thanks to the gauge symmetry, the field $\zeta(x)$ can now be eliminated (\emph{gauged away}) by exploiting the gauge freedom to {\em choose the phase} of $\phi$ at every point of spacetime, 
\begin{equation}
\phi(x) \mapsto \e^{-\ii Q\zeta(x)/v}\phi(x) = \frac{1}{\sqrt{2}}[v+\eta(x)].
\end{equation}
The resulting Lagrangian is
\begin{align}
\lag =& -\frac{1}{4}F_{\mu\nu}F^{\mu\nu}
        +\frac{1}{2}(\d_\mu\eta)(\d^\mu\eta) \nn\\
     & -\lambda v^2\eta^2-\lambda v\eta^3-\frac{\lambda}{4}\eta^4
        +\ \frac{1}{4}\lambda v^4 \nn\\
     & +\frac{1}{2}(eQv)^2 A_\mu A^\mu 
       +\frac{1}{2}(eQ)^2 A_\mu A^\mu(2v\eta+\eta^2)\,.
\end{align}
Observe that we obtain again the same masses $m_{\eta}=\sqrt{2\lambda}\, v$ and  $M_A = |eQv|$. Of course, since the gauge has been `fixed', there is no need to add a $\lag_{\rm GF}$. Actually this corresponds to choosing the so-called  {\em unitary gauge} ($R_\xi$ gauge with $\xi\to\infty$), in which only the physical fields appear:
\begin{equation}
\widetilde D_{\mu\nu}(k) \to \frac{\ii}{k^2-M^2_A+\ii\varepsilon}\left[- g_{\mu\nu}+\frac{k_\mu k_\nu}{M^2_A}\right] \quad
\mbox{and} \quad\widetilde D(k) \to 0.
\end{equation}

The results above are a manifestation of the \emph{Brout-Englert-Higgs mechanism} \cite{Anderson:1963pc,Englert:1964et,Higgs:1964ia,Higgs:1964pj,Guralnik:1964eu,Higgs:1966ev,Kibble:1967sv}:
The {\em gauge bosons} associated with the spontaneously broken generators become massive, the corresponding {\em would-be Goldstone bosons} are unphysical (they can be absorbed), and the remaining massive scalars ({\em Higgs bosons}) are physical.

The existence of Higgs bosons is the smoking gun confirming that this mechanism is responsible for the mass of the gauge bosons associated to broken symmetries. One often says that the would-be Goldstone bosons are `eaten up' by the gauge bosons that `get fat' by acquiring a mass. But keep in mind that the would-be Goldstone bosons only disappear completely in the unitary gauge ($\xi\to\infty$), even though they are unphysical in any gauge. 

Notice also that the number of degrees of freedom (dof) of the physical spectrum remains the same. In the case of the U(1) gauge invariance we have discussed, before spontaneous symmetry breaking ($\mu^2>0$) there are 2 scalars and one massless gauge boson with 2 polarizations ($1+1+2=4$~dof). After spontaneous symmetry breaking ($\mu^2<0$) one of the scalars is physical but the other one is not, and the massive gauge boson has 3 polarizations ($1+0+3=4$~dof).  

Remember that for loop calculations the Feynman-'t Hooft gauge ($R_\xi$ gauge with $\xi=1$) is more convenient because the gauge boson propagators are simpler. However, be aware that in this gauge the Goldstone bosons must be included, in internal lines only.

For completeness, let us mention that, if the gauge group is non-Abelian, the 
(unphysical) Faddeev-Popov ghosts associated to the gauge boson of broken symmetries acquire a gauge-dependent mass. In a general $R_\xi$ gauge the FP propagator is
\begin{equation}
\widetilde D_{ab}(k) = \frac{\ii\delta_{ab}}{k^2-\xi_a M^2_{W^a}+\ii\varepsilon}.
\end{equation}

Finally, it is very important to underline that gauge theories with spontaneous symmetry breaking are renormalizable \cite{tHooft72}. This means that the ultraviolet divergences appearing at loop level can be absorbed by an appropriate redefinition of the parameters and fields in the classical Lagrangian. Since there are a finite number of them, they can all be fixed by the measurement of just a few observables, so these theories are predictive.

\section{The Standard Model}

\subsection{Gauge group and field representations}

The Standard Model (SM) \cite{Glashow:1961tr,Weinberg:1967tq,Salam:1968rm,GellMann:1964nj,Zweig:1964jf,Fritzsch:1973pi} is a gauge theory based on the symmetry group:
\eq{
{\rm SU}(3)_c\otimes{\rm SU}(2)_L\otimes {\rm U}(1)_Y \to
{\rm SU}(3)_c\otimes{\rm U}(1)_Q,
}
where the electroweak symmetry is spontaneously broken to the electromagnetic symmetry by the Brout-Englert-Higgs mechanism. 

The SM particle content, in Table~\ref{tab:smcontent}, consists of three replicas (families or generations) of spin $\frac{1}{2}$ fermions that constitute matter, a set of $8+3+1=12$ gauge vector bosons mediating the fundamental interactions (as many as generators of the gauge group) and one Higgs boson, remnant of the 
Higgs scalar field that triggers the electroweak symmetry breaking (EWSB) giving rise to the masses of elementary particles.

The SM is a chiral theory: left and right-handed components of the fermion fields lay in different representations of the gauge group, as shown in Table~\ref{tab:smmultiplets}. Strong and electroweak interactions can be studied separately and have very different properties. The former, specified by ${\rm SU}(3)_c$, are dubbed {\em quantum chromodynamics} (QCD) because they are only experienced by particles with `color' charges, that is quarks (color triplets) and gluons. The electroweak interactions, described by the group ${\rm SU}(2)_L\otimes {\rm U}(1)_Y$, affect any type (`flavor') of fermions depending on their weak isospin and hypercharge ({\em quantum flavordynamics}). Left/right-handed fermions are isospin doublets/singlets, respectively, and have also different hypercharges. The electric charges $Q$ are associated to the only electroweak symmetry generator that remains unbroken, the sum of the ${\rm SU}(2)_L$ weak isospin $T_3$ and the ${\rm U}(1)_Y$ hypercharge $Y$, leading to {\em quantum electrodynamics} (QED).

\begin{table}
\begin{center}
\begin{tabular}{|l|l|l||c|c|c||r|}
\hline
\multicolumn{3}{|r||}{\bf Fermions} & I & II & III & $Q$ \\
\hline\hline
spin $\frac{1}{2}$ &
Quarks & $f$  & $\uR\uG\uB$ & $\cR\cG\cB$ & $\tR\tG\tB$ & $\frac{2}{3}$  \\
\cline{3-7}
      && $f'$ & $\dR\dG\dB$ & $\sR\sG\sB$ & $\bR\bG\bB$ & $-\frac{1}{3}$ \\
\cline{2-7}
&
Leptons & $f$  & $\nu_e$ & $\nu_\mu$ & $\nu_\tau$ & $0$  \\
\cline{3-7}
        && $f'$ & $\e$    & $\mu$     & $\tau$     & $-1$ \\
\hline
\end{tabular}\qquad
\begin{tabular}{|l|l||r|}
\hline
\multicolumn{2}{|r||}{\bf Bosons} & {\em responsible for} \\
\hline\hline
spin 1 & 8 gluons & strong interaction \\
\cline{2-3}
       & $W^\pm$, $Z$ & weak interaction \\
\cline{2-3}
       & $\gamma$       & em interaction \\
\hline
spin 0 & Higgs          & origin of mass \ \ ~ \\
\hline
\end{tabular}
\end{center}
\caption{SM particle content: 3 fermion families 
of 2 quarks in 3 colors and 2 leptons, 12 gauge bosons and 1 Higgs boson. The electric charges $Q$ of quarks or leptons of the same family ($f$ and $f'$) differ in one unit. \label{tab:smcontent}} 
\end{table}

\begin{table}
\begin{center}
\begin{tabular}{|l|c|c|c|c||r|}
\hline
{\bf Multiplets} &
${\rm SU}(3)_c\otimes {\rm SU}(2)_L\otimes {\rm U}(1)_Y$  & 
{I} & 
{II} & 
{III} & $Q=\ \, T_3+Y\ $ \\
\hline\hline
Quarks & ({\bf 3}, {\bf 2}, $\frac{1}{6}$) & 
\ $\pmat{ u_L \\ d_L }$\ ~ & 
\ $\pmat{ c_L \\ s_L }$\ ~ & 
\ $\pmat{ t_L \\ b_L }$\ ~ & 
$\ba{r} \frac{2}{3}=\phantom{-} \frac{1}{2}+\frac{1}{6} \\ -\frac{1}{3}= -\frac{1}{2}+\frac{1}{6} \ea$ \\
\cline{2-6} & ({\bf 3}, {\bf 1}, $\frac{2}{3}$) & 
$u_R$ & $c_R$ & $t_R$ & $\ba{r} \frac{2}{3}=\phantom{-} 0+\frac{2}{3}\ea$ \\
\cline{2-6} & ({\bf 3}, {\bf 1}, $-\frac{1}{3}$) &
$d_R$ & $s_R$ & $b_R$ & $\ba{r} -\frac{1}{3}=\phantom{-} 0-\frac{1}{3}\ea$ \\ 
\hline\hline
Leptons & ({\bf 1}, {\bf 2}, $-\frac{1}{2}$) & 
$\pmat{ \nu_{e_L} \\ e_L }$ & 
$\pmat{ \nu_{\mu_L} \\ \mu_L }$ & 
$\pmat{ \nu_{\tau_L} \\ \tau_L }$ & $\ba{r} 0=\phantom{-} \frac{1}{2}-\frac{1}{2} \\ -1=-\frac{1}{2}-\frac{1}{2} \ea$ \\
\cline{2-6} & ({\bf 1}, {\bf 1}, $-1$) &
$e_R$ & $\mu_R$ & $\tau_R$ & $\ba{r} -1=\phantom{-} 0-1\ea$ \\
\cline{2-6} & {\GY ({\bf 1}, {\bf 1}, $0$)} &
 ${\GY\nu_{e_R}}$ & ${\GY \nu_{\mu_R}}$ & ${\GY \nu_{\tau_R}}$ & \ \ \ \  
${\GY \ba{r} 0=\ \ \ 0+0\ea}$ \\
\hline
\end{tabular}
\end{center}
\caption{Gauge group representations of left-handed and right-handed fermion fields. They are the same for each family of quarks or leptons (universal). The electric charges $Q$ are fixed by the ${\rm SU}(2)_L$ weak isospin $T_3$ and the ${\rm U}(1)_Y$ hypercharge $Y$. Right-handed neutrinos are sterile (singlets) and were absent in the original SM with massless neutrinos. \label{tab:smmultiplets}}
\end{table}

\subsection{Electroweak interactions}

\subsubsection{One generation of quarks or leptons}

Consider two massless fermion fields $f(x)$ and $f'(x)$ with electric charges $Q_f = Q_{f'} + 1$ and assume their chiral components lay in the following  {\rm SU}(2)$_L\otimes${\rm U}(1)$_Y$ representations:
\eq{
\Psi_1 = \pmat{f_L \\ f'_L} \sim \dis({\bf 2},y_1), \quad
\psi_2 = f_R \sim \dis({\bf 1},y_2)\ , \quad
\psi_3 = f'_R \sim \dis({\bf 1},y_3) ,
}
where $f_{R,L}=P_{R,L}\,f$ with $P_{R,L}=\frac{1}{2}(1\pm\gamma_5)f$ the chiral proyectors, and likewise for $f'_{R,L}$.
Their free Lagrangian, invariant under global transformations, is
\eq{
\lag^0_F&= \ii\overline f \slashed{\d} f 
          +\ii\overline f' \slashed{\d} f'
         = \ii\overline \Psi_1 \slashed{\d} \Psi_1
          +\ii\overline \psi_2 \slashed{\d} \psi_2
          +\ii\overline \psi_3 \slashed{\d} \psi_3.
}
To make it invariant under gauge transformations,
\begin{align}
\Psi_1(x) &\mapsto U_L(x)\e^{-\ii y_1\beta(x)}\Psi_1(x), \quad
           U_L(x)=\e^{-\ii T_i\alpha^i(x)}, \quad
           T_i = \frac{\sigma_i}{2} \\
\psi_2(x) &\mapsto \e^{-\ii y_2\beta(x)}\psi_2(x) 
\\
\psi_3(x) &\mapsto \e^{-\ii y_3\beta(x)}\psi_3(x),
\end{align}
one has to substitute the corresponding covariant derivatives,
\eq{
D_\mu\Psi_1 &= (\d_\mu - \ii g \widetilde W_\mu + \ii g' y_1 B_\mu)\Psi_1
\ , \quad \widetilde W_\mu \equiv \dis\frac{\sigma_i}{2}W^i_\mu, \\
D_\mu\psi_2 & = (\d_\mu + \ii g' y_2 B_\mu)\psi_2, \\
D_\mu\psi_3 & = (\d_\mu + \ii g' y_3 B_\mu)\psi_3,
}
where we have introduced two couplings, $g$ and $g'$, one for each group factor, and four gauge fields, $W^1_\mu(x)$,
$W^2_\mu(x)$, $W^3_\mu(x)$ and $B_\mu(x)$, transforming as:\footnote{The signs of $g$ and $g'$ are conventional, with no effect on physical observables.}
\eq{
\widetilde W_\mu(x) & \mapsto U_L(x)\widetilde W_\mu(x) U_L^\dagger(x)
-\frac{\ii}{g}(\d_\mu U_L(x))U^\dagger_L(x) \\
B_\mu(x) & \mapsto B_\mu(x) + \frac{1}{g'}\d_\mu \beta(x).
}
Then $\lag^0_F$ is replaced by ${\cal L}_F$, which contains charge conjugation ($C$) and parity ($P$) violating interactions. Furthermore, one has to add the Yang-Mills Lagrangian 
\eq{
{\cal L}_{\rm YM} =
- \frac{1}{4} W^i_{\mu\nu} W^{i,\mu\nu}
- \frac{1}{4} B_{\mu\nu} B^{\mu\nu}
\label{eq:YM}
}
with $W^i_{\mu\nu} = \d_\mu W^i_\nu - \d_\nu W^i_\mu + g\epsilon_{ijk} W^j_\mu W^k_\nu$
and $B_{\mu\nu} = \d_\mu B_\nu - \d_\nu B_\mu$, which includes kinetic terms for every vector field and self-interactions for the gauge fields of ${\rm SU}(2)_L$, a non-abelian symmetry.

Note that mass terms for the fermions are incompatible with the symmetry because left and right-handed components do not transform the same under ${\rm SU}(2)_L\otimes {\rm U}(1)_Y$ and 
\eq{
m\overline f f = m(\overline{f_L} f_R + \overline{f_R} f_L).
}
Mass terms for the gauge bosons are not allowed either. Both problems will be solved later. Let us discuss first the different types of interactions that have been generated.

\subsubsubsection{Charged current interactions} 

\begin{figure}
\centering
\begin{tabular}{cc}
\includegraphics[scale=.7]{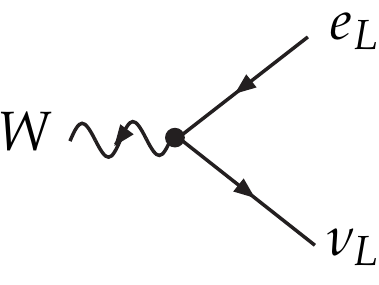} &
\includegraphics[scale=.7]{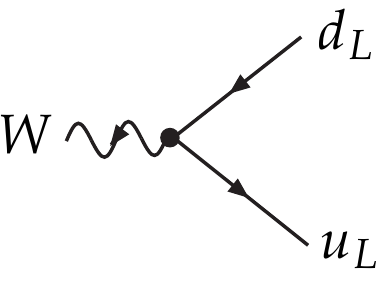} \\
\includegraphics[scale=.7]{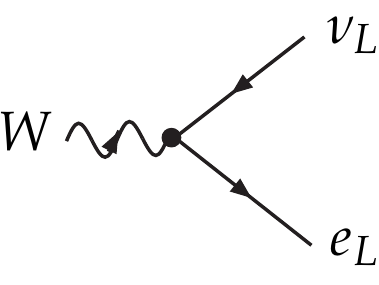} &
\includegraphics[scale=.7]{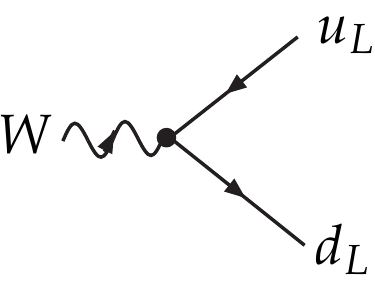}
\end{tabular}
\caption{Weak charged current interactions.
\label{fig:CC}}
\end{figure}

The off-diagonal part of the term $g\overline\Psi_1\gamma^\mu\widetilde W_\mu \Psi_1$ in $ \lag_F$, with
\eq{
\widetilde W_\mu = \dis\frac{1}{2}\pmat {W^3_\mu & \sqrt{2} W^\dagger_\mu \\ 
                                     \sqrt{2} W_\mu & -W^3_\mu}, \quad
W_\mu\equiv\frac{1}{\sqrt{2}}(W^1_\mu + \ii W^2_\mu),
}
gives rise to interactions involving $f_L$ and $f'_L$ and the complex weak field $W_\mu$ (fig.~\ref{fig:CC}),  
\eq{
\lag_F \supset 
\lag_{\rm CC} 
= \frac{g}{\sqrt{2}} \overline{f_L} \gamma^\mu f'_L W^\dagger_\mu + {\rm h.c.}
= \frac{g}{2\sqrt{2}} \overline f \gamma^\mu(1-\gamma_5)f' W^\dagger_\mu + {\rm h.c.} 
}
Note that $W_\mu$, also denoted $W_\mu^-$, annihilates $W^-$ bosons and creates $W^+$ bosons, whereas $W_\mu^\dagger$, also denoted $W_\mu^+$, does the opposite.

\subsubsubsection{Neutral current interactions} 

\begin{figure}
\centering
\begin{tabular}{cc}
\includegraphics[scale=0.7]{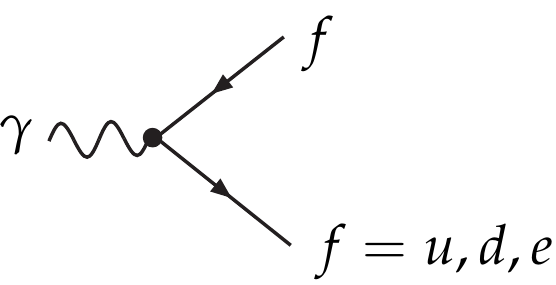} &
\includegraphics[scale=0.7]{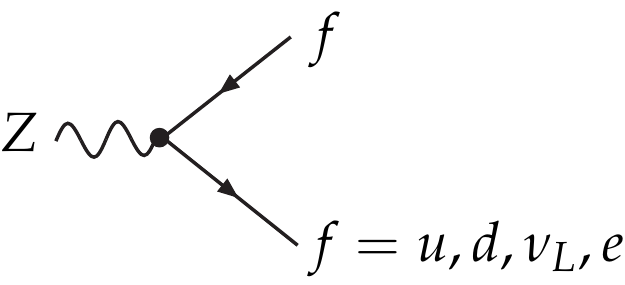}
\end{tabular}
\caption{Electromagnetic and weak neutral current interactions.
\label{fig:NC}}
\end{figure}

The diagonal part of $g\overline\Psi_1\gamma^\mu\widetilde W_\mu \Psi_1$ and the remaining terms,
\eq{
\lag_F \supset
\lag_{\rm NC} = 
  \tfrac{1}{2} g \overline\Psi_1\gamma^\mu \sigma_3 \Psi_1 W_\mu^3 
-       g' (y_1\overline\Psi_1\gamma^\mu\Psi_1
           +y_2\overline\psi_2\gamma^\mu\psi_2
           +y_3\overline\psi_3\gamma^\mu\psi_3) B_\mu ,
}
describe interactions with the vector boson fields $W^3_\mu$ and $B_\mu$ that do not change fermion charge. We are tempted to identify $B_\mu$ with the photon field $A_\mu$ of QED but for that purpose both chiralities of each fermion should couple proportional to the fermion electric charge. However, this is not possible because it would require $y_1 = y_2 = y_3$ and $g' y_j = e Q_j$ simultaneously. Since both $W^3_\mu$ and $B_\mu$ are neutral, one introduces the following orthogonal combinations,
\eq{
\pmat{W^3_\mu \\ B_\mu} \equiv \pmat{c_W & -s_W \\ 
                                     s_W &  c_W}
                               \pmat{Z_\mu \\ A_\mu}, \quad
s_W \equiv \sin\theta_W, \quad c_W \equiv \cos\theta_W,
\label{eq:WBZA}
}
where $\theta_W$ is the weak mixing or Weinberg angle.\footnote{The so-called Weinberg angle was actually introduced by S.~L.~Glashow \cite{Glashow:1961tr}.} Then
\eq{
\lag_{\rm NC} =\sum_{j=1}^3\overline\psi_j\gamma^\mu\left\{ - \left[ gT_3 s_W + g'y_j c_W\right] A_\mu + \left[gT_3 c_W - g'y_j s_W\right] Z_\mu \right\}\psi_j
}
where $T_3=\frac{1}{2}\sigma_3\ (T_3=0)$ is here the third weak isospin component of the doublet (singlet), and we introduced $\psi_1\equiv\Psi_1$ to alleviate the notation. To make $A_\mu$ the photon field is now enough to establish the relations:
\eq{
e = g s_W = g' c_W,  \quad
Q = T_3 + Y.
}
This is the celebrated {\em electroweak unification}, connecting the couplings
$g$ of ${\rm SU}(2)_L$ and $g'$ of ${\rm U}(1)_Y$ to the electromagnetic coupling $e=gg'/\sqrt{g^2+g'^2}$ of ${\rm U}(1)_Q$. 
The electric charges of $f$ and $f'$ are embedded in the operators
\eq{
Q_1 = \pmat{Q_f & 0 \\ 0 & Q_{f'}}, \quad
Q_2 = Q_f, \quad
Q_3 = Q_{f'},
}
so the hyperchages are given in terms of electric charges and weak isospin as shown in Table~\ref{tab:smmultiplets}:
\eq{
y_1 = Q_f-\frac{1}{2} = Q_{f'}+\frac{1}{2}, \quad
y_2 = Q_f, \quad
y_3 = Q_{f'}.
}
As a consequence, $\lag_{\rm NC}=\lag_{\rm QED}+\lag_{\rm NC}^Z$ contains the electromagnetic interactions mediated by the photon ($\gamma$) field (fig.~\ref{fig:NC} left),
\eq{	
\lag_{\rm QED} = -e Q_f \overline f\gamma^\mu f\, A_\mu \quad 
                  +\ (f\to f')
}
and weak neutral current interactions mediated by the $Z$ boson field (fig.~\ref{fig:NC} right),
\eq{
\lag_{\rm NC}^Z = e \overline f \gamma^\mu (v_f - a_f\gamma_5) f\, Z_\mu \quad
                  +\ (f\to f')
}
with
\eq{
v_f = \frac{T_3^{f_L}-2Q_f s^2_W}{2s_Wc_W}, \quad
a_f = \frac{T_3^{f_L}}{2s_Wc_W},
\label{eq:vfaf}
}
where $T_3^{f_L}$ refers to the eigenvalue of $T_3$ that corresponds to $f_L$.
Note that left-handed neutrinos $\nu_L$ have only weak interactions, while right-handed $\nu_R$ would be sterile, hence absent in the original SM with massless neutrinos.

\subsubsubsection{Gauge boson self-interactions} 

\begin{figure}
\centering
\begin{tabular}{cc}
\includegraphics[scale=0.7]{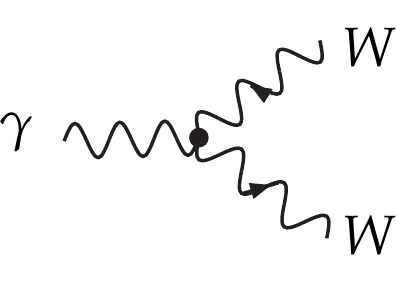} &
\includegraphics[scale=0.7]{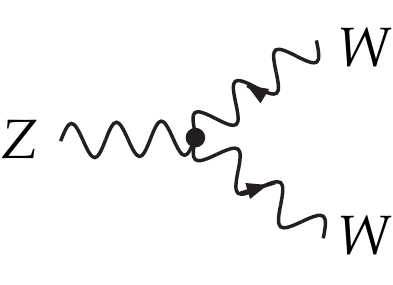}
\end{tabular}
\caption{Triple gauge boson interactions.
\label{fig:TGC}}
\end{figure}

\begin{figure}
\centering
\begin{tabular}{cccc}
\includegraphics[scale=0.7]{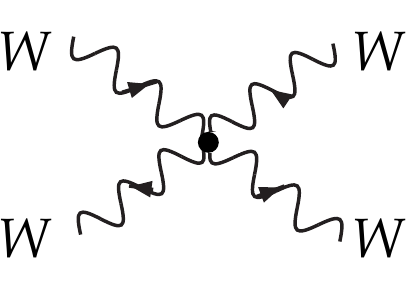} &
\includegraphics[scale=0.7]{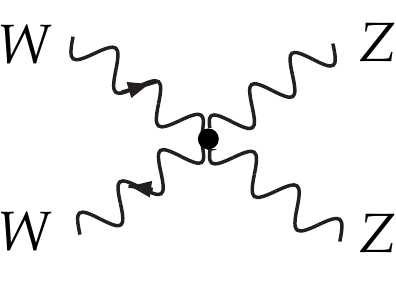} &
\includegraphics[scale=0.7]{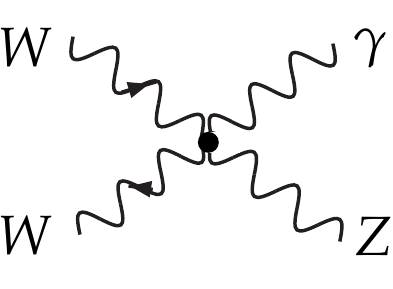} &
\includegraphics[scale=0.7]{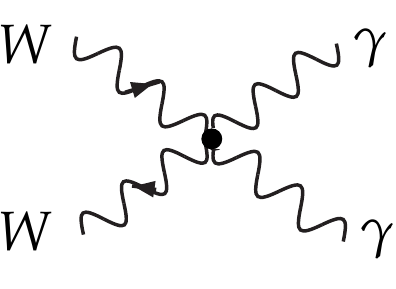}
\end{tabular}
\caption{Quartic gauge boson interactions.
\label{fig:QGC}}
\end{figure}

After some algebra, from the Yang-Mills Lagrangian (\ref{eq:YM}) and the field redefinitions (\ref{eq:WBZA}), one may derive cubic interactions among the gauge boson fields (Fig~\ref{fig:TGC}),
\eq{
\lag_{\rm YM}\supset\lag_3&=-\frac{\ii e c_W}{s_W}\left\{W^{\mu\nu}W_\mu^\dagger Z_\nu-W_{\mu\nu}^\dagger W^\mu Z^\nu - W_\mu^\dagger W_\nu Z^{\mu\nu}\right\} \nn \\
&\quad+\ii e\left\{W^{\mu\nu}W_\mu^\dagger A_\nu - W_{\mu\nu}^\dagger W^\mu A^\nu -  W_\mu^\dagger W_\nu F^{\mu\nu}\right\}
}
with
$F_{\mu\nu} = \d_\mu A_\nu - \d_\nu A_\mu$,
$Z_{\mu\nu} = \d_\mu Z_\nu - \d_\nu Z_\mu$,
$W_{\mu\nu} = \d_\mu W_\nu - \d_\nu W_\mu$, 
and quartic interactions (fig.~\ref{fig:QGC}),
\eq{
\lag_{\rm YM}\supset\lag_4&= -\frac{e^2}{2s^2_W}\left\{\left(W_\mu^\dagger W^\mu\right)^2-W_\mu^\dagger W^{\mu\dagger}W_\nu W^\nu\right\} \nn \\
&\quad-\frac{e^2c^2_W}{s^2_W}\left\{W_\mu^\dagger W^\mu Z_\nu Z^\nu - W_\mu^\dagger Z^\mu W_\nu Z^\nu\right\} \nn \\
&\quad+\frac{e^2c_W}{s_W}\left\{2W_\mu^\dagger W^\mu Z_\nu A^\nu - W_\mu^\dagger Z^\mu W_\nu A^\nu - W_\mu^\dagger A^\mu W_\nu Z^\nu\right\} \nn \\
&\quad-e^2\left\{W_\mu^\dagger W^\mu A_\nu A^\nu - W_\mu^\dagger A^\mu W_\nu A^\nu \right\}.
}
Note that gauge boson self-interactions involve an even number of $W$ and there is no vertex with only $\gamma$ or $Z$.

\subsubsection{Electroweak symmetry breaking: Higgs sector and gauge boson masses}

The weak gauge bosons, $W^\pm$ and $Z$, are massive. To provide them with masses without explicitly breaking gauge invariance one resorts to the Higgs mechanism, that allows to break spontaneously three out of the four generators of {\rm SU}(2)$_L\otimes${\rm U}(1)$_Y$, $T_1,\ T_2,\ T_3,\ Y$, preserving the combination $Q=T_3+Y$ unbroken, so that the photon remains massless.

This cannot be achieved by just introducing one complex scalar field. A complex Higgs doublet of ${\rm SU}(2)$ with the appropriate hypercharge will do the work,
\eq{
\Phi = \pmat{\phi^+ \\ \phi^0}\ , \quad
\bra{0}\Phi\ket{0} \equiv \frac{1}{\sqrt{2}}\pmat{0 \\ v},
}
where $v/\sqrt{2}$ is the Higgs vacuum expectation value, minimum of the Mexican hat potential $V(\Phi)$,
\eq{
V(\Phi) = \mu^2\Phi^\dagger\Phi + \lambda(\Phi^\dagger\Phi)^2
}
and $\mu^2=-\lambda v^2<0$. The Higgs Lagrangian is gauge invariant thanks to the covariant derivative, that leads to interactions with the gauge fields:
\eq{
{\cal L}_\Phi = (D_\mu\Phi)^\dagger D^\mu\Phi - V(\Phi), \qquad
D_\mu\Phi = (\d_\mu - \ii g \widetilde W_\mu + \ii g' y_\Phi B_\mu)\Phi.
\label{eq:lagPhi}
}
By assigning a hypercharge $y_\Phi=\tfrac{1}{2}$ to the Higgs doublet one gets a generator that annihilates the vacuum (associated to the photon field) and three that do not (associated to the massive vector fields), as we wanted:
\eq{
(T_3+Y) \pmat{0 \\ v}=0 \quad\mbox{and}\quad
\{T_1,T_2,T_3-Y\}\pmat{0 \\ v}\ne0.
}

In the unitary gauge one parametrizes the three would-be-Goldstone fields in $\Phi(x)$ as spacetime-dependent phases that can be absorbed (gauged away) thanks to the gauge freedom,
\eq{
\Phi(x) &\equiv \exp\left\{\ii\frac{\sigma_i}{2v}\theta^i(x)\right\}\frac{1}{\sqrt{2}}
\pmat{0 \\ v+H(x)} 
\mapsto\exp\left\{-\ii\frac{\sigma_i}{2v}\theta^i(x)\right\}\Phi(x) =
\frac{1}{\sqrt{2}}\pmat{0 \\ v+H(x)}.
\label{eq:v+H}
}
Only the Higgs field $H(x)$ is physical. The three degrees of freedom apparently lost become the extra (longitudinal) polarizations of $W^\pm$ and $Z$ that are massive particles of spin 1 after the EWSB. Replacing equation (\ref{eq:v+H}) in (\ref{eq:lagPhi}) one gets the gauge boson mass terms:
\eq{
\lag_\Phi \supset \lag_M =
\frac{g^2v^2}{4} W_\mu^\dagger W^\mu + \frac{g^2v^2}{8c^2_W} Z_\mu Z^\mu \quad\Raw\quad
M_W = M_Z c_W = \frac{1}{2} gv.
\label{eq:MWMZ}
}
The fact that the parameter $\rho\equiv M_W^2/(M_Zc_W)^2=1$ is a consequence of the custodial symmetry, a residual global ${\rm SU}(2)$ symmetry of $V(\Phi)$ after EWSB when $\Phi$ is a complex Higgs doublet.\footnote{For instance, if the symmetry breaking is triggered by a complex Higgs triplet one gets $\rho=\frac{1}{2}$.} The $\rho$ parameter measures the relative strength of neutral-current to charged-current interactions, but the tree-level relation $\rho=1$ is slightly broken by quantum corrections (see {\it e.g.} \cite{Schwartz}).

\begin{figure}
\centering
\begin{tabular}{cc}
\includegraphics[scale=0.7]{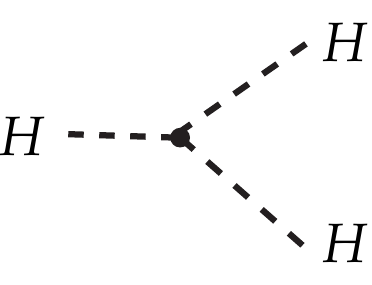} &
\includegraphics[scale=0.7]{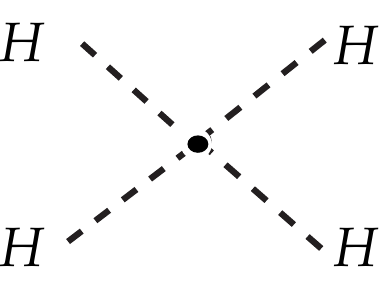}
\end{tabular}
\caption{Higgs boson self-interactions.
\label{fig:Hselfint}}
\end{figure}

\begin{figure}
\centering
\begin{tabular}{cccc}
\includegraphics[scale=0.7]{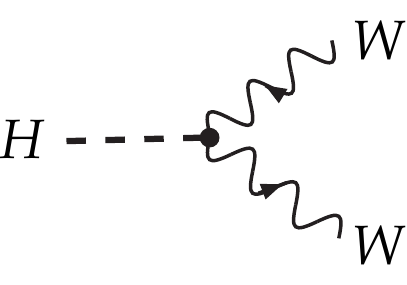}  &
\includegraphics[scale=0.7]{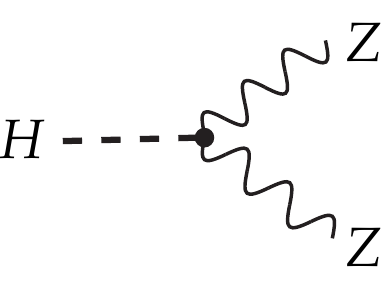}  & 
\includegraphics[scale=0.7]{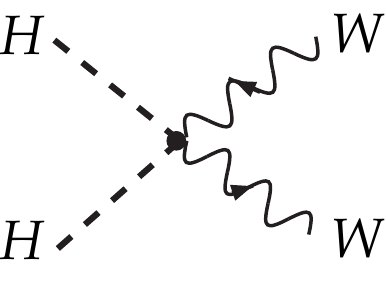} &
\includegraphics[scale=0.7]{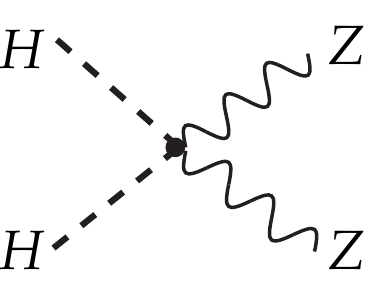}
\end{tabular}
\caption{Higgs-gauge boson interactions.
\label{fig:HVint}}
\end{figure}

In the unitary gauge, where only physical fields are manifest, apart from the gauge boson mass terms, the Higgs Lagrangian contains the (physical) Higgs kinetic terms, its self-interactions  (fig.~\ref{fig:Hselfint}) and the Higgs-gauge boson interactions (fig.~\ref{fig:HVint}):\footnote{An additional constant term $\tfrac{1}{4}\lambda v^4\equiv -\rho_0$ has been omitted. It is irrelevant for the field dynamics but provides a (negative) contribution to the vacuum energy density. See discussion in section~\ref{sec:conrem}.}
\eq{
\lag_\Phi &= \lag_H + \lag_M + \lag_{HV} 
\\
\lag_H &= \frac{1}{2}\partial_\mu H\partial^\mu H - \frac{1}{2}M_H^2 H^2 - \frac{M_H^2}{2v}H^3-\frac{M_H^2}{8v^2}H^4 
\\
\lag_M +
{\cal L}_{HV} &= M^2_W W_\mu^\dagger W^\mu \left\{1 + \frac{2}{v}H + \frac{H^2}{v^2}\right\} + \frac{1}{2} M_Z^2 Z_\mu Z^\mu \left\{1 + \frac{2}{v}H + \frac{H^2}{v^2}\right\},
}
where
\eq{
M_H = \sqrt{-2\mu^2} = \sqrt{2\lambda}\, v.
}


However it is often more convenient to use $R_\xi$ gauges, where the Higgs doublet is parametrized as 
\eq{
\Phi(x) 
\equiv \pmat{ \phi^+(x) \\ \frac{1}{\sqrt{2}}[v+H(x)+\ii\chi(x)] }
}
and $\phi^-(x)\equiv[\phi^+(x)]^\dagger$. Then the Higgs Lagrangian reads
\eq{
\lag_\Phi &= \lag_H + \lag_M + \lag_{HV^2} \nn\\
          &+ (\d_\mu\phi^+)(\d^\mu\phi^-)+\frac{1}{2}(\d_\mu\chi)(\d^\mu\chi) 
           + \ii M_W (W_\mu\d^\mu\phi^+-W_\mu^\dagger\d^\mu\phi^-)+M_Z Z_\mu\d^\mu\chi
           + \dots
}
The omitted terms include trilinear (SSS, SSV, SVV) and quadrilinear (SSSS, SSVV) interactions of vector (V) and scalar (S) fields involving would-be-Goldstone bosons, that can be easily derived. 

In order to define propagators and remove the cross terms $W_\mu\d^\mu\phi^+$, $W_\mu^\dagger\d^\mu\phi^-$, $Z_\mu\d^\mu\chi$ an appropriate gauge-fixing Lagrangian must be added,
\eq{
{\cal L}_{\rm GF}
              = -\frac{1}{2\xi_\gamma}(\d_\mu A^\mu)^2
                -\frac{1}{2\xi_Z}(\d_\mu Z^\mu-\xi_Z M_Z \chi)^2
                -\frac{1}{\xi_W}|\d_\mu W^\mu+\ii\xi_W M_W \phi^-|^2.
}
Then one finds a massless photon propagator, massive propagators for the weak gauge bosons and propagators for the unphysical would-be Goldstone bosons, whose masses are gauge dependent:
\eq{
\widetilde D_{\mu\nu}^\gamma(k)& = \dis\frac{\ii}{k^2+\ii\varepsilon}
\left[- g_{\mu\nu}+(1-\xi_\gamma)\frac{k_\mu k_\nu}{k^2}\right] \\
\widetilde D_{\mu\nu}^Z(k) & = \dis\frac{\ii}{k^2-M^2_Z+\ii\varepsilon}
\left[- g_{\mu\nu}+(1-\xi_Z)\frac{k_\mu k_\nu}{k^2-\xi_Z M^2_Z}\right] 
;\quad \widetilde D^\chi(k) = \dis\frac{\ii}{k^2-\xi_Z M^2_Z + \ii\varepsilon} \\
\widetilde D_{\mu\nu}^W(k) & = \dis\frac{\ii}{k^2-M^2_W+\ii\varepsilon}
\left[- g_{\mu\nu}+(1-\xi_W)\frac{k_\mu k_\nu}{k^2-\xi_W M^2_W}\right]
;\quad \widetilde D^\phi(k) = \dis\frac{\ii}{k^2-\xi_W M^2_W + \ii\varepsilon} .
}
These propagators are much simpler in the Feynman-'t Hooft gauge, where $\xi_\gamma=\xi_Z=\xi_W=1$, which is particularly useful for loop calculations.

Last but not least, the electroweak symmetry group is non-Abelian, so Faddeev-Popov ghosts must be introduced, one per ${\rm SU}(2)$ generator, in order to restore the gauge invariance of the theory at the quantum level. After the EWSB they do not only couple to the ${\rm SU}(2)$ gauge fields but also to the Higgs doublet,
\eq{
{\cal L}_{\rm FP} =
(\d^\mu\overline c_i)(\d_\mu c_i- g\epsilon_{ijk}c_j W^k_\mu) + \mbox{ghost interactions with $\Phi$} .
}
These auxiliary fields $c_i(x)$ $(i=1,2,3)$ are usually written in terms of combinations associated to the ordinary weak and electromagnetic vector fields,
\eq{
c_1 \equiv \frac{1}{\sqrt{2}}(u_++u_-) \ , \quad
c_2 \equiv \frac{\ii}{\sqrt{2}}(u_+-u_-) \ , \quad
c_3 \equiv c_W\ u_Z-s_W\ u_\gamma.
}
For completeness, the full expression of the Faddeev-Popov Lagrangian is as follows:
\eq{
\lag_{\rm FP} =&\quad (\d_\mu\overline u_\gamma)(\d^\mu u_\gamma)
   +(\d_\mu\overline u_Z)(\d^\mu u_Z)
   +(\d_\mu\overline u_+)(\d^\mu u_+)
   +(\d_\mu\overline u_-)(\d^\mu u_-) 
\nn\\
 & + \ii e [(\d^\mu\overline u_+)u_+-(\d^\mu\overline u_-)u_-]A_\mu
   - \frac{\ii e c_W}{s_W} [(\d^\mu\overline u_+)u_+-(\d^\mu\overline u_-)u_-]Z_\mu
\nn\\
 & 
 - \ii e [(\d^\mu\overline u_+)u_\gamma-(\d^\mu\overline u_\gamma)u_-]W^\dagger_\mu
   + \frac{\ii e c_W}{s_W} [(\d^\mu\overline u_+)u_Z-(\d^\mu\overline u_Z)u_-]W^\dagger_\mu 
\nn\\
 & + \ii e [(\d^\mu\overline u_-)u_\gamma-(\d^\mu\overline u_\gamma)u_+]W_\mu
   - \frac{\ii e c_W}{s_W} [(\d^\mu\overline u_-)u_Z-(\d^\mu\overline u_Z)u_+]W_\mu
\nn\\
 & - \xi_Z M_Z^2\ \overline u_Z u_Z
   - \xi_W M_W^2\ \overline u_+ u_+ 
   - \xi_W M_W^2\ \overline u_- u_-
\nn\\
& -e\xi_Z M_Z\ \overline u_Z\left[\frac{1}{2s_Wc_W}Hu_Z               -\frac{1}{2s_W}\left(\phi^+u_-+\phi^-u_+\right)
\right]
\nn\\
& -e\xi_W M_W\ \overline u_+\left[\frac{1}{2s_W}(H+\ii\chi)u_+               -\phi^+\left(u_\gamma-\frac{c^2_W-s^2_W}{2s_Wc_W}u_Z\right)\right]
\nn\\
& -e\xi_W M_W\ \overline u_-\left[\frac{1}{2s_W}(H-\ii\chi)u_-               -\phi^-\left(u_\gamma-\frac{c^2_W-s^2_W}{2s_Wc_W}u_Z\right)\right].
}
From the kinetic terms one can directly see that ghost propagators contain gauge-dependent masses that coincide with those of the partner gauge boson fields in the Feynman-'t Hooft gauge,  
\eq{
\widetilde D^{u_\gamma}(k) = \frac{\ii}{k^2+\ii\varepsilon}\ , \quad
\widetilde D^{u_Z}(k) = \frac{\ii}{k^2-\xi_Z M^2_Z+\ii\varepsilon}\ , \quad
\widetilde D^{u_\pm}(k) = \frac{\ii}{k^2-\xi_W M^2_W+\ii\varepsilon}.
}
The interaction terms include trilinear (UUV) and quadrilinear (SUU) interactions of vector (V) and unphysical ghost fields (U).

\subsubsection{Yukawa interactions: fermion masses}

Masses for quarks and leptons are also needed, without spoiling the gauge symmetry. For that purpose {\em another} interaction is introduced that couples the Higgs doublet $\Phi$ to the fermion fields preserving the ${\rm SU}(2)_L\otimes {\rm U}(1)_Y$ symmetry. Since the left-handed components make a doublet and the right-handed ones are singlets, this can be achieved with the following Yukawa interactions:
\eq{
{\cal L}_{\rm Y} &= -\lambda_d \pmat{\overline u_L & \overline d_L}
\Phi\,
d_R 
- \lambda_u \pmat{\overline u_L & \overline d_L}
\widetilde\Phi\,
 u_R  
\nn\\ &\quad\, 
-\lambda_e \pmat{\overline \nu_{L} & \overline e_L}
\Phi\,
e_R 
-\lambda_\nu \pmat{\overline \nu_{L} & \overline e_L}
\widetilde\Phi\,
\nu_R
\;+\; \mbox{\rm h.c.},
\label{eq:lagyuk}
}
where $\widetilde\Phi\equiv i\sigma_2\Phi^*$ has the appropriate quantum numbers for interactions involving up-type fermion singlets.
\boxexercise{7}{Show that $\widetilde\Phi$ has opposite hypercharge than $\Phi$ but transforms the same under ${\rm SU}(2)$.}
The neutrino Yukawa coupling was not introduced in the original SM with massless neutrinos, but we keep it for further reference. After the EWSB, fermions acquire masses proportional to the corresponding Yukawa couplings,
\eq{
{\cal L}_{\rm Y} \supset -\frac{1}{\sqrt{2}} (v+H)\left\{ \lambda_d\ \overline d d + \lambda_u\ \overline u u + \lambda_e\ \overline e e
{\ +\ \lambda_\nu\ \overline\nu\nu}\right\} \quad\Raw\quad
m_f = \lambda_f\frac{v}{\sqrt{2}},
}
recalling that $\overline f f = \overline{f_L} f_R + \overline{f_R} f_L$.

\subsubsection{Additional generations: fermion mixings}

We know of 3 generations of quarks and leptons in nature. They are identical copies with the same properties under ${\rm {\rm SU}(2)}_L\otimes{\rm {\rm U}(1)}_Y$ differing only in their masses. If one takes $n$ generations and defines
$u^I_i$, $d^I_i$, $\nu^I_i$, $e^I_i$ as the fields corresponding to the $i$-th generation, where the superindex $I$ (standing for `interaction' basis) was omitted so far, the most general gauge-invariant Yukawa Lagrangian is
\eq{
{\cal L}_{\rm Y}
&= -\sum_{ij}\left\{\pmat{\overline u^I_{iL} & \overline d^I_{iL}}
\Phi\,
\lambda^{(d)}_{ij}  d^I_{jR} + \pmat{\overline u^I_{iL} & \overline d^I_{iL}}
\widetilde\Phi\,
\lambda^{(u)}_{ij} u^I_{jR}\right.  
\nn\\
&\left.\quad\quad\quad\ + \pmat{\overline \nu^I_{iL} & \overline e^I_{iL}}
\Phi\,
\lambda^{(e)}_{ij} e^I_{jR} + \pmat{\overline \nu^I_{iL} & \overline e^I_{iL}}
\widetilde\Phi\,
\lambda^{(\nu)}_{ij} \nu^I_{jR}\right\}
 \ + \mbox{\rm h.c.}
\label{eq:lagyuk2}
}
Here $\lambda^{(d)}_{ij}$, $\lambda^{(u)}_{ij}$, $\lambda^{(e)}_{ij}$ (and
$\lambda^{(\nu)}_{ij}$ if present) are $n\times n$ Yukawa matrices in flavor space. After EWSB this Lagrangian contains the following terms in $n$-dimensional matrix form:
\eq{
\lag_{\rm Y} \supset - \left(1 + \frac{H}{v}\right)\,\left\{\,
\overline{\bmd}^I_L \,\bM_d\,
\bmd^I_R \; + \;
\overline{\bmu}^I_L \,\bM_u\,
\bmu^I_R
\; + \;
\overline{\bme}^I_L \,\bM_e\,
\bme^I_R { \; +\;\
\overline{\bmn}^I_L \,\bM_\nu\,
\bmn^I_R} \; +\;
\mathrm{h.c.}\right\} ,
}
where the various mass matrices have the form $(\bM_f)^{}_{ij} =\lambda^{(f)}_{ij} v/\sqrt{2}$. Their diagonalization determines the (physical) mass eigenfields $d_j$, $u_j$, $e_j$, $\nu_j$ in terms of interaction eigenfields $d^I_j$, $u^I_j$, $e^I_j$, $\nu^I_j$, respectively, the latter having well-defined flavor. Each $\bM_f$ can be written as
\eq{
\bM_f=\bH^{}_f\,\cU^{}_f=\bS_f^\dagger\,
\mathbf{\cal M}^{}_f \,\bS^{}_f\,\cU^{}_f 
\quad\Leftrightarrow\quad
\bM_f\bM_f^{\dagger} = \bH^{2}_f = \bS_f^\dagger\,
\mathbf{\cal M}^{2}_f \,\bS^{}_f
}
with $\bH^{}_f\equiv\sqrt{\bM_f\bM_f^{\dagger}}$ a Hermitian positive definite matrix and $\cU^{}_f$ unitary. $\bH^{}_f$ can be diagonalized by a unitary matrix $\bS^{}_f$ and the resulting $\mathbf{\cal M}^{}_f$ is diagonal and positive definite. In the physical basis, where mass matrices are diagonal, $\mathbf{\cal M}^{}_d = \mathrm{diag}(m_d,m_s,m_b,\ldots)$, $\mathbf{\cal M}^{}_u = \mathrm{diag}(m_u,m_c,m_t,\ldots)$, $\mathbf{\cal M}^{}_e = \mathrm{diag}(m_e,m_\mu,m_\tau,\ldots)$, $\mathbf{\cal M}^{}_\nu = \mathrm{diag}(m_{\nu_e},m_{\nu_\mu},m_{\nu_\tau}, \ldots)$,
one finds that fermion couplings to the Higgs are proportional to fermion masses,
\eq{
\lag_{\rm Y}\, \supset\, - \left(1 + \frac{H}{v}\right)\,\left\{\,
\overline{\bmd}\,\mathbf{\cal M}^{}_d\,\bmd \; + \;
\overline{\bmu}\, \mathbf{\cal M}^{}_u\,\bmu \; + \;
\overline{\bme}\,\mathbf{\cal M}^{}_e\,\bme { \; + \;
\overline{\bmn}\,\mathbf{\cal M}^{}_\nu\,\bmn} \,\right\}.
}

\begin{figure}
\centering
\begin{tabular}{ccc}
\raisebox{0\height}{\includegraphics[scale=0.7]{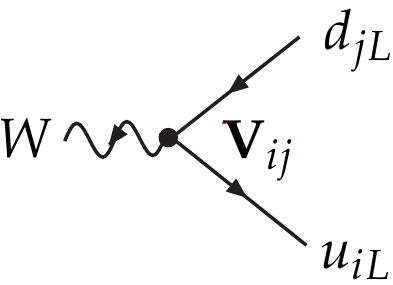}} & \quad
\raisebox{0\height}{\includegraphics[scale=0.7]{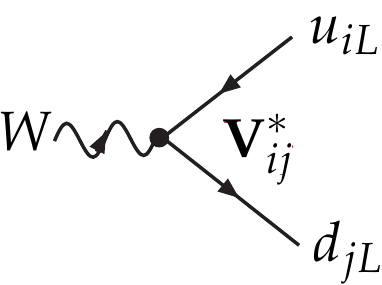}} & \qquad
\raisebox{.02\height}{\includegraphics[scale=0.7]{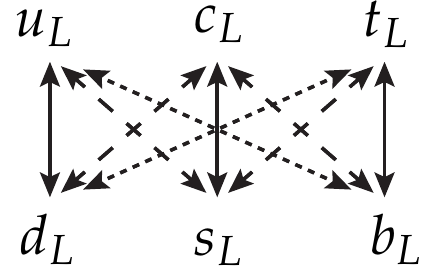}}
\end{tabular}
\caption{Weak charged currents change quark family proportionally to the CKM matrix elements ${\bf V}_{ij}$.
\label{fig:CC-CKM}}
\end{figure}

Replacing now interaction with mass eigenfields,
\eq{
\bmd_L &\equiv \bS_d\ \bmd^I_L\ , &
\bmu_L &\equiv \bS_u\ \bmu^I_L\ , &
\bme_L &\equiv \bS_e\ \bme^I_L\ , &
\bmn_L  &\equiv \bS_\nu\ \bmn^I_L
\label{eq:fermix}
\\
\bmd_R &\equiv \bS_d \cU_d\ \bmd^I_R\ , &
\bmu_R &\equiv \bS_u \cU_u\ \bmu^I_R\ , &
\bme_R &\equiv \bS_e \cU_e\ \bme^I_R\ , &
\bmn_R  &\equiv \bS_\nu \cU_\nu\ \bmn^I_R\ ,
}
it is apparent that neutral-current interactions will keep the same form, because 
$\overline{\bmf}^I_L\gamma^\mu\bmf^I_L = \overline{\bmf}^{}_L\gamma^\mu\bmf^{}_L$ and
$\overline{\bmf}^I_R\gamma^\mu\bmf^I_R = \overline{\bmf}^{}_R\gamma^\mu\bmf^{}_R$, implying that there are no flavor changing neutral currents (FCNC) at tree level. However, the operators involved in charged current interaction terms are not necessarily diagonal in the basis of mass eigenfields. For instance, in the quark sector,
\eq{
\overline{\bmu}^I_L \gamma^\mu \bmd^I_L =
\overline{\bmu}^{}_L \gamma^\mu\,\bS_u\,\bS_d^\dagger\,\bmd^{}_L =
\overline{\bmu}_L\gamma^\mu \bV\bmd_L.
}
The unitary matrix $\bV\equiv\bS_u\,\bS_d^\dagger$ is the Cabibbo-Kobayashi-Maskawa (CKM) mixing matrix \cite{Cabibbo:1963yz,Kobayashi:1973fv} accounting for quark flavor misalignment and inducing inter-family transitions (fig.~\ref{fig:CC-CKM}),
\eq{
\lag_{\rm CC} 
= \frac{g}{\sqrt{2}} \sum_{ij}
\overline u_{Li} \gamma^\mu\ \bV_{ij}\ d_{Lj}\ W^\dagger_\mu +  
\mathrm{h.c.}
= \frac{g}{2\sqrt{2}} \sum_{ij}
\overline u_i \gamma^\mu(1-\gamma_5)\ \bV_{ij}\ d_j\ W^\dagger_\mu +  
\mathrm{h.c.}
}
Thanks to these flavor changes in charged currents, FCNC will appear at the loop level but they are then suppressed (GIM mechanism \cite{Glashow:1970gm}). 

Note that if $u_i$ or $d_j$ had degenerate masses, which is not the case, one could choose $\bS_u=\bS_d$ by field redefinitions and quark families would not mix. Masses and mixings are observable, but the matrix elements of $\bS_u$ and $\bS_d$ are not. Applying the same reasoning, in a lepton sector with massless neutrinos there is no lepton mixing. 

At this point, it is important to discuss how many of the mixing parameters we have introduced are physical. The number of real parameters of a general $n\times n$ unitary matrix, like the CKM, is
\eq{
n^2 = n (n-1)/2 \mbox{ moduli} + n (n+1)/2 \mbox{ phases}.
}
However some phases are unphysical since they can be absorbed by field phase-redefinitions,
\eq{
u_i\to {\rm e}^{\ii\alpha_i}\, u_i\ , \quad
d_j\to{\rm e}^{\ii\beta_j}\, d_j \quad\Raw\quad
\bV_{ij}\to \bV_{ij}\,{\rm e}^{-\ii(\alpha_i-\beta_j)}.
}
Therefore, after removing $2 n-1$ phases, the number of physical parameters is
\eq{
(n-1)^2 = n(n-1)/2 \mbox{ moduli } + (n-1)(n-2)/2 \mbox{ phases}.
}
In particular, for the case of $n=2$ generations, there is only 1 parameter, the Cabibbo angle $\theta_C$:
\eq{
\bV = 
\pmat{\cos{\theta_C} &\sin{\theta_C} \\[2pt] -\sin{\theta_C}& \cos{\theta_C}} .
}
For the actual case of $n=3$ generations, there are 3 angles and 1 phase. In the so-called standard parametrization,
\eq{
\bV &=   \pmat{\bV_{ud} & \bV_{us} & \bV_{ub} \\
               \bV_{cd} & \bV_{cs} & \bV_{cb} \\
               \bV_{td} & \bV_{ts} & \bV_{tb}}
=\pmat{1 & 0 & 0 \\ 0 & c_{23} & s_{23} \\ 0 & -s_{23} & c_{23}} 
 \pmat{c_{13} & 0 & s_{13}{\rm e}^{-\ii\delta} \\ 0 & 1 & 0 \\ -s_{13}{\rm e}^{\ii\delta} & 0 & c_{13}}
\pmat{c_{12} & s_{12} & 0 \\ -s_{12} & c_{12} & 0 \\ 0 & 0 & 1}
\nn \\
&=\pmat{
c_{12} c_{13}  & s_{12} c_{13} & s_{13} {\rm e}^{-\ii\delta} \\[2pt]
-s_{12} c_{23}-c_{12} s_{23} s_{13} {\rm e}^{\ii\delta} &
c_{12} c_{23}- s_{12} s_{23} s_{13} {\rm e}^{\ii\delta} &
s_{23} c_{13}  \\[2pt]
s_{12} s_{23}-c_{12} c_{23} s_{13} {\rm e}^{\ii\delta} &
-c_{12} s_{23}- s_{12} c_{23} s_{13} {\rm e}^{\ii\delta} &
c_{23} c_{13}}
}
with $c_{ij} \equiv \cos\theta_{ij}\geq 0$,
$s_{ij} \equiv \sin{\theta_{ij}}\geq 0$ ($i<j=1,2,3$) and  
$0\leq \delta\leq 2\pi$. The complex phase $\delta$ is the only source of CP violation in the SM Lagrangian, requiring the existence of at least three generations of quarks. Since quarks are confined in hadrons by the strong interaction, the values of the CKM parameters are obtained from a variety of hadronic weak decays \cite{ParticleDataGroup:2020ssz},
\eq{
\theta_{12} \equiv \theta_C \approx 13^\circ, \quad
\theta_{23} \approx 2.3^\circ, \quad 
\theta_{13} \approx 0.2^\circ, \quad
\delta \approx 68^\circ.
}
Interestingly, any CP-violating observable must be proportional to the Jarlskog invariant \cite{Jarlskog:1985ht} given by 
${\rm Im}(V_{ij}V_{kl}V_{il}^*V_{kj}^*)=J\sum_{m,n}\epsilon_{ikm}\epsilon_{jln}$
(phase-convention independent). In the standard parametrization $J=c_{12}c_{23}c^2_{13}s_{12}s_{23}s_{13}\sin\delta$.
The empirical value of $J\approx 3\times10^{-5}$ is small compared with its mathematical maximum of $1/(6\sqrt{3})\approx 0.1$, showing that CP violation is suppressed in the quark sector. 

\begin{figure}
\centering\includegraphics[scale=0.7]{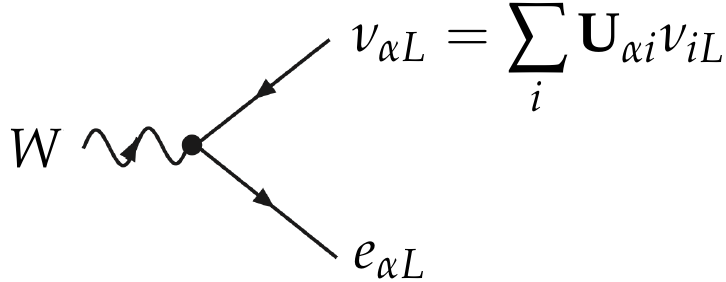}
\caption{A neutrino flavor eigenstate $\nu_\alpha$, produced/detected together with a charged lepton $e_\alpha$, is a coherent superposition of mass eigenstates $\nu_i$, hence the flavor oscillates as it propagates.
\label{fig:nuCC}}
\end{figure}

\begin{figure}
\centering\includegraphics[scale=0.6]{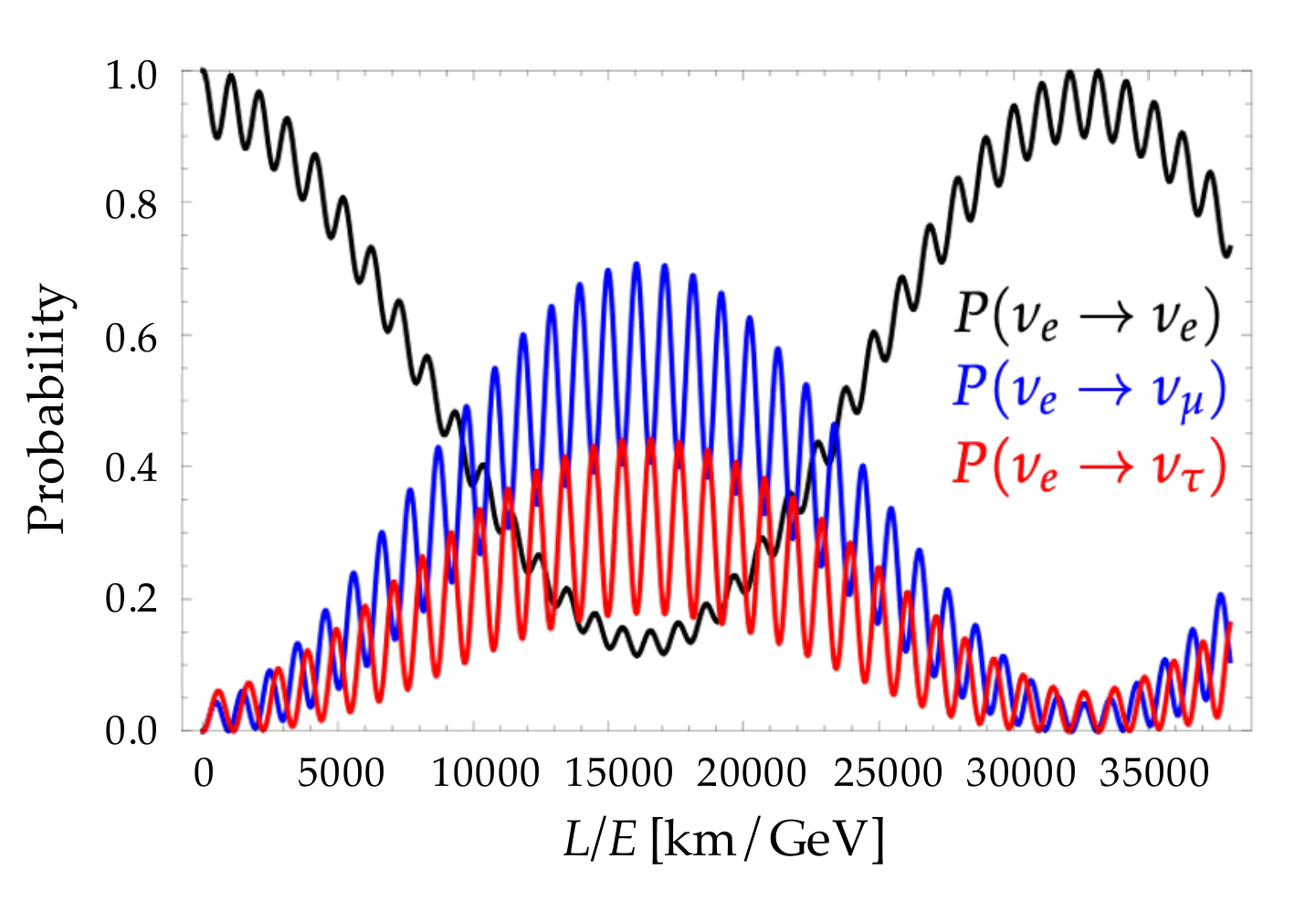}
\caption{Vacuum oscillation probabilities for an initial $\nu_e$ using experimental inputs \eqref{eq:Deltamnu} and \eqref{eq:Thetanu}. 
\label{fig:oscprob}}
\end{figure}

As already mentioned, if neutrinos were massless there would be no lepton mixing. However, the observed phenomenon of neutrino oscillation requires that neutrinos have non-degenerate masses (though very light) and mix. A possible minimal extension of the original SM consists of introducing gauge-singlet neutrinos $\nu_R$ with just Yukawa couplings to the Higgs and the lepton doublet, like the other fermions, as was suggested in equations \eqref{eq:lagyuk} and (\ref{eq:lagyuk2}). This $\nu$SM \cite{Mohapatra:1998rq} is however not very satisfactory: in order to get neutrino masses $m_\nu\lesssim 0.1$~eV one needs tiny Yukawa couplings $\lambda_\nu=\sqrt{2} m_\nu/v\lesssim 10^{-12}$, which apart from being unnatural would predict untestable phenomenology. Alternatively, one can exploit that neutrinos are {\em special} because, in contrast to the other fermions, neutrinos {\it may} be their own antiparticle (Majorana fermions). Then neutrinos can have gauge invariant (but lepton number violating) Majorana mass terms ($m_R$), in addition to the usual Dirac mass terms ($m_D$) from Yukawa interactions with the Higgs doublet, opening the possibility of new mechanisms for the generation of masses and mixings. Particularly interesting is the type-I seesaw mechanism \cite{GellMann:1980vs,Yanagida:1979as} that explains why the active neutrinos are so light by introducing gauge singlets $N_R$ with very large Majorana mass terms $m_R\gtrsim 10^{14}$~GeV and Dirac masses $m_D\sim v/\sqrt{2}\sim 100$~GeV: the resulting mass eigenstates comprise light Majorana neutrinos that are very approximately $\nu=\nu_L+\nu_L^c$, of masses $m_\nu\approx m_D^2/m_R$, and super heavy ones, nearly $N=N_R^c+N_R$, of masses $m_N\approx m_R$, with negligible light-heavy mixings of order $m_D/m_R\sim \sqrt{m_\nu/m_N}$. Majorana fields are self-conjugate ($\nu=\nu^c$), so their chiral components are related.\footnote{
Furthermore, if neutrinos are Majorana particles resulting from the admixture of active and singlet neutrinos, only the active components would experience charged current interactions,
$$
\lag_{\rm CC}\supset \frac{g}{\sqrt{2}}
\overline{e}_i\gamma^\mu P_L {\bf B}_{ij} \nu_j W_\mu + {\rm h.c.}, \quad
{\bf B}_{ij} = \sum_{\alpha=1}^3 \delta_{i\alpha}\bU_{\alpha j}
$$
and there would be FCNC at tree level in the neutrino sector involving both chiral components,
$$
\lag_{\rm NC}^Z \supset \frac{g}{4c_W}
\sum_{ij} \overline{\nu}_j\gamma^\mu(P_L {\bf C}_{ij}-P_R {\bf C}_{ij}^*)\nu_j Z_\mu, \quad
{\bf C}_{ij} = \sum_{\alpha=1}^3 \bU_{\alpha i}^*\bU_{\alpha j}.
$$
See for instance Ref.~\cite{Hernandez-Tome:2020lmh}.
}
As a consequence the intergenerational lepton mixings include additional CP phases that now cannot be absorbed because it is no longer possible to perform neutrino field phase-redefinitions. In any case, global fits to neutrino oscillations are compatible with 3 generations of active neutrino flavors $\nu_{\alpha L}$ ($\alpha=e,\mu,\tau$) that are an admixture of 3 light neutrino mass-eigenstates $\nu_{iL}$ ($i=1,2,3$), 
\eq{
\nu_{\alpha L}=\sum_i\bU_{\alpha i}\nu_{iL},
\label{eq:nualpha}
}
where the Pontecorvo-Maki-Nakagawa-Sakata (PMNS) matrix $\bU$ \cite{Pontecorvo:1957cp,Maki:1962mu,Pontecorvo:1967fh} is the unitary mixing matrix $\bV_\nu^\dagger$ in equation \eqref{eq:fermix}, or perhaps, if neutrinos are Majorana particles, the nearly unitary $3\times3$ block of a larger unitary matrix diagonalizing the Majorana neutrino mass matrix that includes both light and heavy species. The oscillation phenomenon occurs because the mass differences among the various light mass eigenstates are so small that the {\it coherent superposition} $\nu_\alpha$ in equation (\ref{eq:nualpha}) can be produced or detected in a charged current interaction with the corresponding lepton $e_\alpha$ (e, $\mu$, $\tau$), as in fig.~\ref{fig:nuCC}. Then the probability that a (relativistic) neutrino in a quantum state of flavor $\alpha$ is detected as a flavor $\beta$ after traveling (in vacuum) a distance $L=t$ is given by (see fig.~\ref{fig:oscprob})
\eq{
\ket{\nu_\alpha;t}&=\dis\sum_i\bU_{\alpha i}{\rm e}^{-\ii E_i t}\ket{\nu_i},\quad
E_i\approx E+\dis\frac{m^2_i}{2E}
\\
\Raw\quad
\braket{\nu_\beta}{\nu_\alpha;t} &= \dis\sum_i\bU_{\beta i}^*\bU_{\alpha i}{\rm e}^{-\ii E_i t}
\\
\Raw\quad
P(\nu_\alpha\to\nu_\beta;L) &= |\braket{\nu_\beta}{\nu_\alpha;L}|^2 = \dis\sum_{ij}\bU_{\beta i}^*\bU_{\alpha i}
\bU_{\beta j}\bU_{\alpha j}^*\exp\left(-\ii \dis\frac{\Delta m^2_{ij}}{2E} L\right)
\label{eq:oscprob}
}
where $E\approx p$ is the momentum of the relativistic neutrino of mass $m_i$
and $\Delta m^2_{ij}\equiv m^2_i-m^2_j$. Charged lepton flavors do not oscillate because $|\Delta m^2_{ij}| \ll \Delta m^2_{\mu e}$ \cite{Akhmedov:2007fk}, so they can be taken as mass eigenstates. 

In the standard parametrization, the PMNS matrix reads
\eq{
\bU&=\pmat{\bU_{e1} & \bU_{e2} & \bU_{e3} \\
               \bU_{\mu1} & \bU_{\mu2} & \bU_{\mu3} \\
               \bU_{\tau1} & \bU_{\tau2} & \bU_{\tau3}}
\nn \\
&=\pmat{
c_{12}\, c_{13}  & s_{12}\, c_{13} & s_{13}\, {\rm e}^{-i\delta} \\
-s_{12}\, c_{23}-c_{12}\, s_{23}\, s_{13}\, {\rm e}^{i\delta} &
c_{12}\, c_{23}- s_{12}\, s_{23}\, s_{13}\, {\rm e}^{i\delta} &
s_{23}\, c_{13}  \\[2pt]
s_{12}\, s_{23}-c_{12}\, c_{23}\, s_{13}\, {\rm e}^{i\delta} &
-c_{12}\, s_{23}- s_{12}\, c_{23}\, s_{13}\, {\rm e}^{i\delta} &
c_{23}\, c_{13}}
\pmat{1 & 0 & 0 \\ 0 & \e^{i\alpha_{21}/2} & 0 \\ 0 & 0 & \e^{i\alpha_{31}/2}},
}
where additional phases $\alpha_{21}$, $\alpha_{31}$ are needed if neutrinos are Majorana particles, as mentioned above, and the rest are analogous to the CKM mixing parameters, though they have different values. Neutrino mass differences and mixing parameters are constrained by a good number of oscillation experiments using solar, atmospheric, accelerator and reactor neutrinos \cite{deSalas:2020pgw}, 
\eq{
\Delta m^2_{21} \approx 7.5\times 10^{-5}\mbox{ eV}^2, \quad
|\Delta m^2_{31}| \approx 2.5\times 10^{-3}\mbox{ eV}^2
\label{eq:Deltamnu}
\\
\theta_{12}\equiv\theta_\odot \approx 34^\circ, \quad
\theta_{23}\equiv\theta_{\rm atm} \approx 49^\circ, \quad 
\theta_{13} \approx 8^\circ.
\label{eq:Thetanu}
}
The best fit value of the Dirac phase $\delta$ depends on the sign of $\Delta m^2_{31}$, that is whether the ordering of neutrino masses is normal (NO) or inverted (IO): a CP-conserving value $\delta\approx 180^\circ$ is favored by NO but an almost maximal CP-violating $\delta\approx 280^\circ$ is favored by IO. Note that oscillations are not sensitive to Majorana phases as is apparent from equation \eqref{eq:oscprob}. A type of experiments that can elucidate whether neutrinos are Dirac or Majorana fermions would be the observation of neutrinoless double-beta decays \cite{Bilenky:2014uka}.

\subsection{Strong interactions}\label{sec:QCD}

Quantum Chromodynamics (QCD), described by the gauge group ${\rm SU}(3)_c$, is the theory of strong interactions. Quarks and gluons are the fundamental degrees of freedom but they never show up as free states, they are bound in composite systems (hadrons) due to the phenomenon of confinement. Quarks and gluons carry color charges. Hadrons are color neutral systems that can be obtained combining three quarks (baryons) or a quark and an antiquark (mesons). For instance, protons and neutrons are baryons or pions and kaons are mesons (table~\ref{tab:hadrons}). Other (exotic) possible colorless states that have been found are glueballs, tetraquaks and pentaquarks. 

The strong interaction does not only bind quarks in hadrons, it is also responsible for the stability of atomic nuclei: the nucleon-nucleon interaction is an attractive residual strong force, still greater than the electric repulsion of proton charges. Its strength
is independent of the quark flavor, which explains why hadrons made of the lighest quarks have similar masses because actually most of hadron masses comes from the strong interaction energy. In fact, $m_u \approx 2$~MeV, $m_d \approx 5$~MeV and $m_s\approx 100$~MeV, so quark masses account for only 1\% of proton or neutron (nucleon) masses (99\% of the mass of any atom comes from the nucleon binding energy!) and the mass similarities in table~\ref{tab:hadrons} are justified. 

\begin{table}
\centering
\begin{tabular}{cc}
\begin{tabular}[t]{|c|c|r|}
\hline
\multicolumn{3}{|c|}{\bf Mesons} \\
\hline\hline
$J=0$ & quarks & mass
\\
\hline\hline
$\pi^0$ & $u\overline u$, $d\overline d$ & 135
\\
$\pi^+$ & $u\overline d$ & 140
\\
$\pi^-$ & $ d\overline u$ & 140
\\
\hline
$K^+$ & $u\overline s$ & 493
\\
$K^-$ & $s\overline u$ & 493
\\
$K^0$ & $d\overline s$ & 498
\\
$\overline K^0$ & $ s\overline d$ & 498
\\
\hline
$\eta$ & $u\overline u$, $d\overline d$, $s\overline s$ & 547
\\
\hline
\end{tabular} &
\begin{tabular}[t]{|c|c|r|}
\hline
\multicolumn{3}{|c|}{\bf Baryons} \\
\hline\hline
$J=\frac{1}{2}$ & quarks & mass
\\
\hline\hline
$p$ & $uud$ & 938
\\
$n$ & $udd$ & 939
\\
\hline
$\Lambda$ & $uds$ & 1115
\\
\hline
$\Sigma^+$ & $uus$ & 1189
\\
$\Sigma^0$ & $uds$ & 1192
\\
$\Sigma^-$ & $dds$ & 1197
\\
\hline
$\Xi^0$ & $uss$ & 1314
\\
$\Xi^-$ & $dss$ & 1321
\\
\hline
\end{tabular}
\end{tabular}
\caption{Lightest hadrons made of quarks $u,d$ and $s$. Masses in MeV.
\label{tab:hadrons}}
\end{table}

The Lagrangian, gauge invariant under ${\rm SU}(3)_c$ transformations, reads
\eq{
\lag_{\rm QCD} = \overline\Psi_{fi}\left(\ii\slashed{D}_{ij}-m_f\delta_{ij}\right)\Psi_{fj}
-\frac{1}{4}F_{\mu\nu}^a F^{a\,\mu\nu}
\quad\mbox{(flavor diagonal)},
\label{eq:lagQCD}
}
where $F_{\mu\nu}^a=\d_\mu{\cal A}_\nu^a-\d_\nu{\cal A}_\mu^a+g_s f^{abc}{\cal A}_\mu^b{\cal A}_\nu^c$ and $\Psi_f$ is any quark field of flavor $f$ whose color components $\Psi_{fi}$ lay in the fundamental representation of ${\rm SU}(3)$,
\eq{
\Psi_{fi}\quad ({\bf 3}): \quad
f=u, d, s, c, b, t \quad\mbox{(flavor index)}; \quad
i=1,\dots,N_c=3 \quad \mbox{(color index)}. 
}
Gluons ${\cal A}^a_\mu$ come in $N_c^2-1=8$ combinations of color and anticolor transform in the adjoint representation,
\eq{
{\cal A}_\mu^a\quad ({\bf 8}): \quad 
a=1,\dots,N_c^2-1=8 \quad \mbox{(color index)}.
}
The quark kinetic terms and quark-gluon interactions come from the covariant derivative,
\eq{
(D_\mu)_{ij} &= \delta_{ij}\d_\mu-\ii g_s t_{ij}^a {\cal A}_\mu^a\ , \quad
t^a_{ij}=\frac{1}{2}\lambda^a_{ij} \quad\mbox{(8 Gell-Mann matrices $3\times3$)}.
}
The Yang-Mills part contains the gluon kinetic terms and their self-interactions fixed by the ${\rm SU}(3)$ structure constants $f^{abc}$,
\eq{
\lag_{\rm kin} &= 
-\frac{1}{4}(\d_\mu {\cal A}_\nu^a - \d_\nu {\cal A}_\mu^a)
            (\d^\mu {\cal A}^{a\,\nu} - \d^\nu {\cal A}^{a\,\mu}), \nn\\
\lag_{\rm cubic} &= -\frac{1}{2} g_s f_{abc}\ (\d_\mu {\cal A}_\nu^a-\d_\nu {\cal A}_\mu^a)
{\cal A}^{b\,\mu} {\cal A}^{c\,\nu},
\nn\\
\lag_{\rm quartic} &=-\frac{1}{4} g_s^2 f_{abe} f_{cde}\ {\cal A}^a_\mu {\cal A}^b_\nu {\cal A}^{c\,\mu} {\cal A}^{d\,\nu}.
}
All interactions depend on a single coupling constant $g_s$ and are fully determined by the symmetry.\footnote{The quantum Lagrangian also includes gauge-fixing and Faddeev-Popov terms.} The quark-gluon and gluon-gluon interaction vertices and the corresponding Feynman rules are the following (momenta are taken incoming):
\eq{
\raisebox{-.4\height}{\includegraphics[scale=0.6]{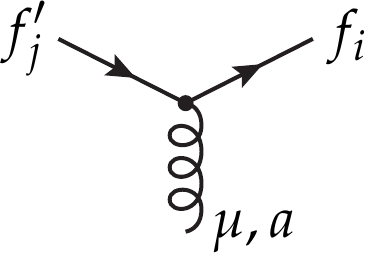}}\quad\!\!\! &=
\ii g_s t^a_{ij} \gamma^\mu \delta_{ff'}
\label{eq:QCD-FR1}
\\
\raisebox{-.4\height}{\includegraphics[scale=0.6]{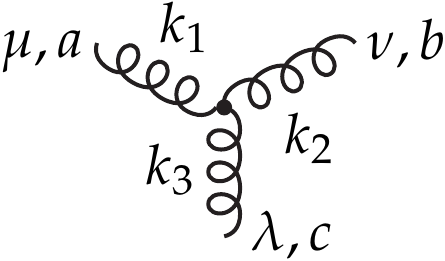}} &=
 g_s f_{abc}\ [g_{\mu\nu}(k_1-k_2)_\lambda +g_{\nu\lambda}(k_2-k_3)_\mu +g_{\lambda\mu}(k_3-k_1)_\nu]
\label{eq:QCD-FR2}
\\
\raisebox{-.4\height}{\includegraphics[scale=0.6]{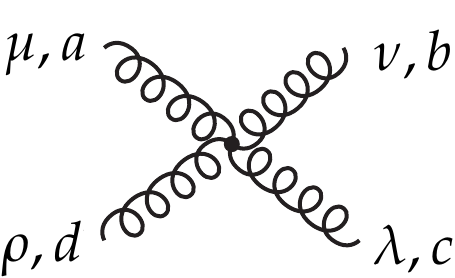}} &=
-\ii g_s^2 \left[\begin{array}{r}
    f_{abe}f_{cde}\ (g_{\mu\lambda}g_{\nu\rho}-g_{\mu\rho}g_{\nu\lambda})
\\+ f_{ace}f_{dbe}\ (g_{\mu\rho}g_{\nu\lambda}-g_{\mu\nu}g_{\lambda\rho})
\\+ f_{ade}f_{bce}\ (g_{\mu\nu}g_{\lambda\rho}-g_{\mu\lambda}g_{\nu\rho})
\end{array}\right].
\label{eq:QCD-FR3}
}

The 3 color charges of quarks are usually dubbed red ($R$), green ($G$) and blue ($B$). Antiquarks have opposite color charges ($\overline{R}$, $\overline{G}$, $\overline{B}$). 
The gluons are 8 independent non-singlet combinations of color and anticolor in ${\bf 3}\otimes\overline{\bf 3}={\bf 1}\oplus{\bf 8}$, associated to the generators
\eq{
\lambda_1 &=G\overline{R}+R\overline{G},&
\lambda_2 &=\ii(G\overline{R}-R\overline{G}),&
\lambda_3 &=R\overline{R}-G\overline{G},&
\lambda_4 &=B\overline{R}+R\overline{B}, \\
\lambda_5 &=\ii(B\overline{R}-R\overline{B}),&
\lambda_6 &=B\overline{G}+G\overline{B},&
\lambda_7 &=\ii(B\overline{G}-G\overline{B}),&
\lambda_8 &=\frac{1}{\sqrt{3}}(R\overline{R}+G\overline{G}-2B\overline{B}).
}
A strong interaction {\it repaints} the quark. For example,
\eq{
\raisebox{-0.4\height}{\includegraphics[scale=0.6]{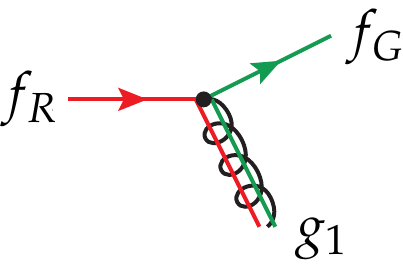}}
\qquad \overline\Psi_i t^1_{ij}\Psi_j \sim \frac{1}{2} \pmat{0&1&0}\pmat{0&1&0\\1&0&0\\0&0&0}\pmat{1\\0\\0}.
}
If a color-singlet massless gluon state existed (associated to the generator $\frac{1}{\sqrt{3}}(R\overline{R} + G\overline{G} + B\overline{B})$), it would give rise to a strong force of infinite range!
Conversely, only color-singlet states can appear as free particles:
\eq{
q\bar{q}' &\quad ({\bf 1}\in{\bf 3}\otimes{\bf \bar{3}}={\bf 1}\,{ \oplus\,{\bf 8}})  &\Raw\quad& \mbox{mesons}\quad
\dis\frac{1}{\sqrt{3}}\delta^{ij}\ket{q_i \bar{q}'_j}
\\[1ex]
qq'q'' &\quad ({\bf 1}\in{\bf 3}\otimes{\bf 3}\otimes{\bf 3}={\bf 1}\,{ \oplus\,{\bf 8}\,\oplus\,{\bf 8}\,\oplus{\bf 10}}) &\Raw\quad&  \mbox{baryons}\quad
\dis\frac{1}{\sqrt{6}}\epsilon^{ijk}\ket{q_i q'_j q''_k}
}
where $i,j,k\in\{R,G,B\}$.
However $qq'$ bound states do no exist because there are no color singlets in ${\bf 3}\otimes{\bf 3}={\bf\bar 3}\oplus{\bf 6}$.

Computations in QCD make an extensive use of the color algebra, for which the following identities result very useful:
\eq{
\raisebox{-.45\height}{\includegraphics[scale=.6]{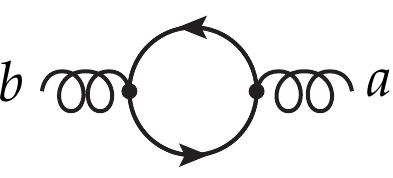}}\quad
&\propto {\rm Tr}(t^at^b)=T_R\delta_{ab},\quad T_R=\dis\frac{1}{2},
\\
\raisebox{-.45\height}{\includegraphics[scale=.6]{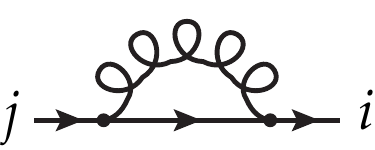}}\quad\
&\propto t^a_{ik}t^a_{kj}=C_F\delta_{ij},\quad C_F=\dis\frac{N_c^2-1}{2N_c}=\frac{4}{3},
\\
\raisebox{-.45\height}{ \includegraphics[scale=.6]{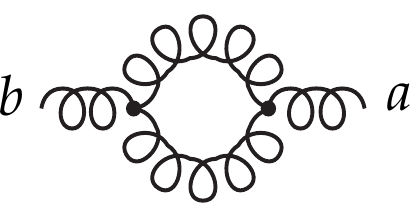}}\quad
&\propto f_{acd}f_{bcd}=C_A\delta_{ab},\quad C_A=N_c=3 .
}
These numbers indeed indicate that the probability of a gluon to create a quark-antiquark pair is smaller than that of a quark to radiate a gluon, that in turn is smaller than the probability of a gluon to create a pair of gluons. 

\begin{figure}
\centering
\begin{tabular}{cc}
\includegraphics[scale=0.6]{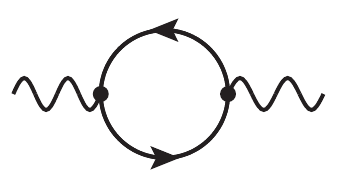} &\hfill
\includegraphics[scale=0.6]{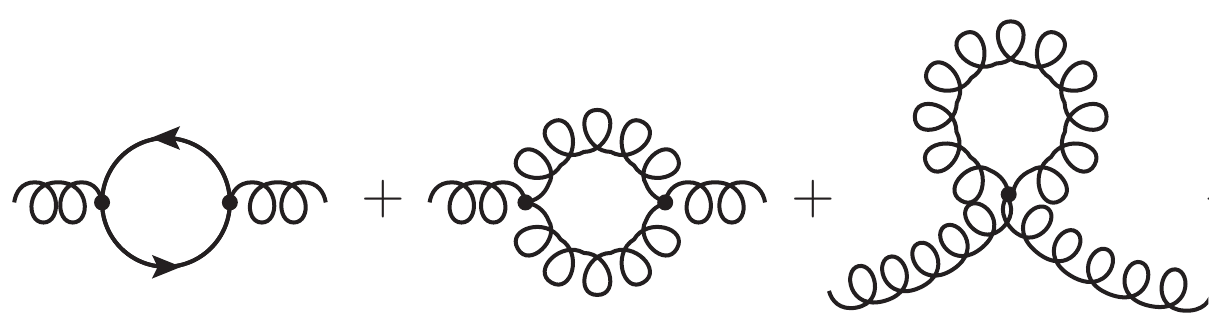}
\\
\includegraphics[scale=0.45]{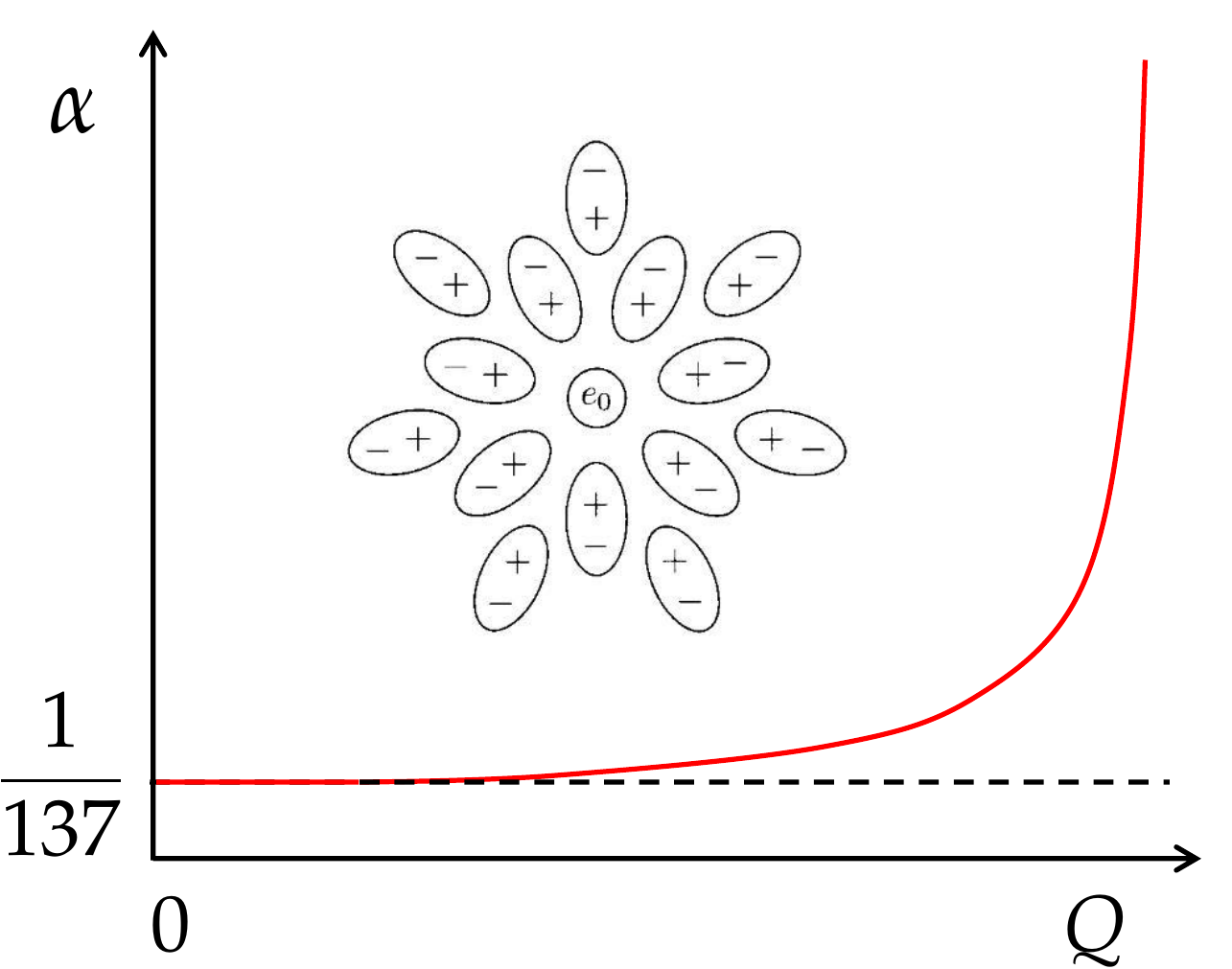} &
\includegraphics[scale=0.45]{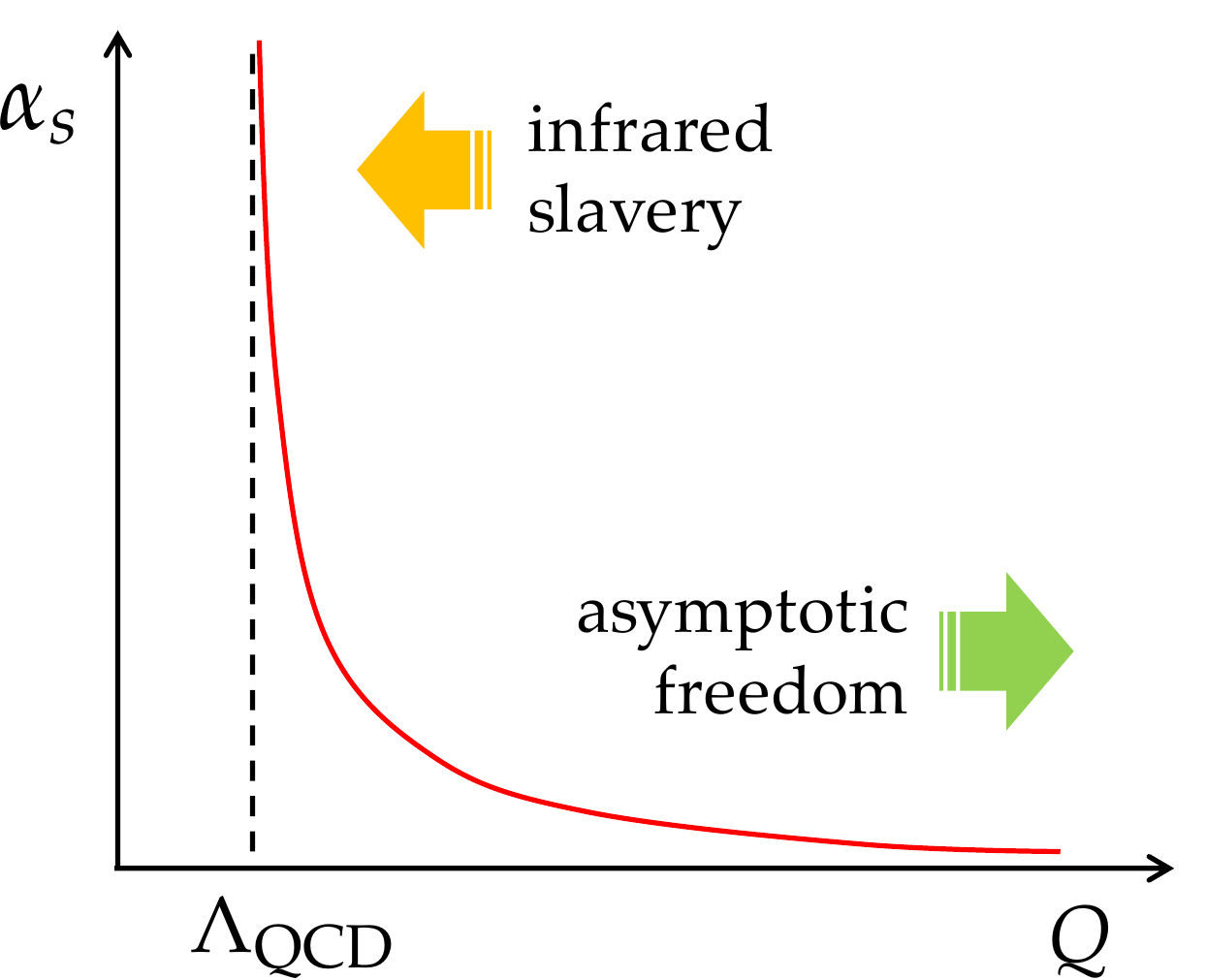}
\end{tabular}
\caption{Vacuum polarization diagrams in QED (left) and QCD (right). They cause different running of the electromagnetic and strong couplings, shown in the lower panel.
\label{fig:vacpol}}
\end{figure}

It is now time to explain a special feature of the strong interaction, that becomes stronger as the distance between quarks increases, in contrast to the electromagnetic interaction between charges. Remember that quantum corrections make the coupling `constants' depend on the energy (momentum transfer) scale at which they are probed. Photon and gluons are massless mediators so the strength of both interactions is proportional to the square of the corresponding couplings. The running of any $\alpha\equiv g^2/(4\pi)$ obeys the renormalization group equation
\eq{
Q^2\frac{\d\alpha}{\d Q^2} = \beta(\alpha)\ , & \quad
\beta(\alpha) \equiv -\alpha^2(\beta_0+\beta_1\alpha+\beta_2\alpha^2+\dots),
}
where the beta function $\beta(\alpha)$ can be written order by order in perturbation theory as an expansion in powers of $\alpha$, whose coefficients $\beta_n$ are derived from the ultraviolet behavior of loop corrections to the photon or gluon propagator (vacuum polarization diagrams). Solving this differential equation one can relate the values of $\alpha$ at two scales $Q$ and $Q_0$. At leading order,
\eq{
\alpha(Q^2)=\frac{\alpha(Q_0^2)}{1+\beta_0\alpha(Q_0^2)\ln\dis\frac{Q^2}{Q_0^2}}.
}
The running of $\alpha$ with $Q$ crucially depends on the sign of $\beta$, dominated by the sign of $\beta_0$, that is physically related to the screening or antiscreening of the fundamental charges by quantum fluctuations as illustrated in figure~\ref{fig:vacpol}. 

In QED the fluctuating vacuum behaves like a dielectric medium, screening the electric charges at increasing distances $R\sim 1/Q$ due to the excitation of virtual $\e^+\e^-$ pairs yielding a negative $\beta_0$,
\eq{
\mbox{[QED]}\quad
\beta_0=-\dis\frac{1}{3\pi}.
}
Therefore the electromagnetic $\alpha$ increases with $Q$. For instance, $\alpha(0)\approx 1/137<\alpha(M_Z^2)\approx 1/128$. 

\begin{figure}
\centering
\includegraphics[scale=0.5]{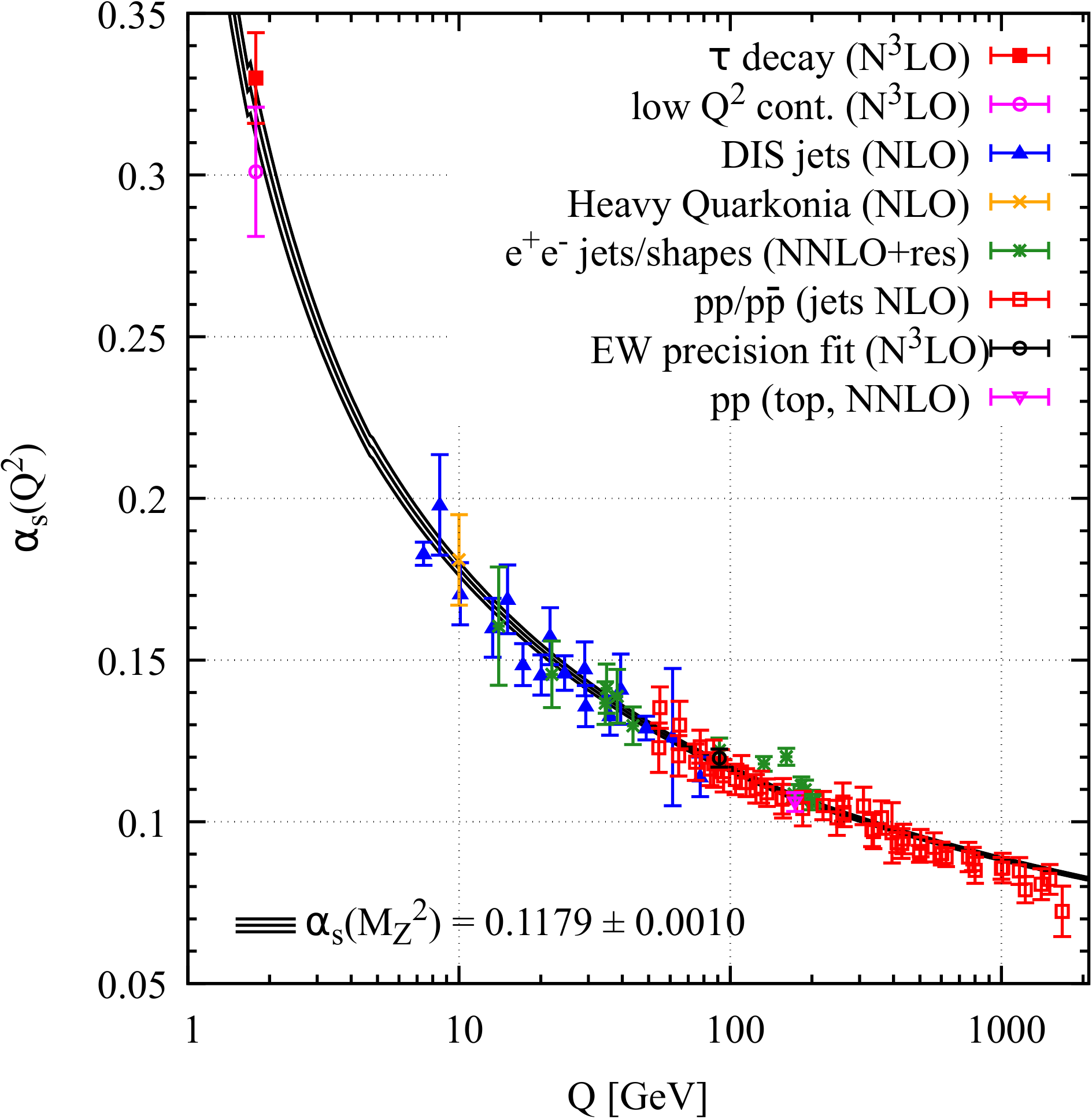}
\caption{Compilation of measurements of $\alpha_s$ as a function of the energy scale $Q$ from \cite{ParticleDataGroup:2020ssz}.
\label{fig:alphas-running}}
\end{figure}

However, the QCD beta function has the opposite sign, because similar negative contributions to the gluon vacuum polarization from virtual quark pairs are overcome by antiscreening effects from gluon self-interactions, 
\eq{
\mbox{[QCD]}\quad
\beta_0=\frac{11 C_A - 4 T_R N_f}{12\pi}=\frac{33-2N_f}{12\pi},
}
that is positive as long as the number of active quark flavors $N_f\le 16$ (there are 6). It is remarkable that there is a scale, $\Lambda_{\rm QCD}$, for which $\alpha_s$ blows up (infrared Landau pole) given at leading order by 
\eq{
\Lambda^2_{\rm QCD} = Q^2\exp\left\{-\frac{1}{\beta_0\alpha_s(Q^2)}\right\}
\quad\Leftrightarrow\quad
\alpha_s(Q^2)=\frac{1}{\beta_0\ln\dis\frac{Q^2}{\Lambda^2_{\rm QCD}}}.
}
This expression for the running of $\alpha_s$ makes sense only at scales $Q>\Lambda_{\rm QCD}$. Note that we have traded the dimensionless QCD coupling for a fundamental scale in nature (dimensional transmutation).\footnote{
Conversely, there is an ultraviolet Landau pole in QED at a ridiculously high energy, way above the Planck scale.} The value of $\Lambda_{\rm QCD}$ depends on $N_f$, and also on the renormalization scheme for more than 2 loops. It has been determined perturbatively using up to 4-loop expressions for the beta function \cite{Deur:2016tte} from a wide variety of $\alpha_s$ measurements (figure~\ref{fig:alphas-running}). The best fit value at $Q=M_Z$ with $N_f=5$ in the $\overline{\rm MS}$ scheme is  \cite{ParticleDataGroup:2020ssz} 
\eq{
\alpha_s(M_Z^2)\approx 0.1179 \quad\Leftrightarrow\quad
\Lambda_{\rm QCD}\approx 210\mbox{ MeV}.
}
This result is compatible with (non-perturbative) lattice calculations \cite{FlavourLatticeAveragingGroup:2019iem}. 

Note that the peculiar properties of the strong interaction based on the non-Abelian gauge group ${\rm SU}(3)_c$ for quarks and gluons encode the existence of a cutoff scale $R\sim 1/\Lambda_{\rm QCD}\approx 1$~fm (the size of a proton!) defining its boundary of applicability. For larger distances, $R>1/\Lambda_{\rm QCD}$ ($Q<\Lambda_{\rm QCD}$), one cannot resolve color charges (quarks and gluons), there only exist colorless hadrons. 
This is the origin of color confinement and explains why strong interactions are short-range despite of the gluon being massless. One can distinguish two domains:
\begin{itemize}
\item
At short distances ($Q\gg\Lambda_{\rm QCD}$) quarks and gluons are almost free. The interaction is so weak that the theory can be treated perturbatively, like the electroweak interaction. The coupling tends to vanish in the limit of high energy scales. This is the {\em asymptotic freedom} \cite{Gross:1973id,Politzer:1973fx}.
\item
At long distances ($Q\sim\Lambda_{\rm QCD}$) quarks and gluons interact so strongly that they cannot be separated, a manifestation of the {\em infrared slavery}. If you try to put them apart they combine with quarks and antiquarks spontaneously created from the vacuum to form hadrons (hadronization). This is a non-perturbative regime.
\end{itemize}


\subsection{Anomalies?}

An anomaly is a symmetry of the classical Lagrangian broken by quantum corrections. They appear when {\em both} axial ($\psi\gamma^\mu\gamma_5\psi$) and vector ($\psi\gamma^\mu\psi$) currents are involved. 

Anomalies of global symmetries are not a problem, rather just the contrary. For example, the process $\pi^0\to\gamma\gamma$ occurs thanks to the coupling of an axial current $j^\mu_A=(\overline{u}\gamma^\mu\gamma_5u-\overline{d}\gamma^\mu\gamma_5d)$ to two electromagnetic (vector) currents, breaking the conservation of the axial current ($\d^\mu j^\mu_A\ne 0$) at one loop, even in the limit of massless quarks.

However, gauge anomalies are a disaster. They break Ward-Takahashi identities spoiling renormalizability. The gauge anomalies (${\cal A}^{abc}$) are generated by triangle diagrams connecting three gauge bosons $V^a$, $V^b$, $V^c$, each coupled to fermions by $(\overline{\Psi}_L\gamma^\mu T_L^a \Psi_L + \overline{\Psi}_R\gamma^\mu T_R^a \Psi_R) V^a_\mu$ with $T^a_L$ an $T^a_R$ the associated generators:
\eq{
\raisebox{-.5\height}{\includegraphics[scale=0.6]{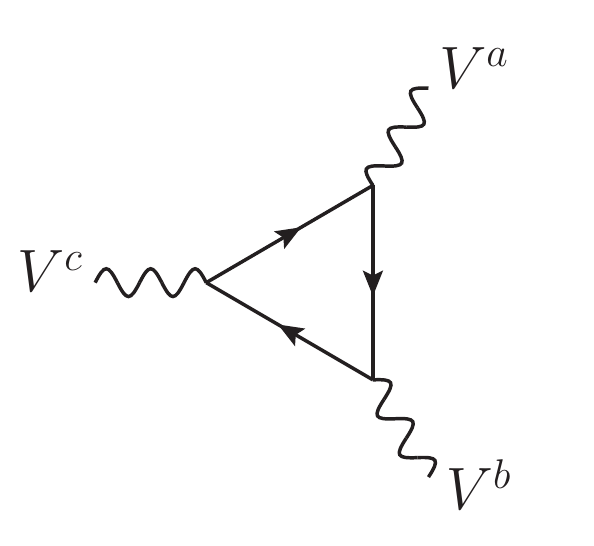}}
\hspace{-4mm}+\mbox{crossed}\quad\Raw\quad {\cal A}^{abc} = {\rm Tr}\left(\{T^a_L,T^b_L\}T^c_L\right)
                 -{\rm Tr}\left(\{T^a_R,T^b_R\}T^c_R\right).
\label{eq:Aabc}
}
Here the traces include summation over all fermions running in the loop. The gauge symmetry is preserved at the quantum level if every ${\cal A}^{abc}=0$. 

The generators of the SM gauge group ${\rm SU}(3)_c\times {\rm SU}(2)_L\times {\rm U}(1)_Y$ in the fundamental representations are 
$T^a\in\{\tfrac{1}{2}\lambda^i,\tfrac{1}{2}\sigma^i,Y\}$ that verify the following identities,
\eq{
{\rm Tr}(\lambda^i\lambda^j)=2\delta^{ij}, \quad
\{\sigma^i,\sigma^j\}=2\delta^{ij}\,\mathbf{1}, \quad
{\rm Tr}(\lambda^i)={\rm Tr}(\sigma^i)=0.
}
Since ${\rm SU}(3)_c$ is non-chiral (not anomalous), the only possible combinations of gauge bosons with non-trivial contributions to the anomaly in equation \eqref{eq:Aabc} are
\eq{
{\rm SU}(3)^2{\rm U}(1):\quad& {\rm Tr}(\{\lambda^i,\lambda^j\}Y\}) &\Raw&\quad
{\cal A}^{abc} \propto \dis\sum_{\rm quarks} (Y_L - Y_R) = 0 
\label{eq:cancel1}
\\
{\rm SU}(2)^2{\rm U}(1):\quad& {\rm Tr}(\{\sigma^i,\sigma^j\}Y) &\Raw&\quad
{\cal A}^{abc} \propto \dis\sum_{\rm leptons} Y_L + N_c \dis\sum_{\rm quarks} Y_L
= 0
\label{eq:cancel2}
\\
{\rm U}(1)^3:\quad& {\rm Tr}(Y^3) &\Raw&\quad
{\cal A}^{abc} \propto \dis\sum_{\rm leptons} (Y_L^3 - Y_R^3) + N_c \dis\sum_{\rm quarks} (Y_L^3 - Y_R^3) 
= 0
\label{eq:cancel3}
}
where $Y_{L,R}$ are the hypercharges of the chiral components in table~\ref{tab:smmultiplets}, related to the electric charges of the two fermions $f$ and $f'$ (quarks or leptons) per generation, with $Q_f-Q_{f'}=1$. They verify  
\eq{
\sum Y_L = \sum Y_R = Q_f+Q_{f'}, \quad
\sum (Y_L^3-Y_R^3) = -\frac{3}{4}(Q_f+Q_{f'}).
}
Then the first anomaly cancelation is trivial and the other two require that the number of colors is precisely $N_c=3$,
\eq{
Q_\nu + Q_e + N_c (Q_u + Q_d) = -1 + \frac{1}{3} N_c 
= 0.
}
This is very striking because the triangle diagrams in equations \eqref{eq:cancel2}-\eqref{eq:cancel3} involve only electroweak interactions. We conclude that the necessary cancelation of gauge anomalies in the electroweak SM requires complete generations of quarks {\it and} leptons, and furthermore every quark must come in exactly 3 colors: somehow the electroweak interactions need that QCD exists!

\subsection{Electroweak phenomenology}

\subsubsection{Feynman rules for all vertices}

The full Lagrangian of the electroweak SM is
\eq{
\lag_{\rm EW} = \lag_F + \lag_{\rm YM} + \lag_\Phi + \lag_{\rm Y} + \lag_{\rm GF} + \lag_{\rm FP}.
\label{eq:lagEW}
}
It provides a number of interactions for fermions (F), vector bosons (V) and scalar particles (S), including the physical Higgs and unphysical, would-be Goldstone bosons. And it also involves unphysical Faddeev-Popov ghost fields (U) that are auxiliary, anticommuting scalar fields. 
All these interactions can be cast into the following set of Lorentz-invariant Lagrangians, written for convenience in terms of generic couplings normalized to appropriate powers of the electromagnetic coupling $e$,
\eq{
{\cal L}_{\rm FFV} &= e\, \overline\psi_i\gamma^\mu(g_V-g_A\gamma_5)\psi_j\,V_\mu
                    = e\, \overline\psi_i\gamma^\mu(g_LP_L+g_RP_R)\psi_j\,V_\mu
\\
{\cal L}_{\rm FFS} &= e\, \overline\psi_i(g_S-g_P\gamma_5)\psi_j\, \phi
                    = e\, \overline\psi_i(c_LP_L+c_RP_R)\psi_j\, \phi
\\
{\cal L}_{\rm VVV} &= - \ii e\, c_{VVV}\left(W^{\mu\nu}W_\mu^\dagger V_\nu-W_{\mu\nu}^\dagger W^\mu V^\nu - W_\mu^\dagger W_\nu V^{\mu\nu}\right) 
\\
{\cal L}_{\rm VVVV} &= e^2\, c_{VVVV}\, \left(2 W_\mu^\dagger W^\mu V_\nu V'^\nu 
- W_\mu^\dagger V^\mu W_\nu V'^\nu - W_\mu^\dagger V'^\mu W_\nu V^\nu\right)
\\
{\cal L}_{\rm SSV} &= - \ii e\,c_{SSV}\,\phi\overleftrightarrow{\partial_\mu}\phi'\, V^\mu
\\
{\cal L}_{\rm SVV} &= e\, c_{SVV}\,\phi\, V^\mu V'_\mu
\\
{\cal L}_{\rm SSVV} &= e^2\, c_{SSVV}\, \phi\phi' V^\mu V'_\mu 
\\
{\cal L}_{\rm SSS} &= e\, c_{SSS}\, \phi\phi'\phi''
\\
{\cal L}_{\rm SSSS} &= e^2\, c_{SSSS}\, \phi\phi'\phi''\phi''',
}
where $g_{L,R}=g_V\pm g_A$, $c_{L,R}=g_S\pm g_P$, $\phi\overleftrightarrow{\partial_\mu}\phi' \equiv \phi\partial_\mu\phi'-(\partial_\mu\phi)\phi'$ and $V_\mu\in\{A_\mu,Z_\mu,W_\mu,W_\mu^\dagger\}$. 
Applying the general Feynman rules for the computation of Green functions or scattering amplitudes,\footnote{
To each vertex, assign a weight composed of the coupling constant appearing in $\ii\lag$, the possibe tensors in internal indices ($\gamma_\mu$, $g_{\mu\nu}$, etc.), a factor $-\ii p_\mu$ for each field derivative $\d_\mu\phi$ of any type of field $\phi$ where $p_\mu$ is the corresponding incoming momentum, and a factor coming from the degeneracy of identical particles in the vertex (e.g. $2\times HHZ$ or $3!\times HHH$).} the different types of interaction vertices read (momenta are taken incoming):
\eq{
\mbox{[FFV$_\mu$]} &=
\ii e\gamma^\mu(g_LP_L+g_RP_R)
\\
\mbox{[FFS]} &=
\ii e(c_LP_L+c_RP_R)
\\
\mbox{[V$_\mu(k_1)$V$_\nu(k_2)$V$_\rho(k_3)$]} &=
\ii e\, c_{VVV} \left[ g_{\mu\nu}(k_2-k_1)_\rho +  g_{\nu\rho}(k_3-k_2)_\mu +  g_{\mu\rho}(k_1-k_3)_\nu\right]
\\
\mbox{[V$_\mu$V$_\nu$V$_\rho$V$_\sigma$]} &=
\ii e^2\, c_{VVVV} \left[2 g_{\mu\nu} g_{\rho\sigma}- g_{\mu\rho} g_{\nu\sigma}- g_{\mu\sigma} g_{\nu\rho}\right]
\\
\mbox{[S$(p)$S$(p')$V$_\mu$]} &=
\ii e\, c_{SSV}\, (p_\mu-p'_\mu)
\\
\mbox{[SV$_\mu$V$_\nu$]} &=
\ii e\, c_{SVV} g_{\mu\nu}
\\
\mbox{[SSV$_\mu$V$_\nu$]} &=
\ii e^2\, c_{SSVV}\, g_{\mu\nu}
\\
\mbox{[SSS]} &= \ii e\, c_{SSS}
\\
\mbox{[SSSS]} &= \ii e^2\, c_{SSSS}.
}
The interactions for [UUV$_\mu$V$_\nu$] and [SUU] are analogous to those of [SSV$_\mu$V$_\nu$] and [SSS], respectively. Potential symmetry factors from the degeneracy of identical particles are absorbed in the coefficients of expressions above, as well as in tables~\ref{tab:FFV}, \ref{tab:FFS}, \ref{tab:SelfV}, \ref{tab:S-V} and \ref{tab:SelfS}, that collect the values of all these generic couplings in the electroweak SM with massless neutrinos. The couplings for would-be Goldstone bosons and Faddeev-Popov ghosts in [SSVV], [SSS], [SUUU], [SSSS] and [UUVV] are omitted. All vertices can be generated by the computer package {\em FeynArts} \cite{Hahn:1998yk}, that uses the same sign conventions. The Feynman rules for QCD vertices are obtained analogously and were already given in equations~(\ref{eq:QCD-FR1}, \ref{eq:QCD-FR2}, \ref{eq:QCD-FR3}).

\boxexercise{8}{Derive the Feynman rules of the SM interaction vertices. Particularly instructive are [VVV] and [VVVV].}

\begin{table}
\centering
\begin{tabular}{|c|c|c|c|c|c|c|}
\hline
FFV & $\overline f_if_j\gamma$ & $\overline f_if_jZ$ & $\overline u_id_jW^+$ & $\overline d_ju_iW^-$ & $\overline \nu_ie_jW^+$ & $\overline e_j\nu_iW^-$
\\
\hline\hline
$g_L$ & $-Q_{f_i}\delta_{ij}$ & $g^f_+\delta_{ij}$ & $\dis\frac{1}{\sqrt{2}s_W}\bV_{ij}$ & $\dis\frac{1}{\sqrt{2}s_W}\bV^*_{ij}$ & $\dis\frac{1}{\sqrt{2}s_W}\delta_{ij}$ &  $\dis\frac{1}{\sqrt{2}s_W}\delta_{ij}$
\\
\hline
$g_R$ &$-Q_{f_i}\delta_{ij}$ & $g^f_-\delta_{ij}$ & 0 & 0 & 0 & 0
\\
\hline 
\end{tabular}
\caption{Fermion-vector boson vertices. Here
$g^f_\pm= v_f\pm a_f$ with
$v_f=\dis\frac{T_3^{f_L}-2Q_fs_W^2}{2s_Wc_W}$ and
$a_f=\dis\frac{T_3^{f_L}}{2s_Wc_W}$.
\label{tab:FFV}
}
\bigskip
\centering
\begin{tabular}{|c|c|c|c|c|}
\hline
FFS & $\overline f_if_jH$ & $\overline f_if_j\chi$ 
    & $\overline u_id_j\phi^+$ & $\overline d_ju_i\phi^-$ 
\\
\hline\hline
$c_L$ & $-\dis\frac{1}{2s_W}\frac{m_{f_i}}{M_W}\delta_{ij}$ &   
        $-\dis\frac{\ii}{2s_W}2T_3^{f_L}\frac{m_{f_i}}{M_W}\delta_{ij}$ &
        $+\dis\frac{1}{\sqrt{2}s_W}\frac{m_{u_i}}{M_W}\bV_{ij}$ &
        $-\dis\frac{1}{\sqrt{2}s_W}\frac{m_{d_j}}{M_W}\bV_{ij}^*$ 
\\
\hline
$c_R$ & $-\dis\frac{1}{2s_W}\frac{m_{f_i}}{M_W}\delta_{ij}$ & 
        $+\dis\frac{\ii}{2s_W}2T_3^{f_L}\frac{m_{f_i}}{M_W}\delta_{ij}$ &
        $-\dis\frac{1}{\sqrt{2}s_W}\frac{m_{d_j}}{M_W}\bV_{ij}^*$ &
        $+\dis\frac{1}{\sqrt{2}s_W}\frac{m_{u_j}}{M_W}\bV_{ij}^*$
\\
\hline
\end{tabular}

\bigskip
\centering
\begin{tabular}{|c|c|c|}
\hline
FFS & $\overline \nu_ie_j\phi^+$ & $\overline e_j\nu_i\phi^-$
\\
\hline\hline
$c_L$ & 0 & $-\dis\frac{1}{\sqrt{2}s_W}\frac{m_{e_j}}{M_W}\delta_{ij}$
\\
\hline
$c_R$ & $-\dis\frac{1}{\sqrt{2}s_W}\frac{m_{e_j}}{M_W}\delta_{ij}$ & 0
\\
\hline 
\end{tabular}
\caption{Fermion-scalar vertices.
\label{tab:FFS}
}
\bigskip
\centering
\begin{tabular}{|c|c|c|}
\hline
VVV & $W^+W^-\gamma$ & $W^+W^-Z$ \\
\hline\hline
$c_{VVV}$ & $-1$ & $\dis\frac{c_W}{s_W}$
\\
\hline 
\end{tabular}

\bigskip
\centering
\begin{tabular}{|c|c|c|c|c|}
\hline
VVVV & $W^+W^+W^-W^-$ & $W^+W^-ZZ$ &  $W^+W^-\gamma Z$ & $W^+W^-\gamma\gamma$ \\
\hline\hline
$c_{VVVV}$ & $\dis\frac{1}{s^2_W}$ & $ -\dis\frac{c^2_W}{s^2_W}$ & $\dis\frac{c_W}{s_W}$
& $-1$
\\
\hline 
\end{tabular}
\caption{Gauge boson self-interaction vertices (symmetry factors are included).
\label{tab:SelfV}
}
\bigskip
\centering
\begin{tabular}{|c|c|c|c|c|c|}
\hline
SSV & $\chi HZ$ & $\phi^\pm\phi^\mp\gamma$ & $\phi^\pm \phi^\mp Z$ & $\phi^\mp H W^\pm$ & $\phi^\mp \chi W^\pm$\\
\hline\hline
$c_{SSV}$ & $-\dis\frac{\ii}{2s_Wc_W}$ & $\mp1$ & $\pm\dis\frac{c_W^2-s_W^2}{2s_Wc_W}$ & $\mp\dis\frac{1}{2s_W}$ & $-\dis\frac{\ii}{2s_W}$
\\
\hline 
\end{tabular}

\bigskip
\centering
\begin{tabular}{|c|c|c|c|c|}
\hline
SVV & $HZZ$ & $H W^+W^-$ & $\phi^\pm W^\mp \gamma$ & $\phi^\pm W^\mp Z$ 
\\
\hline\hline
$c_{SVV}$ & $\dis\frac{M_W}{s_Wc_W^2}$ & $\dis\frac{M_W}{s_W}$ & $-M_W$ & $-\dis\frac{M_W s_W}{c_W}$ 
\\
\hline 
\end{tabular}
\;\;
\begin{tabular}{|c|c|c|}
\hline
SSVV & $HHW^+W^-$ & $HHZZ$\\
\hline\hline
$c_{SSVV}$ & $\dis\frac{1}{2s^2_W}$ & $\dis\frac{1}{2s^2_Wc^2_W}$
\\
\hline 
\end{tabular}
\caption{Scalar-vector boson vertices (symmetry factors are included).
\label{tab:S-V}
}
\bigskip
\centering
\begin{tabular}{|c|c|}
\hline
SSS & $HHH$ \\
\hline\hline
$c_{SSS}$ & $-\dis\frac{3M_H^2}{2M_Ws_W}$ 
\\
\hline 
\end{tabular}
\;\;
\begin{tabular}{|c|c|}
\hline
SSSS & $HHHH$ \\
\hline\hline
$c_{SSSS}$ & $-\dis\frac{3M_H^2}{4M_W^2s^2_W}$ 
\\
\hline 
\end{tabular}
\caption{Scalar self-interaction vertices (symmetry factors are included).
\label{tab:SelfS}
}
\end{table}

\subsubsection{Input parameters}

The QCD Lagrangian (\ref{eq:lagQCD}) for massless quarks requires one input,\footnote{In principle one may add another parameter $\overline\theta$ which would be responsible for CP violation in the QCD sector through an extra term in the Lagrangian \eqref{eq:lagQCD} of the form $\overline\theta g_s^2/(64\pi^2)\epsilon^{\mu\nu\alpha\beta}F^a_{\mu\nu}F^a_{\alpha\beta}$. From experimental contraints to the neutron EDM one derives $|\overline\theta|\lesssim 10^{-10}$ \cite{Dragos:2019oxn} so it can be ignored. The absence of strong CP violation in the QCD sector is called the strong CP problem.} the coupling $g_s$ (or $\alpha_s$). The electroweak gauge group introduces two couplings, $g=e s_W$ and $g'=e c_W$ (or $\alpha$ and $\theta_W$). The electroweak symmetry breaking is parametrized by two more, $\mu^2=-\lambda v^2$ and $\lambda$ (or $M_W$ and $M_H$). And the gauge-invariant Yukawa interactions of the Higgs doublet with fermions introduce most of the free parameters of the SM: 3 charged-lepton masses, 6 quark masses and 4 quark mixings. Therefore the electroweak Lagrangian (\ref{eq:lagEW}) depends on 17 parameters.\footnote{
If light neutrino masses and mixings are included, add 3 more masses and 4 (or 6 for the Majorana case) parameters in the PMNS matrix.} A practical set is:
\eq{
\alpha = \frac{e^2}{4\pi}\ , \quad
M_W = \frac{1}{2}gv\ , \quad
M_Z = \frac{M_W}{c_W}\ , \quad
M_H = \sqrt{2\lambda}\, v\ , \quad
m_f = \lambda_f\frac{v}{\sqrt{2}}\ , \quad
\bU_{\rm CKM}.
}
Fortunately this not so small number of free parameters can be determined from very many different experiments, so the model is overconstrained and its predictions and self-consistency can be checked. It is only after the Higgs boson was discovered that all parameters have been measured. We present below what are the current experimental values of the most `influential' parameters, and in the next section we elaborate on how this information is extracted from processes at increasing energy scales.

\begin{itemize}

\item
{\em Fine structure constant}.
The asymptotic value of the running $\alpha$ at zero momentum transfer can be estimated by several independent methods. One of the most precise determinations is based on the very accurate measurement of the electron anomalous magnetic moment ($g_e$) in a quantum cyclotron at Harvard, that is compared with a very accurate QED theoretical calculation \cite{Aoyama:2017uqe},
\eq{
[g_e]\quad \alpha^{-1} = 137.035\,999\,150\,(33).
}
This is compatible with the value of $\alpha$ that can be measured directly using the quantum Hall effect with larger uncertainty. Even more precise are other recent measurements based on photon recoil in atom interferometry with Cesium \cite{Parker:2018vye} and Rubidium \cite{Morel:2020dww}, that are at present in conflict with one another,
\eq{
\mbox{[Cs]}\quad \alpha^{-1} &= 137.035\,999\,046\,(27) \\
\mbox{[Rb]}\quad \alpha^{-1} &= 137.035\,999\,206\,(11).
}

\item
{\em Weak boson masses}.
The SM predicts $M_W<M_Z$ (\ref{eq:MWMZ}) in agreement with measurements. The weak gauge bosons were discovered at the Sp$\bar{\text{p}}$S collider (CERN) in 1983 \cite{UA1:1983crd,UA2:1983tsx,UA1:1983mne,UA2:1983mlz}. Today the weak boson masses are known with a precision of 0.1 per mille or better form combined measurements at the $e^+e^-$ colliders LEP (CERN) and SLC (SLAC), and at the hadron colliders Tevatron (Fermilab) and LHC (CERN). The current world averages \cite{ParticleDataGroup:2020ssz} are
\eq{
M_Z & = 91.1876\pm0.0021\mbox{ GeV} \quad\mbox{[LEP1/SLC]} \\
M_W & = 80.379\pm0.012\mbox{ GeV}   \quad\mbox{[LEP2/Tevatron/LHC]} .
}

\item
{\em Top quark mass}.
The top is the only quark that is not confined in hadrons because being so heavy it weakly decays into a $W$ boson and a $b$ quark before hadronizing. It was discovered at the Tevatron in 1995 \cite{CDF:1995wbb,D0:1995jca}. Direct measurements of the kinematics of $t\bar{t}$ events are sensitive to what is usually interpreted as the pole mass. The current average \cite{ParticleDataGroup:2020ssz} is:
\eq{
m_t = 172.76\pm0.30\mbox{ GeV} \quad\mbox{[Tevatron/LHC]} .
}

\item
{\em Higgs boson mass}.
The Higgs boson was discovered at the LHC in 2012 \cite{ATLAS:2012yve,CMS:2012qbp} and its mass is already known at the permille level \cite{ParticleDataGroup:2020ssz},
\eq{
M_H = 125.25\pm0.17\mbox{ GeV} \quad\mbox{[LHC]} .
}

\end{itemize}

\subsubsection{Observables and experiments}

\subsubsubsection{Low energy observables}

\begin{figure}
\centering
\begin{tabular}{cc}
\includegraphics[scale=0.55]{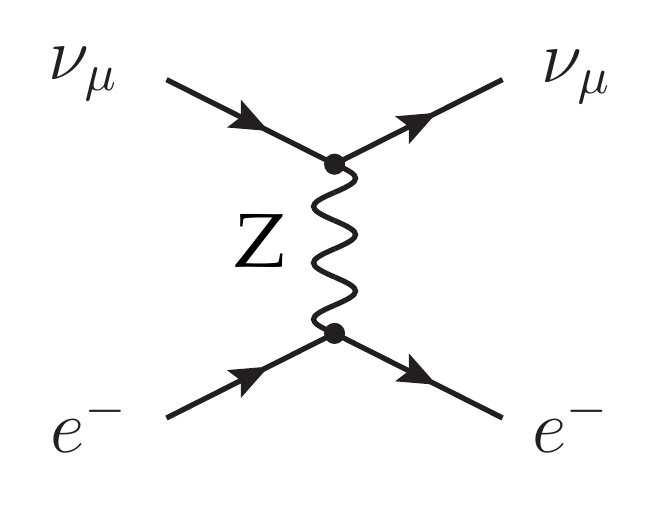} &\qquad
\raisebox{-0.15\height}{\includegraphics[height=4cm]{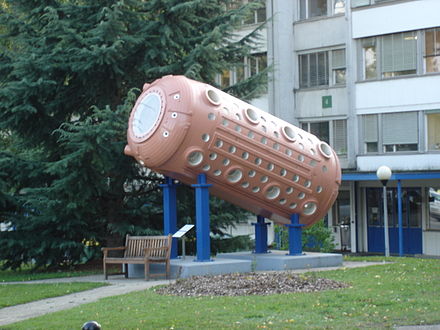}}
\end{tabular}
\caption{Weak neutral currents (left) discovered in the CERN bubble chamber detector Gargamelle (right).
\label{fig:WNC}}
\end{figure}

At low momentum transfer $Q^2\ll M_Z^2$ one can already get relevant information about the electroweak interactions. For example, the weak neutral currents were discovered by the observation of the elastic neutrino-electron scattering in the CERN bubble chamber detector Gargamelle in 1973 \cite{GargamelleNeutrino:1973jyy} (fig.~\ref{fig:WNC}). The source of muon neutrinos, of energies less than 10 GeV, was a proton beam of 26 GeV from the PS accelerator. This was the confirmation of a cornerstone of the SM that won Glashow, Salam and Weinberg their Nobel prize, even before the $W$ and the $Z$ were found in  $p\bar{p}$ collisions at a center-of-mass energy of 540~GeV ten years later. At present, very accurate measurements of the weak mixing angle $\theta_W$ come from the ratio of cross-sections $\sigma_{\bar\nu_\mu e}/\sigma_{\nu_\mu e}$ of neutrinos and antineutrinos in neutrino-electron scattering, and from the ratios of neutral to charged current cross-sections $\sigma_{\nu N}^{\rm NC}/\sigma_{\nu N}^{\rm CC}$ in neutrino-nucleon scattering at CERN and Fermilab.

The weak mixing angle can also be obtained from the left-right asymmetry (parity violation) in the cross-sections of polarized electrons off nucleons, $e_{R,L}N\to e X$,
and from tiny parity violating effects induced by the weak interactions between electrons and quarks in heavy atoms (atomic parity violation), due to $Z$ boson exchange, that grow with roughly the third power of the atomic number.

\begin{figure}
\centering
\includegraphics[scale=0.55]{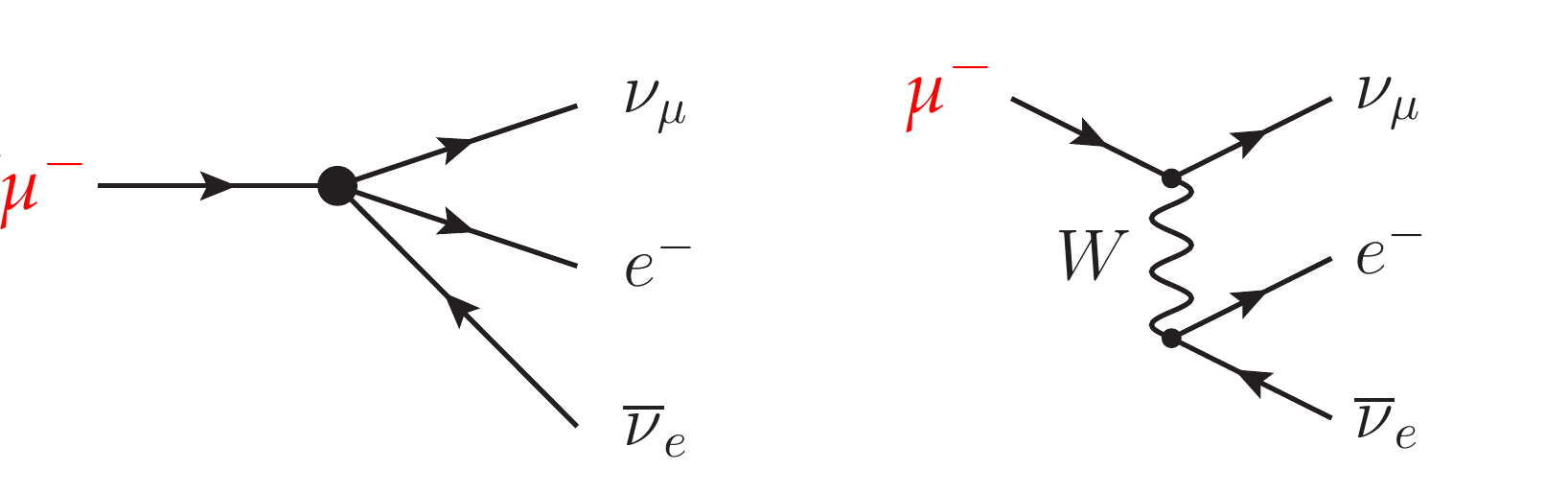}
\caption{Muon decay in the 4-Fermi model (left) and tree-level contribution in the SM (right).
\label{fig:mudcy}}
\end{figure}

Valuable information comes from the measurement of the muon lifetime. The muon decay \cite{Garwin:1957hc}, together with the beta decay in Cobalt \cite{Wu:1957my}, provided the first confirmation of the violation of parity shortly after the seminal work of Yang and Lee \cite{Lee:1956qn} in 1956. The process $\mu\to \e\, \overline\nu_e\nu_\mu$ proceeds at tree level in the SM through the exchange of a $W$ boson with very low momentum transfer ($-q^2\equiv Q^2\le m_\mu^2\ll M_W^2$), and can be described by the effective 4-Fermi theory (proposed to explain the $\beta$ decay in 1934 \cite{Fermi:1934hr}) (fig.~\ref{fig:mudcy}),
\eq{
\ii{\cal M} = 
-\ii\frac{4G_F}{\sqrt{2}}\ (\overline{e}\gamma^\rho\nu_L)(\overline{\nu_L}\gamma_\rho \mu)
= \left(\frac{\ii e}{\sqrt{2}s_W}\right)^2
\overline{e_L}\gamma^\rho\nu_L\ \frac{-\ii g_{\rho\delta}}{q^2-M_W^2}
\overline{\nu_L}\gamma^\delta \mu_L ,
}
from which the Fermi constant $G_F$ can be derived in terms of parameters of the fundamental theory,
\eq{
\frac{G_F}{\sqrt{2}} = \frac{\pi\alpha}{2s_W^2 M_W^2}.
\label{eq:GF0}
}
The muon lifetime $\tau=\Gamma^{-1}$ is the inverse of its total decay width,\footnote{
The process $\mu\to \e\, \overline\nu_e\nu_\mu$ is by far the dominant decay channel. The decays $\mu\to \e\overline\nu_e\nu_\mu e^+e^-$ and
$\mu\to \e\overline\nu_e\nu_\mu\gamma$ with branching ratios $\sim10^{-5}$ and $10^{-8}$, respectively, must be taken into account when accuracy requires it.}
\eq{
\Gamma = \dis\frac{G_F^2 m_\mu^5}{192\pi^3}\ f(m_e^2/m_\mu^2), \quad
f(x) = 1-8x+8x^3-x^4-12x^2\ln{x},
}
where $f(m_e^2/m_\mu^2) = 0.99981295$ is a kinematic factor from phase space integration.
The Fermi constant is measured very precisely from the muon lifetime at PSI in Villigen \cite{ParticleDataGroup:2020ssz},
\eq{
G_F=1.166\,378\,7(6)\times 10^{-5}\mbox{ GeV}^{-2}.
\label{eq:GF}
}
It provides the value of the Higgs VEV (the electroweak scale),
\eq{
v = \left(\sqrt{2}G_F\right)^{-1/2}\approx246\;\mbox{\rm GeV}
}
and constrains the product $M_W^2s_W^2$, which implies
\eq{
M_Z^2 >
M_W^2 = \frac{\pi\alpha}{\sqrt{2} G_F s_W^2} > 
        \frac{\pi\alpha}{\sqrt{2} G_F} \approx (37.4\mbox{ GeV})^2,
}
providing a lower limit of the weak boson masses, before their discovery. On the other hand, since we have now independent measurements of $G_F$, $\alpha$, $M_W$ and $M_Z$ one can attempt  a first consistency check of the model by comparing the value of $G_F$ in (\ref{eq:GF}) with the prediction using the {\em tree-level} expression \eqref{eq:GF0},
\eq{
s_W^2 = 1-M_W^2/M_Z^2 \quad\Raw\quad
G_F=\frac{\pi\alpha}{\sqrt{2}(1-M_W^2/M_Z^2) M_W^2} \approx 1.125\times 10^{-5} .
\label{eq:GFhigh}
}
The glaring discrepancy will disappear when quantum corrections are included (see section~\ref{sec:precision}).

\subsubsubsection{Fermion-pair production in e$^+$e$^-$ colliders}

\begin{figure}
\centering
\begin{tabular}{cc}
\includegraphics[scale=0.25]{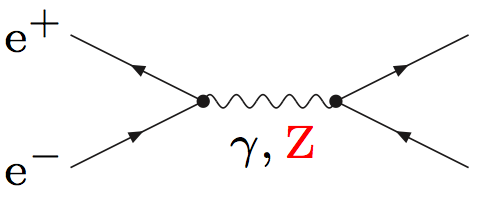} &\quad
\raisebox{-0.4\height}{\includegraphics[scale=0.4]{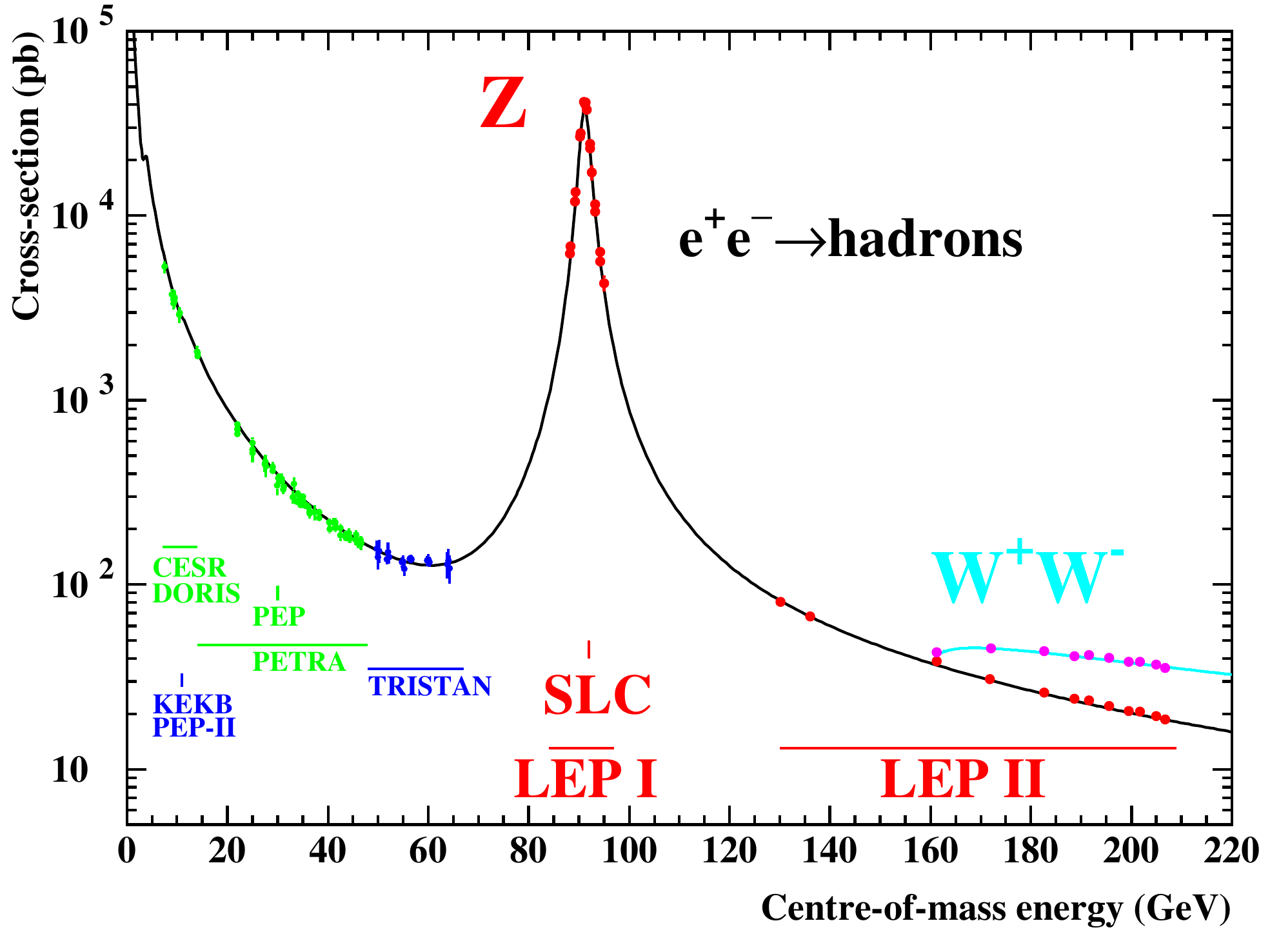}}
\end{tabular}
\caption{Left: Tree-level contributions to fermion-pair production in $e^+e^-$ collisions. Right: Hadronic cross-section as a function of the center-of-mass energy.
The solid line is the SM prediction, and the points are the experimental measurements at different colliders, whose energy ranges are also indicated. From Ref.~\cite{ALEPH:2005ab}.
\label{fig:epemffbar}}
\end{figure}

Lepton colliders provide a clean environment to study the electroweak interactions. In particular, the $e^+e^-$ annihilation into a fermion-antifermion pair is given, at tree level, by the exchange of a photon and a $Z$ boson in the $s$-channel. At increasing center-of-mass energies the cross-section falls like $1/s$ dominated by the virtual photon exchange while the $Z$ exchange becomes more important until it reaches a maximum right at $s=M_Z^2$ where it presents a resonance peak (fig.~\ref{fig:epemffbar}). 

It is a good exercise to try and reproduce the differential cross-section for $e^+e^-\to \bar{f}f$ (in the case of unpolarized fermions), 
\eq{
\frac{\dd\sigma}{\dd\Omega} &= N_c^f\frac{\alpha^2}{4s}\beta_f\Big\{
\left[1+\cos^2\theta + (1-\beta_f^2)\sin^2\theta\right] G_1(s) 
+ 2(\beta_f^2-1) G_2(s)+2\beta_f\cos\theta G_3(s)
\Big\} 
\label{eq:epemxs1}
\\
&G_1(s)=Q_e^2Q_f^2+2Q_eQ_fv_ev_f{\rm Re}\chi_Z(s)+(v_e^2+a_e^2)(v_f^2+a_f^2)|\chi_Z(s)|^2
\\
&G_2(s)=(v_e^2+a_e^2)a_f^2|\chi_Z(s)|^2
\\
&G_3(s)=2Q_eQ_fa_ea_f{\rm Re}\chi_Z(s)+4v_ev_fa_ea_f|\chi_Z(s)|^2
\label{eq:epemxs4}
}
where $\chi_Z(s)\equiv s/(s-M_Z^2+\ii M_Z\Gamma_Z)$ contains the $Z$ propagator including an imaginary part relevant in the vicinity of the resonance, $N_c^f=1$ (3) for $f=$ lepton (quark), $v_f$ and $a_f$ are the vector and axial-vector couplings (\ref{eq:vfaf}) and $\beta_f= \sqrt{1-4m^2_f/s}$ is the final fermion velocity in the center of mass frame (the electron mass can be safely neglected). The contribution of each diagram and their interference is evident and the parity violation due to the $Z$ exchange manifests itself as a forward-backward asymmetry: the term proportional to $\cos\theta$ involving both vector and axial vector couplings. Integrating over the solid angle, the total cross-section reads
\eq{
\sigma(s)=N_c^f\frac{2\pi\alpha^2}{3s}\beta_f\left[
(3-\beta_f^2)G_1(s)-3(1-\beta_f^2)G_2(s)\right]
}
that gives the profile of fig.~(\ref{fig:epemffbar}). The (inclusive) hadronic cross-section is obtained by summing over all quark flavors above threshold at a given energy, essentially five in the displayed range. 

\subsubsubsection{Z pole observables}

On the resonance peak ($s=M^2_Z$) the $Z$ propagator becomes purely imaginary, the interference of the photon and $Z$ exchange diagrams vanishes and the cross-section is dominated by the weak interaction (the QED contribution is suppressed by a factor $(\Gamma_Z/M_Z)^2\lesssim10^{-3}$). This was the energy domain of the first phase (1989-1995) of the circular $e^+e^-$ collider LEP at CERN and the linear collider SLC (1992-1998) at SLAC. The former collected 17 million $Z$ decays at center-of-mass energies within plus or minus 3 GeV of the $Z$ pole, and the latter only 600 thousand but with a longitudinally polarized electron beam. At these colliders very precise measurements of various $Z$ pole observables have been performed. These include the $Z$ mass $M_Z$, the total width $\Gamma_Z$, and partial widths $\Gamma_{\bar{f}f}$ for $Z\to \bar{f}f$. It is customary to introduce
\eq{
\sigma^0_{\rm had} \equiv 12\pi
\frac{ \Gamma_{e^+e^-} \Gamma_{\rm had} }
{M_Z^2\Gamma_Z^2}, \quad
R_\ell \equiv \frac{\Gamma_{\rm had}}{\Gamma_{\ell^+\ell^-}}, \quad
R_q \equiv \frac{\Gamma_{q\bar{q}}}{\Gamma_{\rm had}}
\label{eq:Zobs}
}
where $\ell=\e,\mu,\tau$, $q=b$ or $c$ and $\Gamma_{\rm had}$ is the partial width into hadrons.\footnote{The three measured values for $R_\ell$ are consistent with lepton universality.} The effects of the photon-exchange diagram are subtracted in $\sigma^0_{\rm had}$. Very useful constraints follow from various $Z$ pole (forward-backward and left-right) asymmetries,
\eq{
A^f_{\text{FB}} = \frac{\sigma(\cos\theta>0)-\sigma(\cos\theta<0)}
                {\sigma(\cos\theta>0)+\sigma(\cos\theta<0)}
         = \frac{3}{4} A_f \frac{A_e+P_e}{1+P_eA_e} \qquad
  A_{LR} = \frac{\sigma_L-\sigma_R}{\sigma_L+\sigma_R}
         = A_e P_e
}
where $P_e$ is the initial electron polarization and
\eq{
A_f \equiv \frac{2v_fa_f}{v_f^2+a_f^2}.
}

\boxexercise{9}{Show that, if fermion masses are neglected,
\eq{
\Gamma_{\bar{f}f}&\equiv\Gamma(Z\to \bar{f}f) = N_c^f\frac{\alpha M_Z}{3}(v_f^2+a_f^2).
}
Then, using equations (\ref{eq:epemxs1}-\ref{eq:epemxs4}) at $s=M_Z^2$, check that for unpolarized electrons
\eq{
\sigma^0_{\rm had}&= 12\pi\frac{\Gamma_{e^+e^-}\Gamma_{\rm had}}{M_Z^2\Gamma_Z^2} \qquad
A^f_{\text{FB}} = \frac{3}{4}A_fA_e .
}
}

\begin{figure}
\centering
\includegraphics[width=6cm]{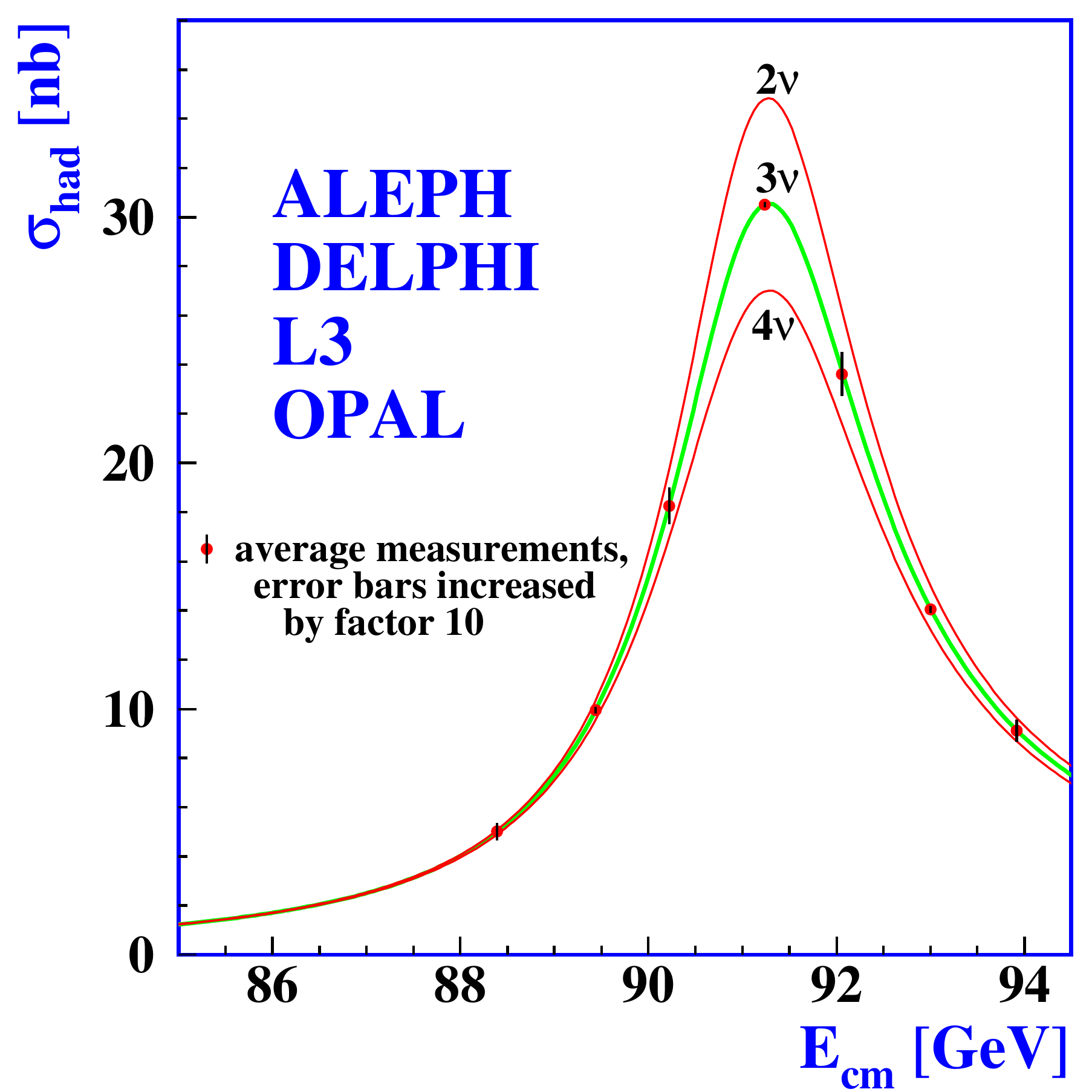}
\caption{Measurements of the hadron production cross-section around the $Z$ resonance (lineshape) at LEP. The curves indicate the predicted cross-section for two, three and four neutrino species with SM couplings and negligible mass. From Ref.~\cite{ALEPH:2005ab}.
\label{fig:Nnu}
}
\end{figure}

By measuring the $Z$ pole observables (\ref{eq:Zobs}) one can estimate the $Z$ invisible width, $\Gamma_{\text{inv}} = \Gamma_Z-\Gamma_{e^+e^-}-\Gamma_{\mu^+\mu^-}
-\Gamma_{\tau^+\tau^-}-\Gamma_{\rm had}$, that can be used to deduce the number of light neutrino species, $N_\nu = \Gamma_{\text{inv}}/\Gamma_{\nu\bar{\nu}}$, from the partial width to neutrinos predicted by the SM. The overall scale of the $Z$ lineshape is fixed by the peak cross-section $\sigma_{\rm had}$, whose experimental value is extracted from the number of observed hadronic events given the collider luminosity, that in turn is measured from the rate of $e^+e^-\to e^+e^-$ events at low angle provided the (accurate enough) theoretical prediction of the Bhabha scattering cross-section. The combination of the measurements made by the four LEP experiments (fig.~\ref{fig:Nnu}) led to $N_\nu = 2.9840 \pm 0.0082$ \cite{ALEPH:2005ab}, two standard deviations away from 3.0, the number of fermion generations in the SM. Very recently the prediction for the Bhabha cross-section was found to be overestimated, and consequently the luminosity underestimated \cite{Janot:2019oyi}. The new analysis of the $Z$ lineshape fit, reducing $\sigma_{\rm had}$ while slightly increasing $\Gamma_Z$, yields the result $N_\nu = 2.9963\pm 0.0074$, hence putting an end to the $2\sigma$ tension with the SM. 

\subsubsubsection{W boson production}

\begin{figure}
\centering
\begin{tabular}{cc}
\includegraphics[scale=0.25]{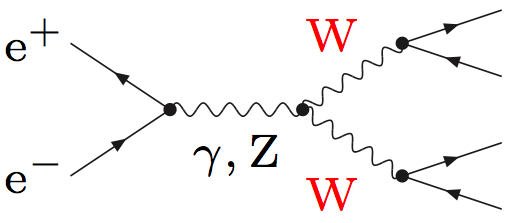}
\raisebox{-1.1mm}{\includegraphics[scale=0.25]{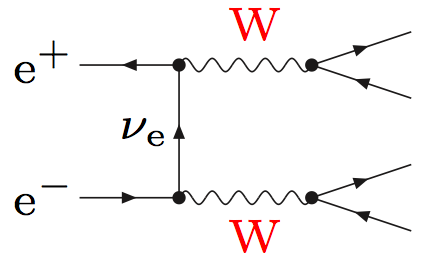}} &\quad
\raisebox{-0.4\height}{\includegraphics[width=6cm]{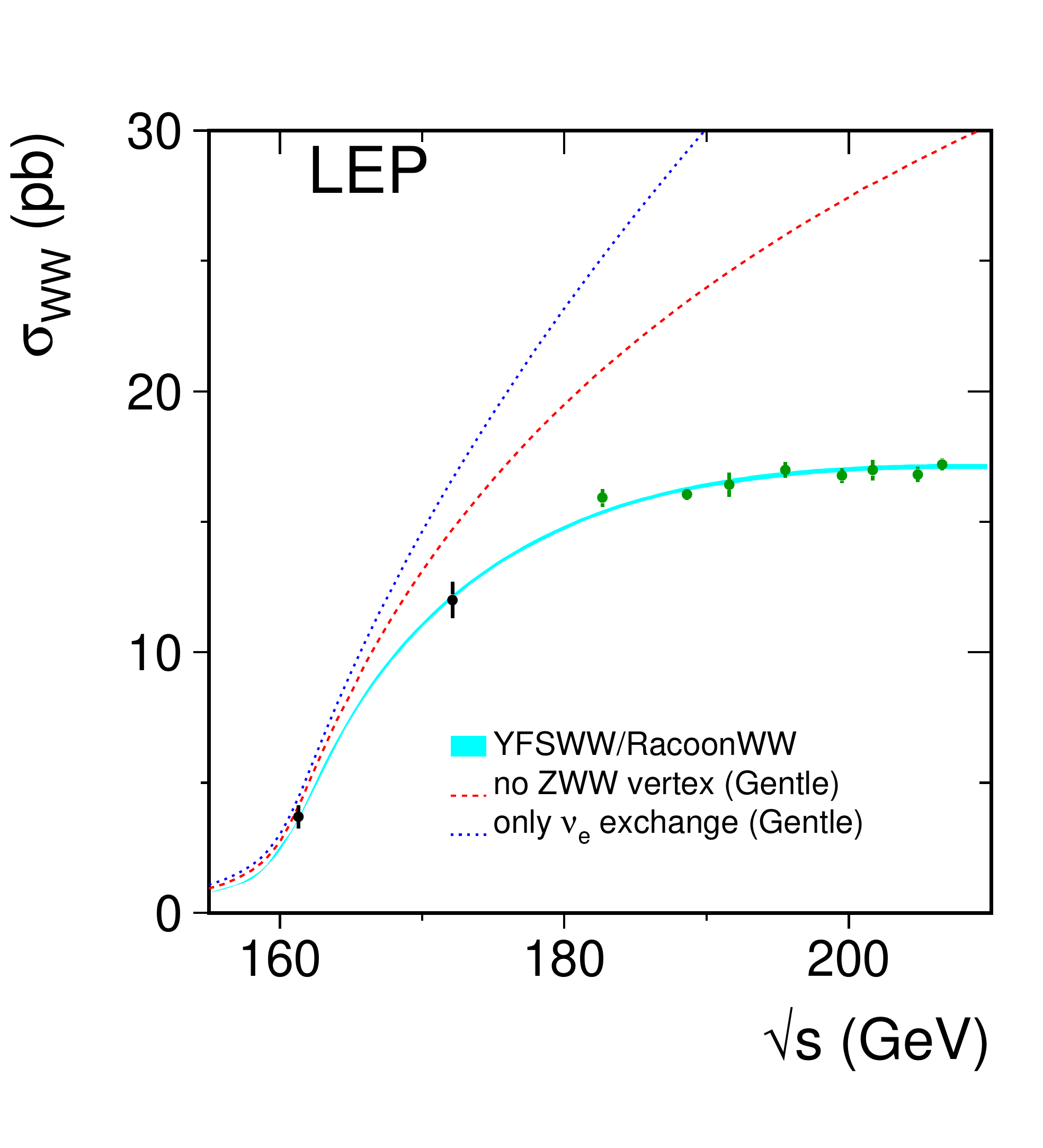}}
\end{tabular}
\caption{Left: Tree-level contributions to $e^+e^-\to W^+W^-$. Right: Measurements of the $W$-pair production cross-section at LEP2, compared to theoretical predictions (taking $M_W=80.35$~GeV) including all diagrams (cyan), removing the $ZWW$ vertex (red), and assuming only the $\bar{\nu}_e$ exchange (blue). From Ref.~\cite{ALEPH:2013dgf}.
\label{fig:WW}
}
\end{figure}

LEP2 (1996-2000) operated at higher center-of-mass energies (fig.~\ref{fig:epemffbar}) to study  $W$-pair production (fig.~\ref{fig:WW}), and in part also to search (unsuccessfully) for the Higgs boson \cite{ALEPH:2013dgf}. Particularly important was the exploration of the $W^+W^-$ threshold (161~GeV), where the dependence of the cross-section with the $W$ mass is large, that allowed to determine $M_W$ very precisely. At higher energies (172 to 209~GeV) this dependence is much weaker and $W$ bosons were directly reconstructed and their mass determined from the invariant mass of the decay products. LEP2 was also the first to probe the triple gauge couplings $WW\gamma$ and $WWZ$, predicted by the non-Abelian gauge symmetry (fig.~\ref{fig:WW}), another milestone of the SM. 

\begin{figure}
\centering
\includegraphics[scale=0.25]{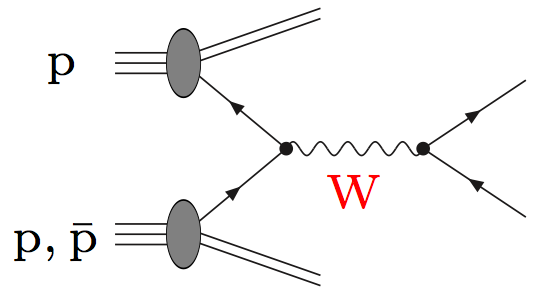} \quad
\includegraphics[scale=0.25]{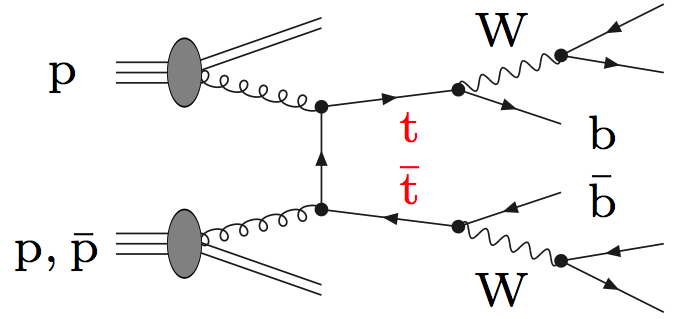}
\caption{W production and top-quark production at hadron colliders.
\label{fig:hadcoll}}
\end{figure}

In hadron colliders, on-shell $W$ bosons are tagged by their decay into charged leptons with high transverse momentum (fig.~\ref{fig:hadcoll}). The values of the $W$ mass from Tevatron and LHC are compatible with the measurements from LEP2 and have at present very similar precision.

\subsubsubsection{Top quark production}

Top quarks are produced in hadron colliders dominantly in pairs through the strong processes $q\bar{q}\to t\bar{t}$ (fig.~\ref{fig:hadcoll}) and $gg\to t\bar{t}$ at leading order. At Tevatron ($p\bar{p}$, $\sqrt{s}=1.96$~TeV) 85\% of the producction cross-section is from $q\bar{q}$ annihilation, while at LHC ($pp$) about 90\% ($\sqrt{s}=7$~TeV) or 80\% ($\sqrt{s}=14$~TeV) comes from from gluon fusion. Single-top quarks are also produced in electroweak processes, $q\bar{q}'\to t\bar{b}$, $qb\to q' t$, $b g\to W t$, with somewhat smaller cross-sections.
The top-quark mass is kinematically reconstructed from invariant mass distributions of the final states in different decay channels.

\subsubsubsection{Higgs boson production}

The Higgs boson is the smoking gun providing evidence that the spontaneous breaking of the electroweak symmetry {\it does} generate the masses of weak bosons and fermions. The Higgs mechanism is essential not only because the renormalizability of the SM is then guaranteed \cite{tHooft72}, a requirement that is nowadays not considered so crucial as in former times \cite{Weinberg:2021exr}, but also because it ensures the unitarity of the model \cite{LlewellynSmith:1973yud}: the scattering amplitudes have a good behavior at high energy because of `miraculous' cancellations that follow when the electroweak boson self-interactions are of the Yang-Mills form,\footnote{Note the steep growth of the $e^+e^-\to W^+W^-$ cross-section in fig.~\ref{fig:WW} when the gauge self-interactions are ignored.} as prescribed by the gauge symmetry ({\it e.g.} $e^+e^-\to W^+W^-$) and scalar-exchange diagrams of the Higgs type are included ({\it e.g.} $W^+W^-\to W^+W^-$). After its long awaited discovery, the predicted properties of the SM Higgs boson \cite{Djouadi:2005gi} can finally be checked against experiment.

\begin{figure}
\centering
\includegraphics[width=\linewidth]{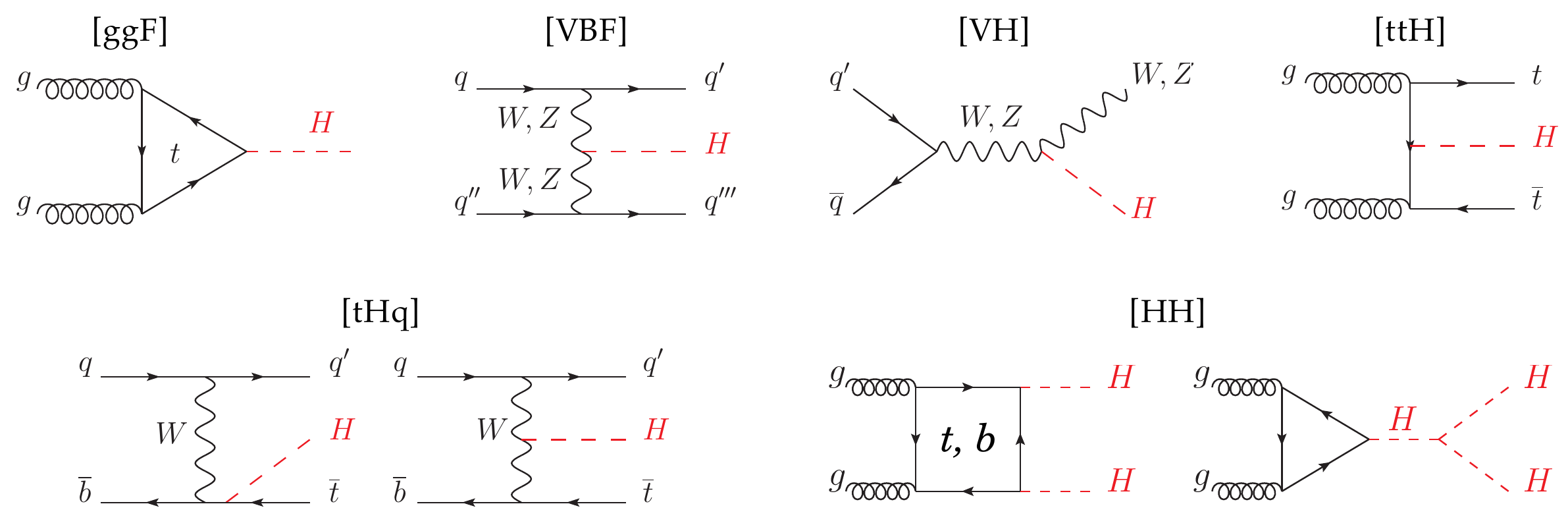}
\caption{Leading-order diagrams for Higgs production mechanisms at hadron colliders: gluon fusion [ggF], vector boson fusion [VBF], Higgs-strahlung [VH], associated with a pair of top quarks [ttH] or a single top quark [tH] and Higgs boson pair production [HH]. From Ref.~\cite{ParticleDataGroup:2020ssz}.
\label{fig:HprodFD}}
\end{figure}

\begin{figure}
\centering
\includegraphics[width=.8\linewidth]{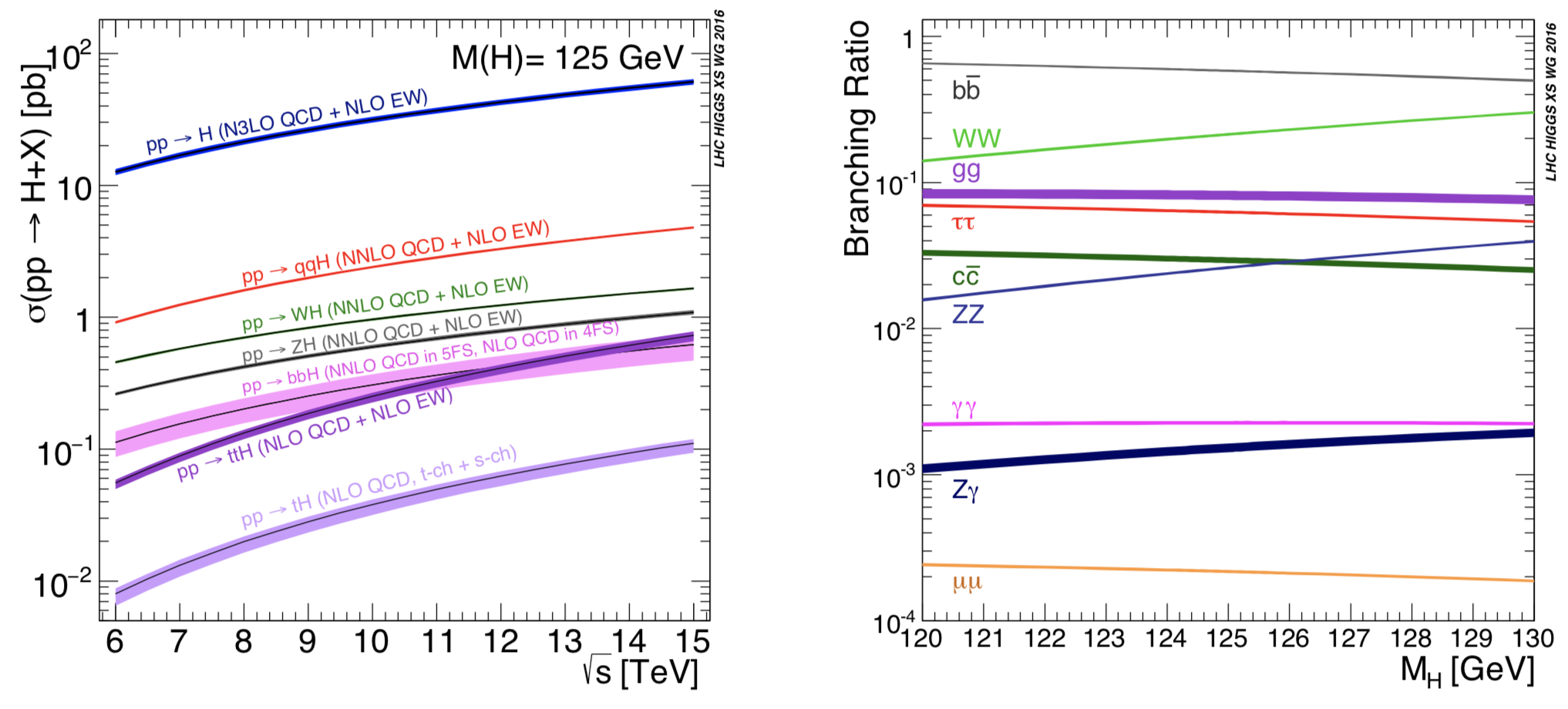}
\caption{Higgs boson production cross-sections as a function of the LHC center of mass energy (left) and Higgs boson branching ratios for the mass range around 125 GeV (right). From Ref.~\cite{LHCHiggsCrossSectionWorkingGroup:2016ypw}.
\label{fig:HprodandBR}}
\end{figure}

The main production mechanisms at the Tevatron\footnote{Tevatron did not have enough statistical significance to claim  `discovery' of the Higgs boson.} and the LHC are gluon fusion, weak-boson fusion, associated production with a gauge boson, and associated production with a pair of top quarks or with a single top quark (see figure~\ref{fig:HprodFD}). The Higgs boson pair production in the SM is more rare but very important because it allows to check the trilinear Higgs boson self-coupling. The production cross-sections in $pp$ collisions at LHC energies and the branching ratios for the decay of a Higgs boson with a mass around 125~GeV are shown in figure~\ref{fig:HprodandBR}. The Higgs boson is mostly produced by gluon fusion (gluons are the most abundant parton in the proton at low $x\sim M_H/\sqrt{s}$) mediated by a top-quark loop (whose heavy mass enhances the effective coupling). The dominant decay channel is $H\to b\bar{b}$ (about 58\%) but it suffers from large backgrounds. Less probable are $H\to ZZ^*,WW^*$ (with one of the gauge bosons off-shell) and $H\to\gamma\gamma$ but they provide cleaner signals and played an important role in the Higgs discovery. In fact, the decay into $b\bar{b}$ has been discovered (significance above $5\sigma$) as recently as 2018 \cite{ATLAS:2018kot}. At the other end, there is `evidence' for the $\mu\mu$ channel (significance above $3\sigma$) from 2020 \cite{CMS:2020xwi}.

\begin{figure}
\centering
\includegraphics[width=0.9\linewidth]{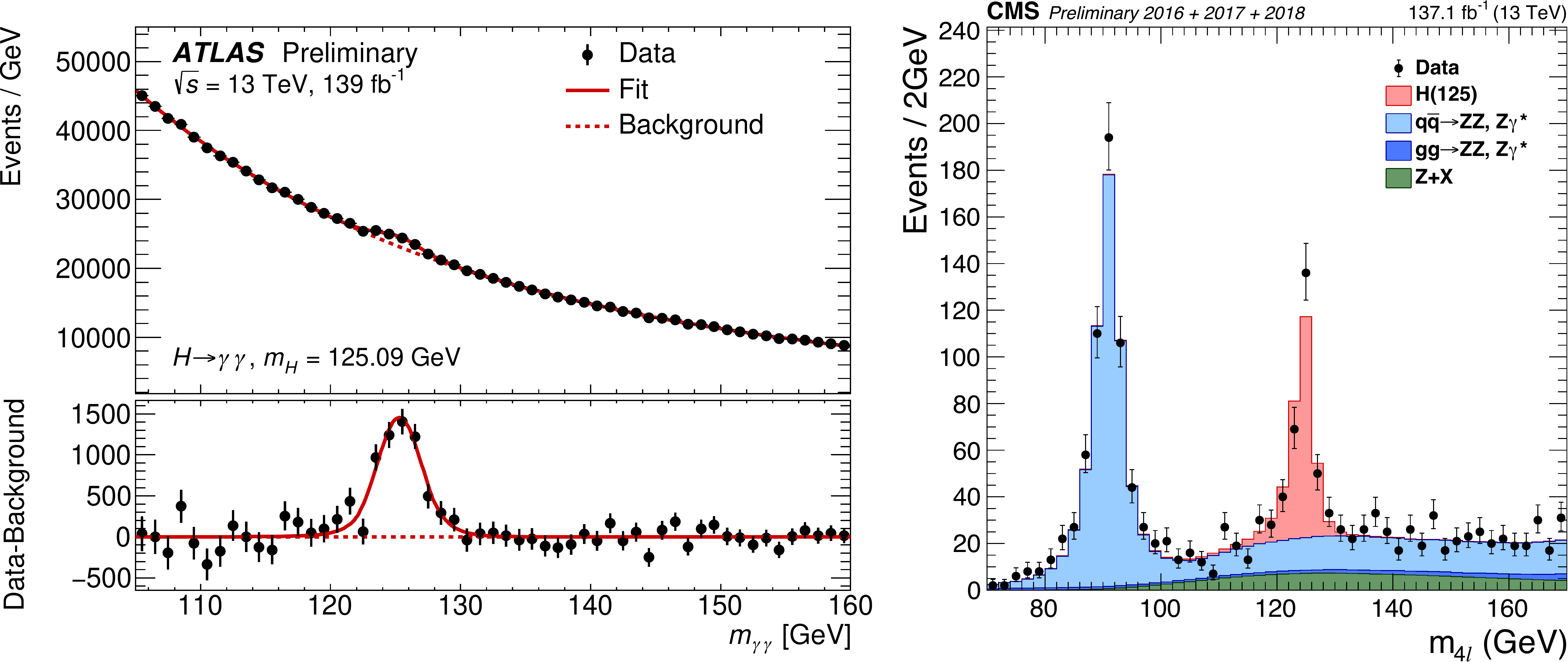}
\caption{Left: Invariant diphoton mass distribution observed by ATLAS \cite{ATLAS:2017myr}. Right: invariant $m_{4l}$ distribution from CMS \cite{CMS:2019chr}. They exhibit clear signals of $H\to\gamma\gamma$ and $H\to ZZ^*\to4l$, respectively, allowing to measure the Higgs boson mass.
\label{fig:Hinvmass}}
\end{figure}

The current average value of the Higgs boson mass comes from the combination of mass measurements in the $\gamma\gamma$ and $ZZ$ channels (figure~\ref{fig:Hinvmass}).

\boxexercise{10}{Compute the Higgs partial widths 
\eq{
\Gamma(H\to f\bar f) &= N_c^f\frac{G_F M_H}{4\pi\sqrt{2}}m_f^2\left(1-\frac{4m_f^2}{M_H^2}\right)^{3/2}
}
and for on-shell weak bosons (this overestimates $H\to VV^*$),
\eq{
\Gamma(H\to W^+W^-) &= \frac{G_F M_H^3}{8\pi\sqrt{2}}\sqrt{1-\frac{4M_W^2}{M_H^2}}
\left(1-\frac{4M_W^2}{M_H^2}+\frac{12M_W^4}{M_H^4}\right)
\\[1ex]
\Gamma(H\to ZZ) &= \frac{G_F M_H^3}{16\pi\sqrt{2}}\sqrt{1-\frac{4M_Z^2}{M_H^2}}
\left(1-\frac{4M_Z^2}{M_H^2}+\frac{12M_Z^4}{M_H^4}\right)
}
}

\begin{figure}
\centering
\includegraphics[width=0.8\linewidth]{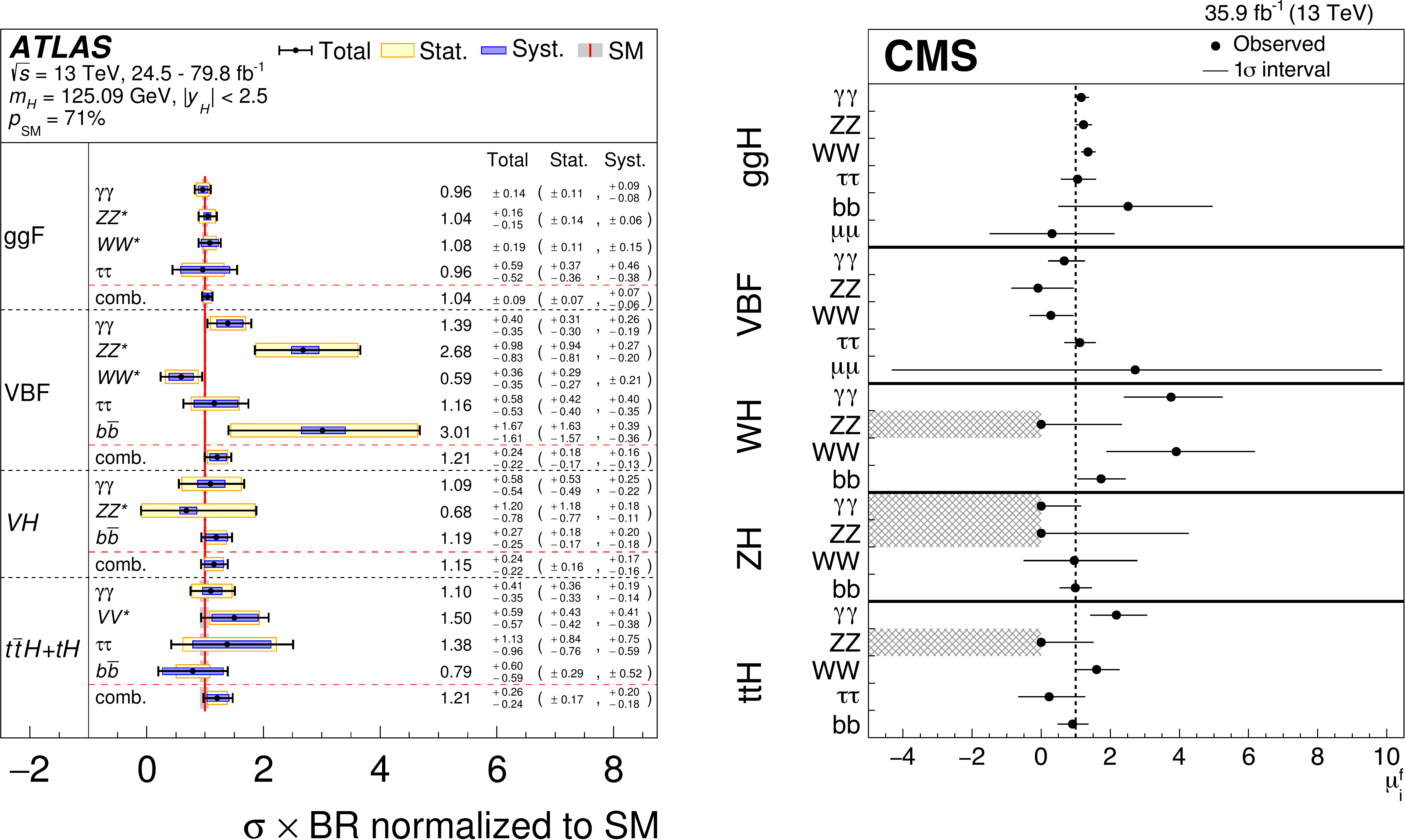}
\caption{Combined measurements of the signal strengths for the five main production and five main decay modes. The hatched combinations require more data. From Ref.~\cite{ParticleDataGroup:2020ssz}.
\label{fig:HsigmaxBR} 
}
\end{figure}

The Higgs event rates are proportional to the production cross-sections times the branching ratios (BR). Experimental results are often normalized to the SM predictions and expressed in terms of signal strengths $\mu=(\sigma\times{\rm BR})_{\rm obs}/(\sigma\times{\rm BR})_{\rm SM}$. Figure~\ref{fig:HsigmaxBR} shows that data are in fair agreement with predictions for a good number of channels and production mechanisms. 

\begin{figure}
\centering
\includegraphics[height=0.45\linewidth]{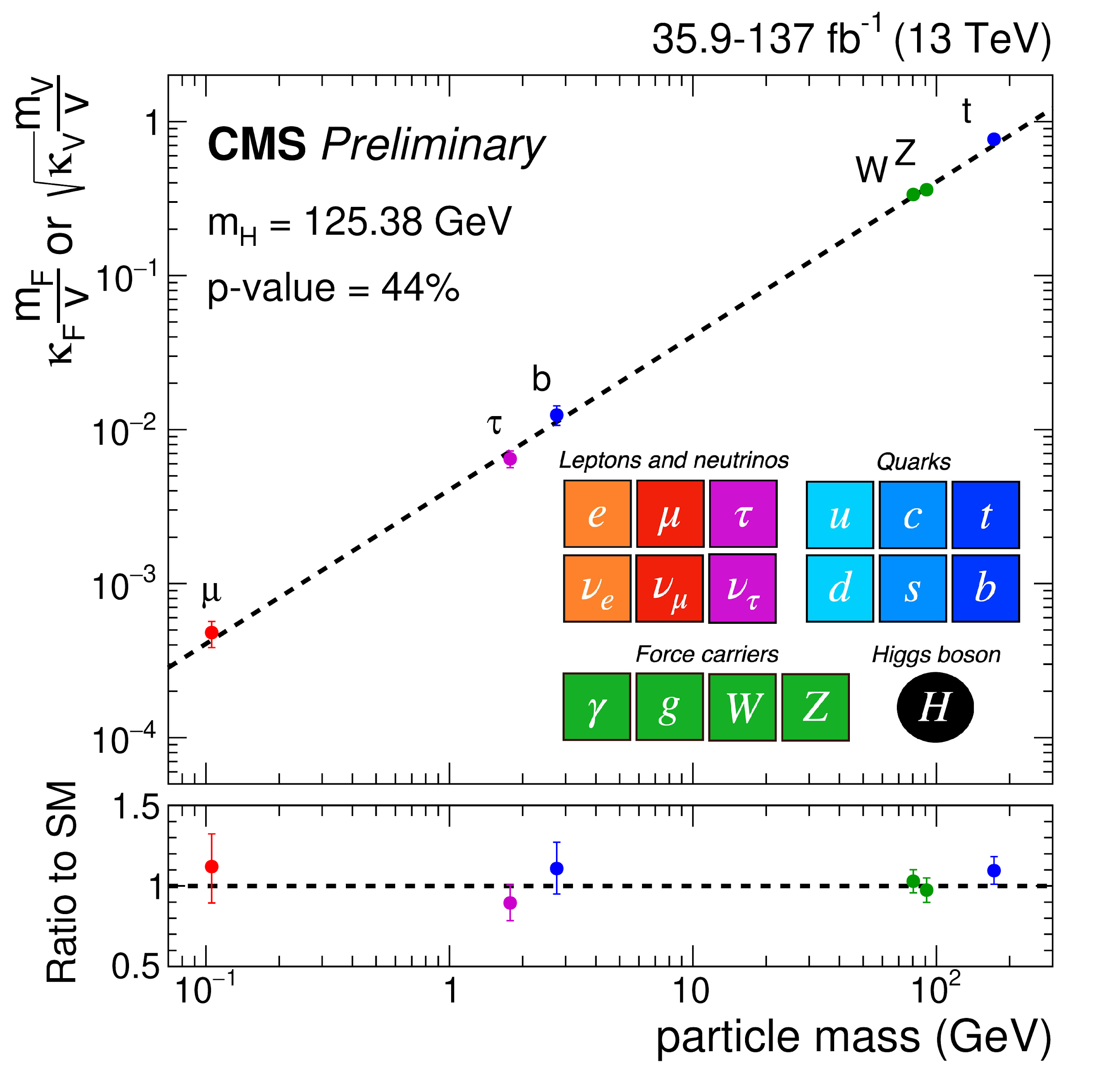}
\caption{Best fit estimates for the Higgs coupling strengths to fermions and gauge bosons. From Ref.~\cite{CMS:2020xwi}.
\label{fig:Hcouplings}}
\end{figure}

As for the tests of Higgs couplings, recall that in the SM the Yukawa coupling between the Higgs boson and the fermions is proportional to the fermion mass ($m_F$), while the coupling to weak bosons is proportional to the square of the vector boson masses ($m_V$). Then one may define $y_F\equiv\kappa_F m_F/v$ for fermions and $y_V\equiv\sqrt{\kappa_V} m_V/v$ for weak bosons where $\kappa_F$ and $\kappa_V$ are coupling strengths that measure the ratio of observations to SM predictions. The Higgs couplings to fermions and gauge bosons have been probed over more that three orders of magnitude with no significant deviations from the SM (figure~\ref{fig:Hcouplings}).

\subsubsection{Precise determination of parameters \label{sec:precision}}

Experimental precision requires accurate theoretical predictions, that are based on calculations beyond the tree-level approximation. The trouble is that the computation of loop corrections is laborious and plagued of infinities which involves the extra complication of renormalization.

\begin{figure}
\centering
\includegraphics[width=45mm]{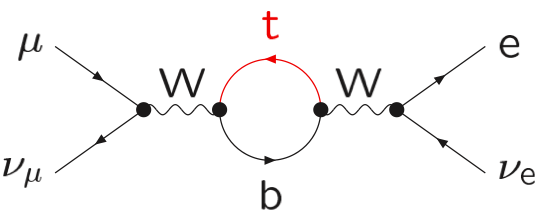} \raisebox{8.2mm}{$+$}
\includegraphics[width=45mm]{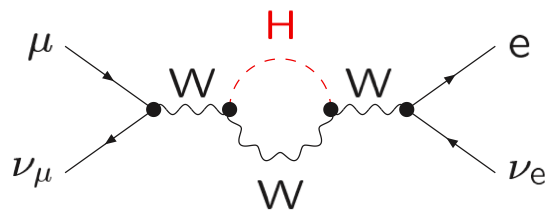} \raisebox{8.2mm}{$+$}
\includegraphics[width=45mm]{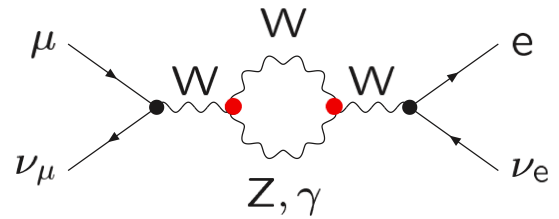}
\caption{One-loop corrections to the muon decay amplitude.
\label{fig:mudcy1loop}}
\end{figure}

A good example of the need for quantum corrections is the derivation of the Fermi constant from the measurement of the muon lifetime that follows from the identification
\eq{
G_F = \frac{\pi\alpha}{\sqrt{2}(1-M_W^2/M_Z^2)M_W^2}[1+\Delta r(m_t,  M_H)]
\label{eq:GFloop}
}
where $\Delta r$ depends on the masses and couplings of virtual particles exchanged in the the loop as in figure~\ref{fig:mudcy1loop}. This correction will fill the gap between \eqref{eq:GF} and \eqref{eq:GFhigh}. 

\begin{figure}
\centering\includegraphics[width=0.41\linewidth]{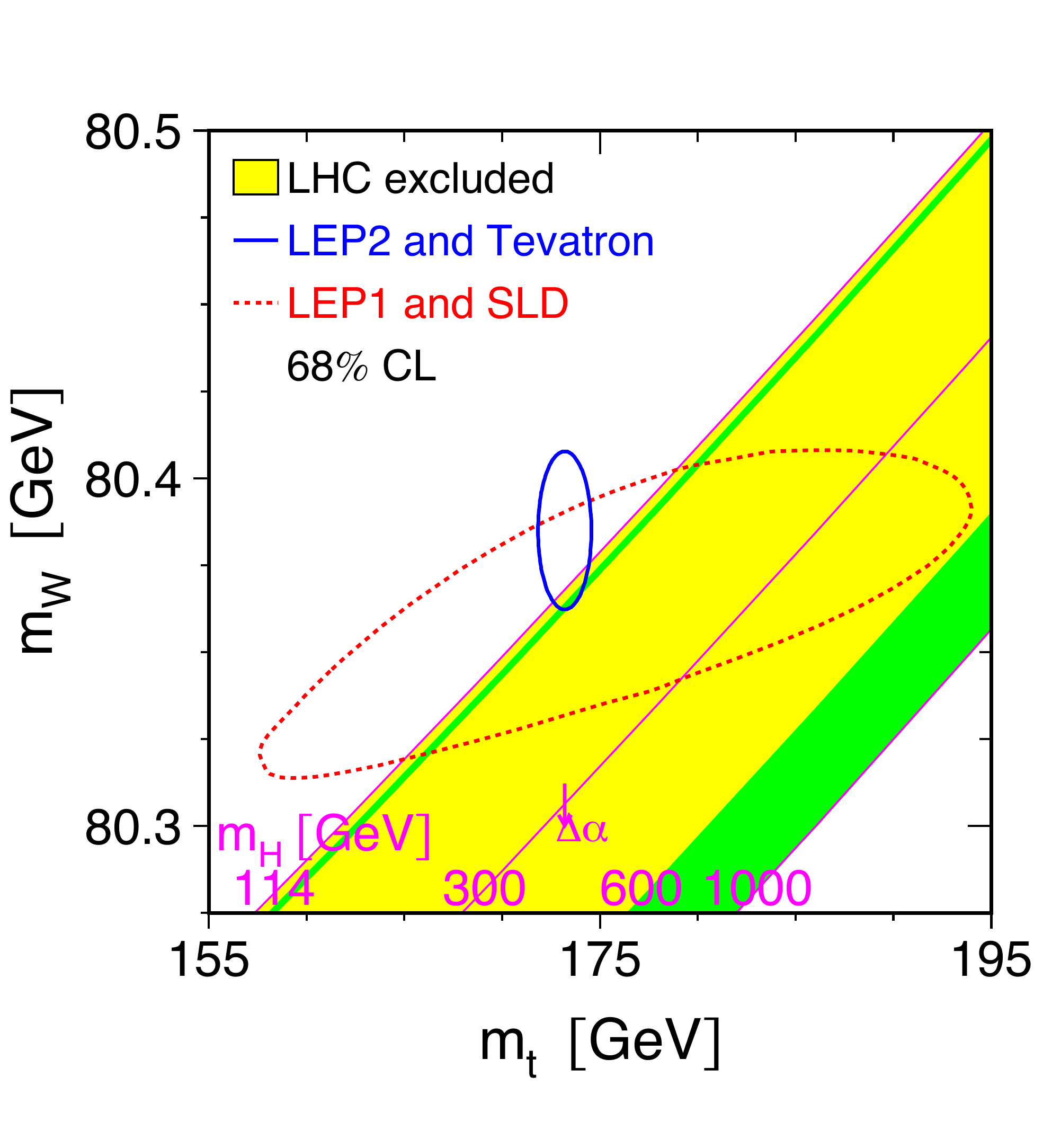}
\caption{Indirect constraints on $M_W$ and $m_t$ from LEP1/SLC data (dashed contour)
and direct measurements from LEP2/Tevatron data (solid contour). Also shown is the relation between both masses and the Higgs mass (solid lines), the region allowed by direct Higgs searches (dark green bands) and the region excluded by the LHC right before the Higgs boson discovery. From Ref.~\cite{ALEPH:2013dgf}.
  \label{fig:MWmt2013}
  }
\end{figure}

Actually, since the muon lifetime is measured more precisely than $M_W$, the $W$ mass can be independently obtained from the expression of $G_F$ in equation~(\ref{eq:GFloop}) that implies
\eq{
M_W^2(\alpha, G_F, M_Z, m_t, M_H) = \frac{M_Z^2}{2}\left(1+ \sqrt{1-\frac{4\pi\alpha}{\sqrt{2}G_FM_Z^2}[1+\Delta r(m_t, M_H)]}\right)
}
introducing a correlation between $M_W$, $m_t$ and $M_H$, given $\alpha$, $G_F$ and $M_Z$. This correlation has historically served as a handle to constrain yet unknown parameters from the value of others. As an example, figure~\ref{fig:MWmt2013} by the LEP Electroweak Working Group \cite{ALEPH:2013dgf} shows the comparison of indirect and direct constraints on $M_W$ and $m_t$ from LEP and Tevatron together with the region of Higgs masses consistent with precision tests before the Higgs boson was found at the LHC.

\begin{figure}
\centering
\includegraphics[width=0.58\linewidth]{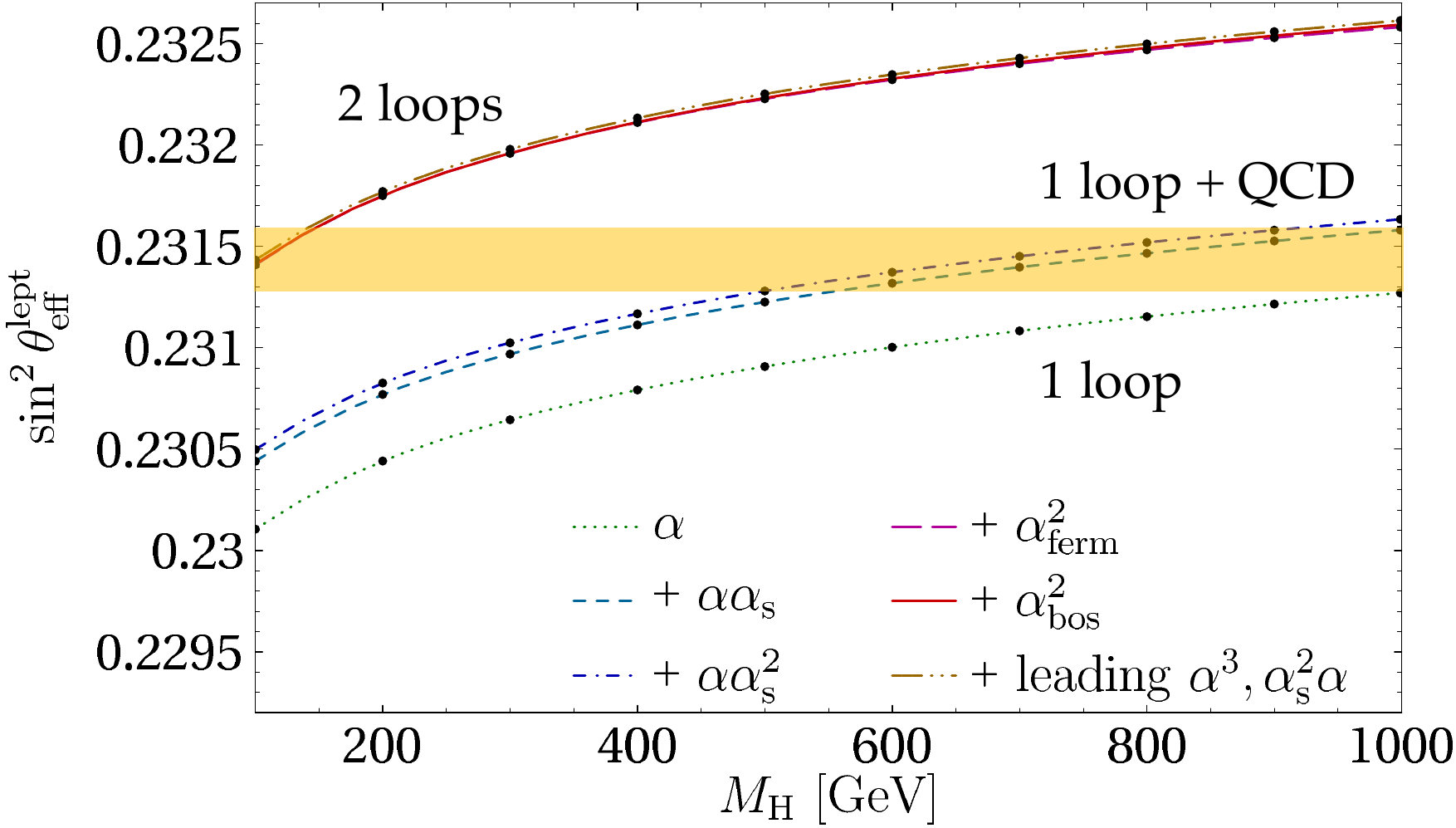}
\caption{Contribution of several orders of radiative corrections to the effective leptonic weak mixing angle $\sin^2\theta^{\rm lept}_{\rm eff}$ as a function of the Higgs mass. The tree-level value $s_W^2=1-M_W^2/M_Z^2\approx0.2230$ is below the range shown. The yellow band is the experimental accuracy at the time, $\sin^2\theta^{\rm lept}_{\rm eff}= 0.23147\pm0.000017$. From Ref.~\cite{Awramik:2006uz}.
\label{fig:sw2lept}}
\end{figure}

Another example is the corrections to vector and axial-vector couplings from $Z$ pole observables, 
\eq{
v_f \to g_V^f = v_f + \Delta g_V^f \qquad
a_f \to g_A^f = a_f + \Delta g_A^f
}
that lead to a fermion-dependent effective weak mixing angle given by
\eq{
\sin^2\theta_{\rm eff}^f \equiv \frac{1}{4|Q_f|}\left|1-{\rm Re}(g_V^f/g_A^f)\right| &\equiv s_W^2 (1+\Delta\kappa_Z^f)
}
where $\Delta\kappa_Z^f$ is the quantum correction in the $\overline{\rm MS}$ renormalization scheme and $s_W^2=1-M_W^2/M_Z^2$ is the tree-level value. As shown in figure~\ref{fig:sw2lept}, the effective leptonic weak mixing angle has been measured with high precision and at least two-loop calculations are needed \cite{Awramik:2006uz} to get a prediction compatible with experiment, already pointing to a light Higgs mass (the remaining theoretical uncertainty from unknown higher-order corrections was estimated to be $4.7\times10^{-5}$).

\begin{figure}
\centering
\includegraphics[width=0.6\linewidth]{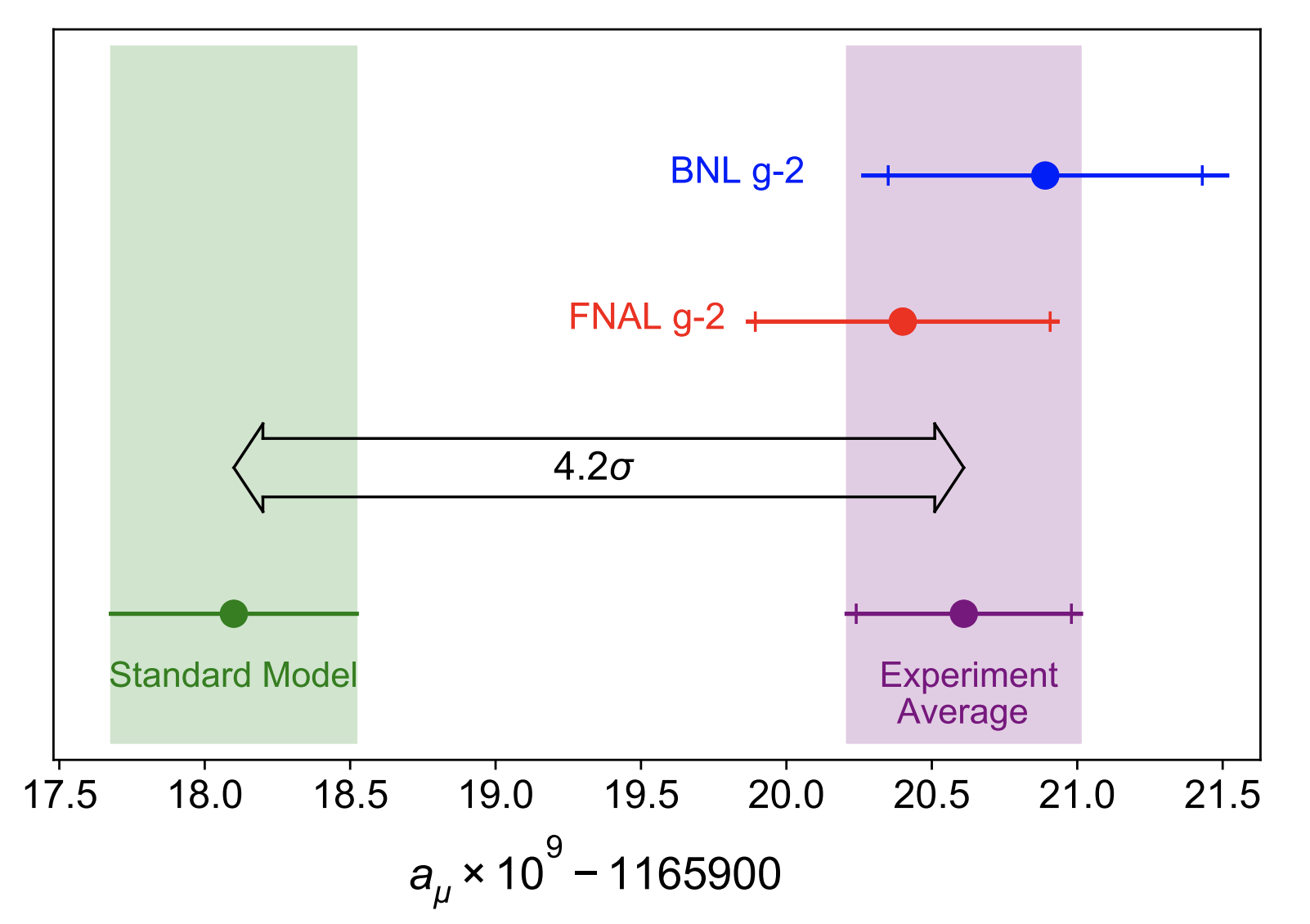}
\caption{
Experimental values of $a_\mu$ from Brookhaven, Fermilab and combined average. The inner tick marks indicate the statistical contribution to the total uncertainties. The recommended value for the standard model prediction \cite{Aoyama:2020ynm} is also shown.
\label{fig:amu}}
\end{figure}

There are also experiments and observables testing the flavor structure of the SM, either
flavor-conserving, like dipole moments, or flavor-changing, like $B_s\to X_s\gamma$ and many other hadron and lepton decays. They are very sensitive to new physics through loop corrections. 
As already mentioned, the extremely precise measurement of the electron anomalous magnetic moment $a_e=(g_e-2)/2$, 
\eq{
a_e^{\rm exp} = 0.001\,159\,652\,182\,032\,(720)
}
is used to estimate the fine structure constant $\alpha$ from the QED prediction at 5 loops \cite{Aoyama:2017uqe}. 
On the other hand, the anomalous magnetic moment of the muon was measured at Brookhaven with very high precision \cite{Muong-2:2006rrc}
\eq{
a_\mu^{\rm exp} = 116\,592\,089\,(63)\times10^{-11}
\qquad\mbox{[BNL]}
} 
but it does not match the SM prediction, a puzzle that has survived for almost two decades. The most recent calculation by the Muon $g-2$ Theory Initiative  \cite{Aoyama:2020ynm} yields
\eq{
a_\mu^{\rm SM} =  116\,591\,810\,(43)\times10^{-11}
}
that gives $a_\mu^{\rm exp}-a_\mu^{\rm SM} = 279\,(76)\times10^{-11}$, a $3.7\sigma$ deviation. Very recently a new experiment at Fermilab has released its first results \cite{Muong-2:2021ojo}, 
\eq{
a_\mu^{\rm exp} = 116\,592\,061\,(41)\times10^{-11}
\qquad\mbox{[FNAL]}
}
compatible with the previous measurements and increasing the discrepancy to $4.2\sigma$ (figure~\ref{fig:amu}). This is nowadays considered a very compelling evidence of physics beyond the SM.

\begin{figure}
\centering
\begin{tabular}{cc}
\includegraphics[width=0.33\linewidth]{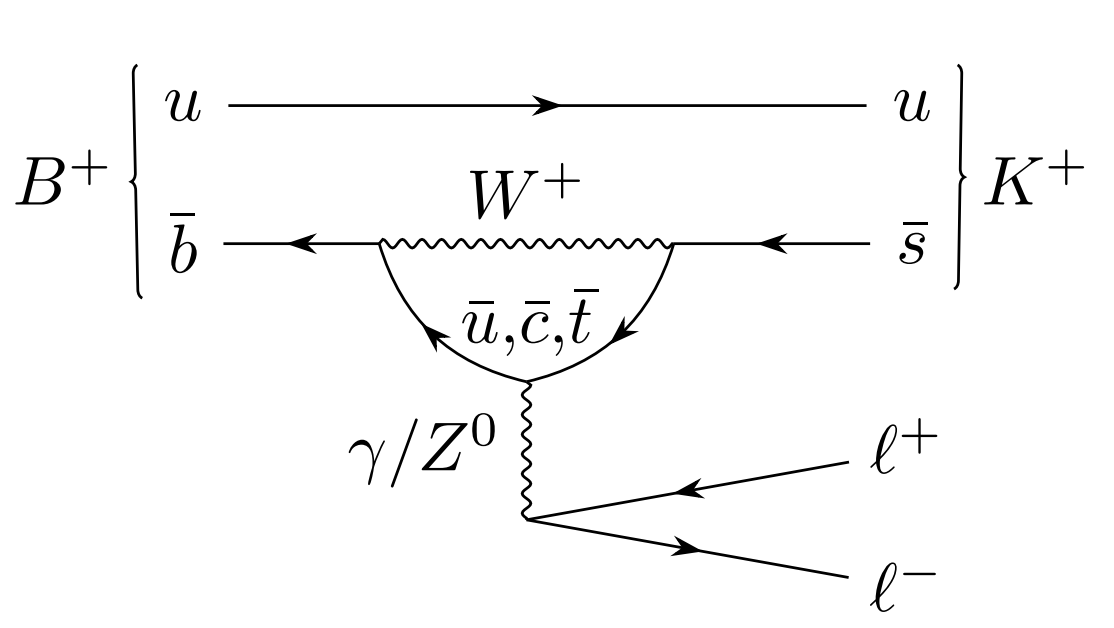} &
\raisebox{-0.25\height}{\includegraphics[width=0.5\linewidth]{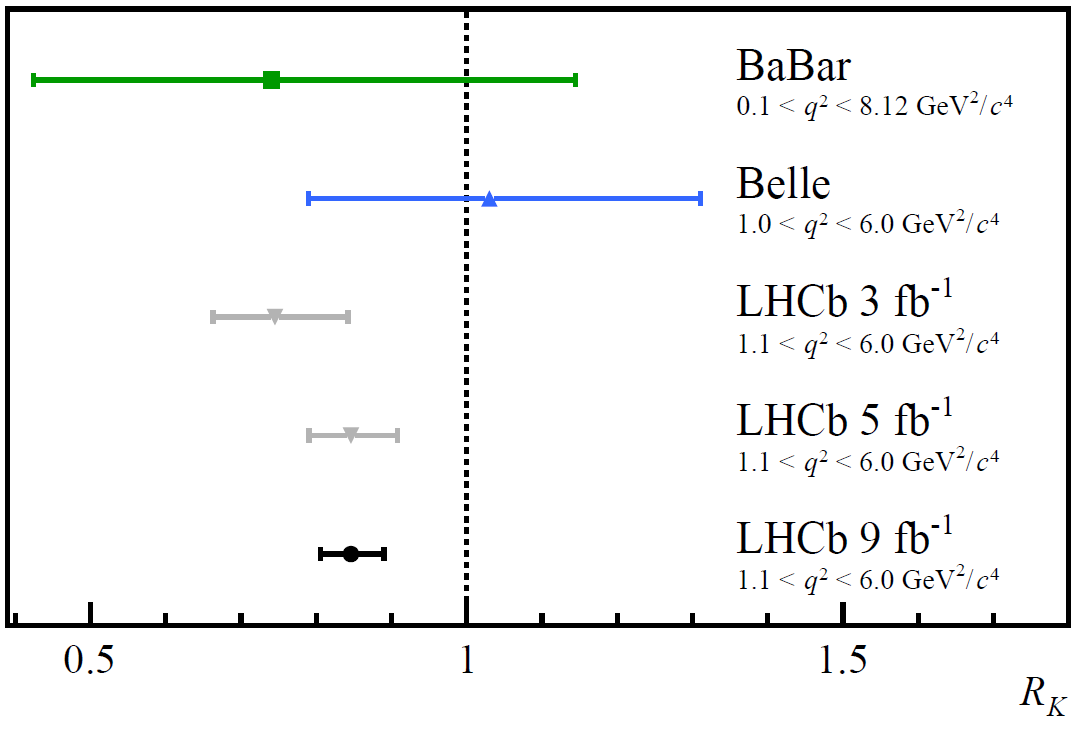}}
\end{tabular}
\caption{Fundamental processes contributing to $B^+\to K^+\ell^+\ell^-$ decays in the SM
and comparison between $R_K$ measurements. From Ref.~\cite{LHCb:2021trn}.
\label{fig:LFUtest}}
\end{figure}

Another playground where precision physics has revealed departures from the SM predictions is $b$-hadron decays, with tensions in rare flavor-changing neutral currents and in tree-level semileptonic decays that constitute the so-called flavor anomalies in B-physics (see \cite{Albrecht:2021tul} for a recent review). They have been observed in measurements of branching fractions and angular observables, as well as in lepton flavor universality tests. A good example of the latter is the measurement by LHCb \cite{LHCb:2021trn} of the ratio
\eq{
R_K = \dis\frac{{\rm BR}(B^+\to K^+\mu^+\mu^-)}{{\rm BR}(B^+\to K^+e^+e^-)} = 0.846^{+0.044}_{-0.041}
}
that is about $3\sigma$ from the SM prediction, $1.00\pm0.01$, providing evidence for the violation of lepton universality in these decays. This tension, that was not significant in previous measurements at BaBar (SLAC) and Belle (KEK), has survived and even grown with increasing statistics at LHCb (figure~\ref{fig:LFUtest}). More data from LHCb and the forthcoming Belle II experiment \cite{Belle-II:2018jsg} will establish whether this anomaly must be taken seriously.

\subsubsection{Global fits}

As we have seen, precision measurements test the SM at the quantum level, which allows to perform consistency checks among the results. The global fits consist of finding the values of a set of input parameters that minimize the $\chi^2$ accounting for the deviation between a number of precision observables and their SM predictions. The predictions are given by theoretical expressions that are functions of the input parameters. The precision observables are sometimes more appropriately named `pseudo-observables' because they are not directly experimental observables but derived quantities depending on the order of perturbation theory and on the choice of renormalization scheme.  

The latest electroweak global fit by Gfitter \cite{Haller:2018nnx}, using the observables
$M_H$, $M_W$, $\Gamma_W$, $M_Z$, $\Gamma_Z$, $\sigma^0_{\rm had}$, $R_{\ell,c,b}$, 
$A_{\text{FB}}^{0,\ell}$, $A_{\text{FB}}^{0,c}$, $A_{\text{FB}}^{0,b}$, $A_{\ell}$, $A_{c}$, $A_{b}$, $\sin^2\theta_{\rm eff}^\ell$, $m_{c}$, $m_{b}$, $m_{t}$, $\alpha(M_Z^2)$ and $\alpha_s(M_Z^2)$, converges to a $\chi^2_{\rm min}=18.6$ for 15 degrees of freedom (number of fit observables minus number of free parameters). This corresponds to a $p$-value of 0.23. The $p$-value tests the likelihood of the null-hypothesis, the probability of obtaining data equal or less compatible with the theory, so the lower the better.  

\begin{figure}
\centering
\begin{tabular}{cc}
\includegraphics[height=0.7\linewidth]{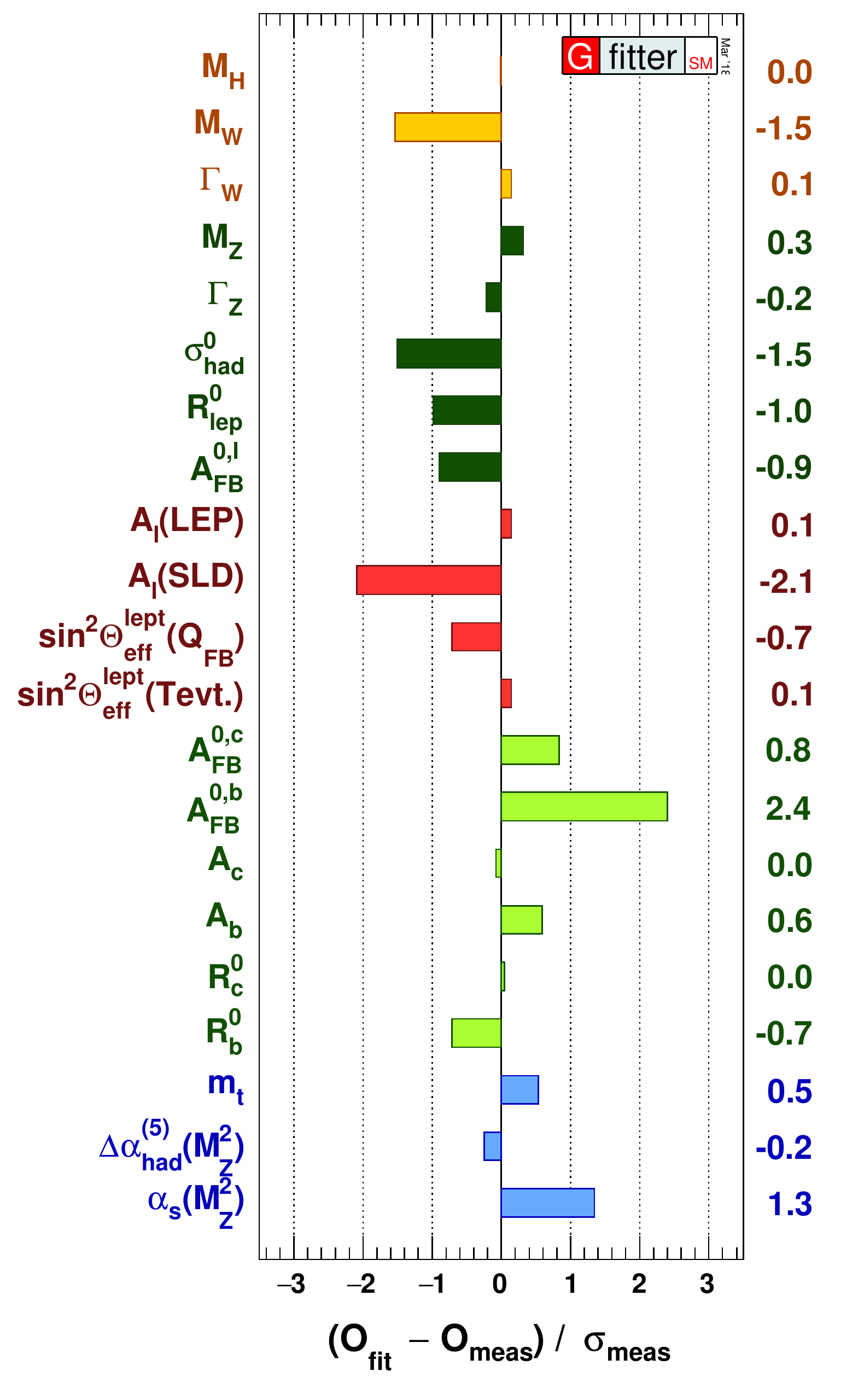} &
\includegraphics[height=0.7\linewidth]{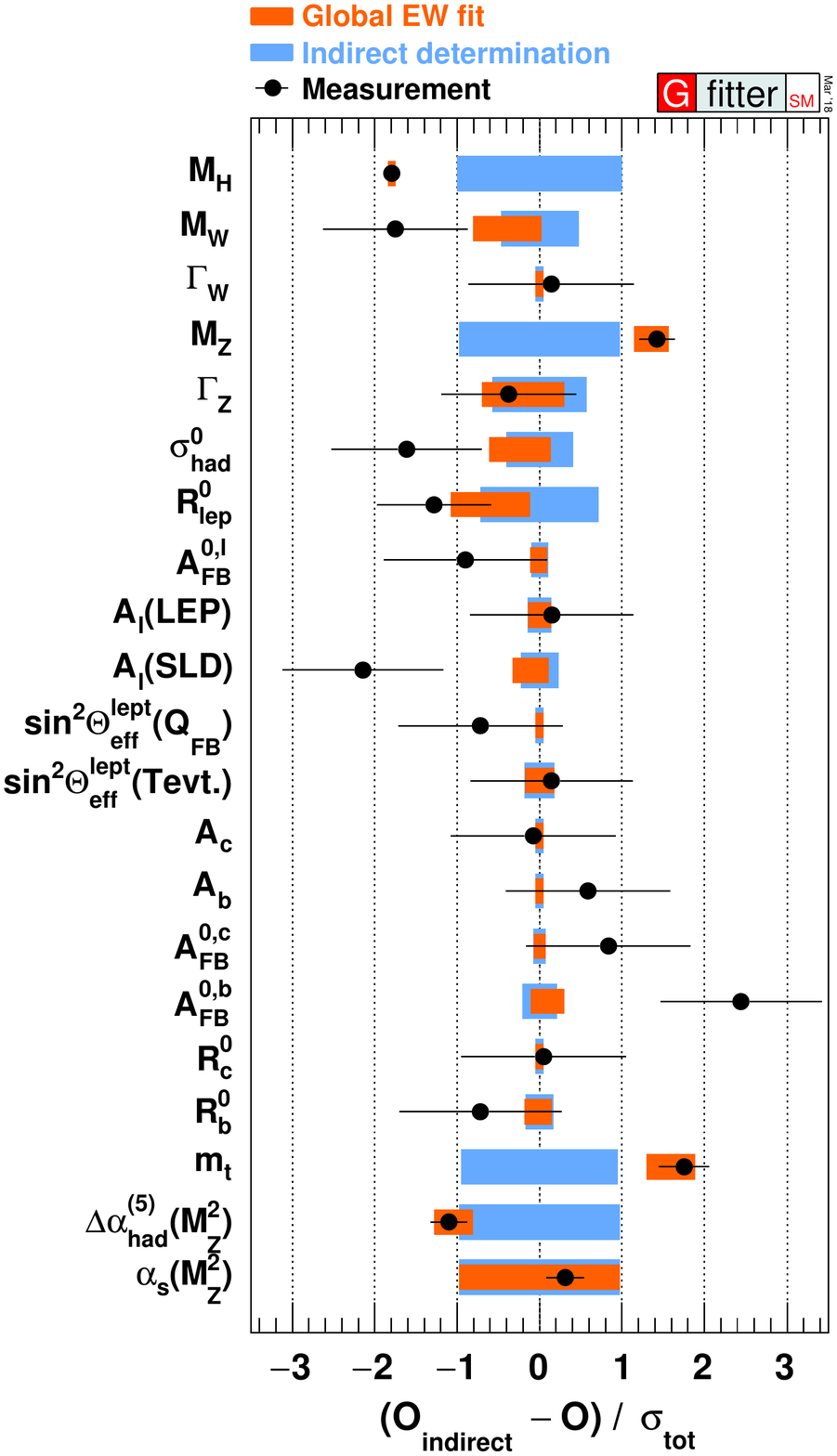}
\end{tabular}
\caption{Left: Comparing fit results with direct measurements.
Right: Comparing fit results (orange bars) with indirect determinations (blue bars) and direct measurements (data points). The total error is taken to be the error of the direct measurement added in quadrature with the error from the indirect determination.
From Ref.~\cite{Haller:2018nnx}.
\label{fig:fitsvsmeasurements}
}
\end{figure}

\begin{figure}
\centering
\includegraphics[width=0.49\linewidth]{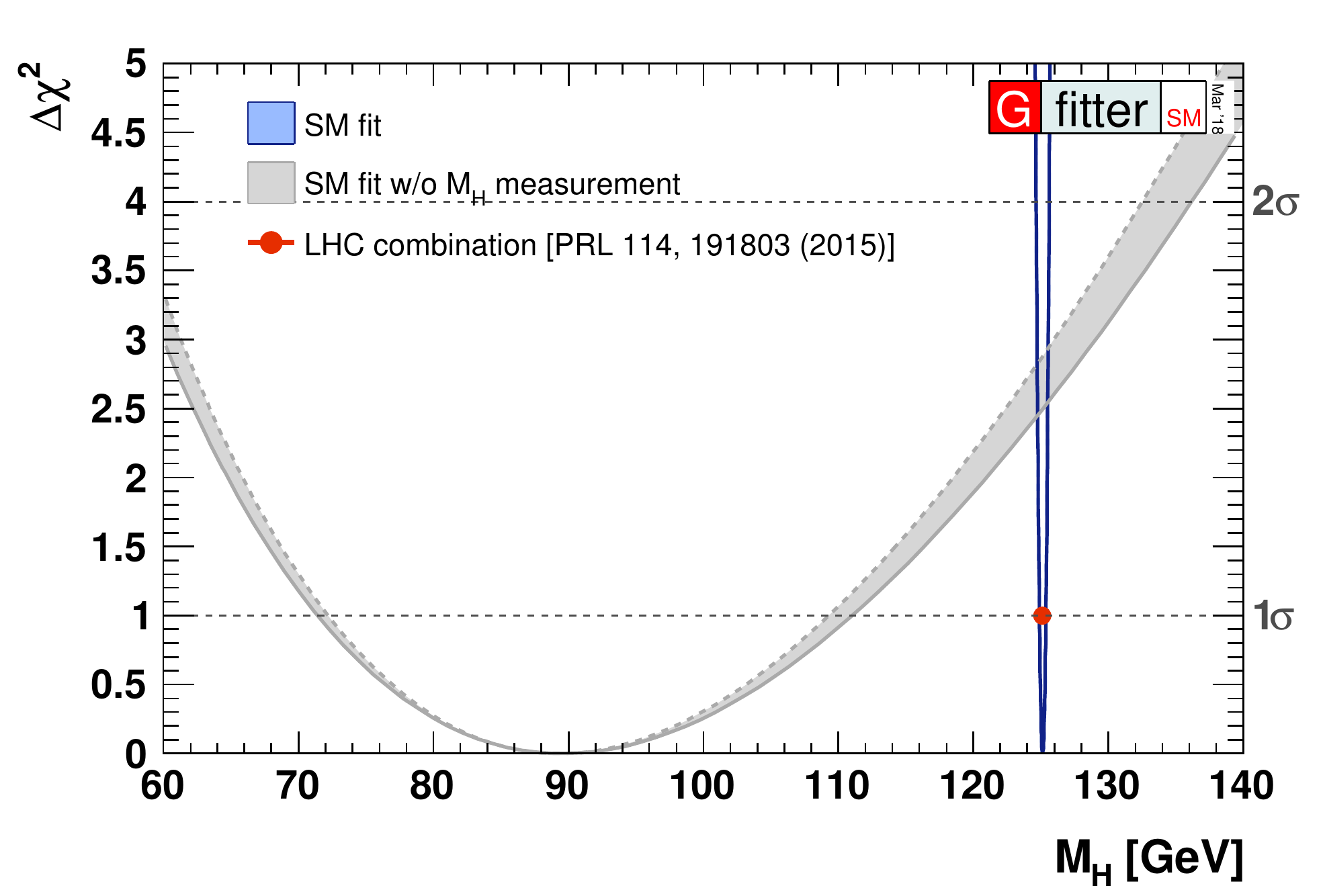}
\includegraphics[width=0.49\linewidth]{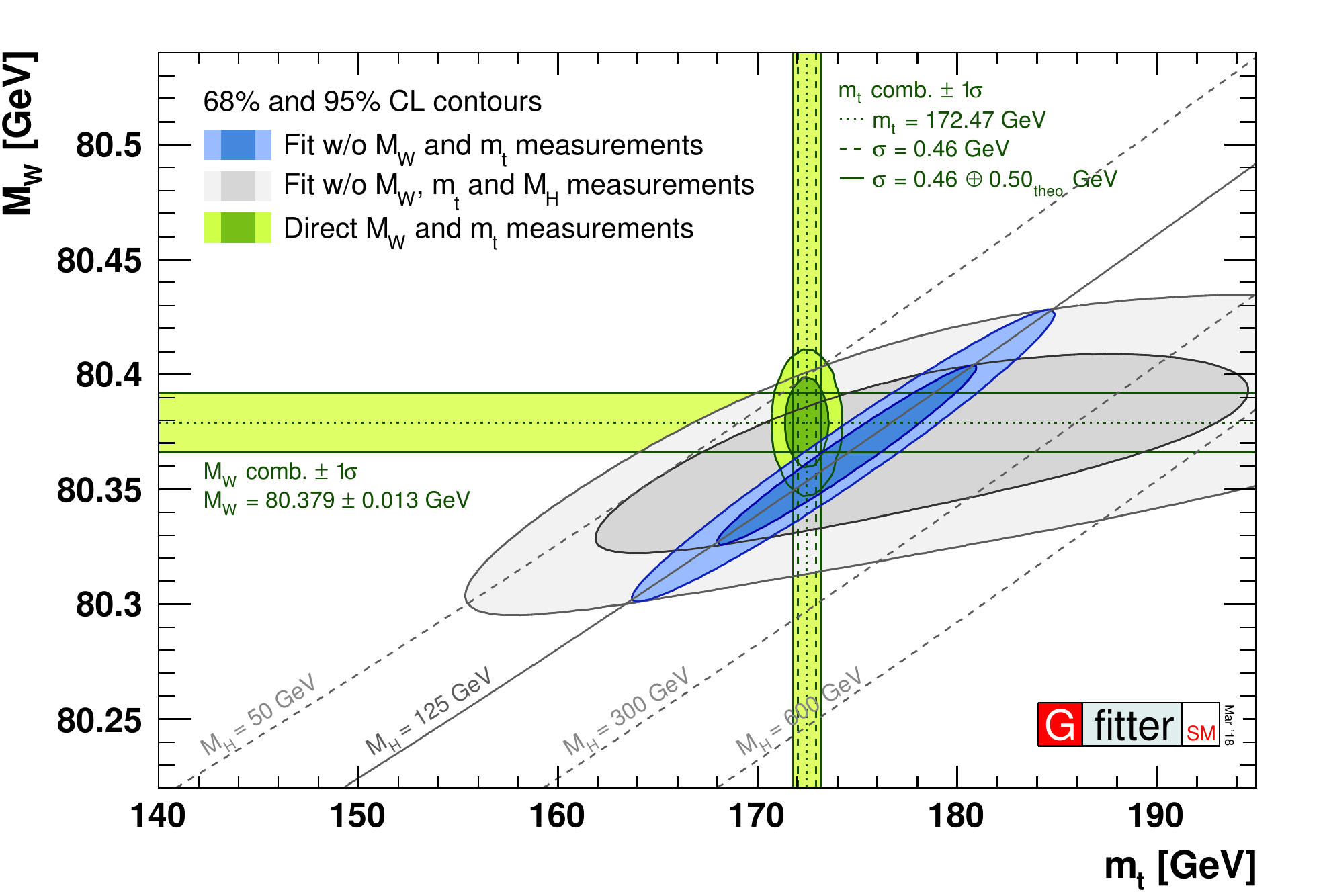}
\caption{Left: $\Delta\chi^2$ as a function of Higgs boson mass for a global SM fit with and without the $M_H$ measurement (blue and grey bands). 
Right: Contours of 68\% and 95\% confidence level obtained from scans of fits with fixed variable pairs $M_W$, $m_t$. The narrower blue and larger grey allowed regions are the results of the fit including and excluding the $M_H$ measurement, respectively. 
From Ref.~\cite{Haller:2018nnx}.
\label{fig:fitMH-MW-mt}}
\end{figure}

It is also interesting to compare the fit results with the input measurements \cite{Haller:2018nnx}. The left panel of fig.~\ref{fig:fitsvsmeasurements} shows the deviations between global fit values and direct measurements in units of the experimental uncertainty. There are some tensions but none above $3\sigma$. The right panel of fig.~\ref{fig:fitsvsmeasurements} shows the difference between the global fit results (orange bars) as well as the input
measurements (data points) with the indirect determinations (blue bars). The indirect determinations are the best fit values without using the constraint from the corresponding input measurement. This illustrates the impact of indirect uncertainties on total uncertainties.
Finally, the left panel of fig.~\ref{fig:fitMH-MW-mt} shows that the global fit to the SM prefers a somewhat lighter Higgs boson. The right panel is an updated version of the confidence level profile of $M_W$ versus $m_t$ in fig.~\ref{fig:MWmt2013} where the $M_H$ measurement at LHC is included in the fit or not (blue or grey contour). The good agreement of both contours with the direct measurements (green bands and ellipse for two degrees of freedom) is the ultimate confirmation of the consistency of the SM.

\section{Concluding remarks\label{sec:conrem}}

The Standard Model of the electroweak and strong interactions of particle physics is a relativistic quantum field theory based on a gauge symmetry that is spontaneously broken by the Brout-Englert-Higgs mechanism. As a consequence it is renormalizable and fully predictive. It has been confirmed by a plethora of low and high energy experiments with remarkable accuracy, at the level of quantum corrections, with (almost) no significant deviations.

However, in spite of its tremendous success, the SM leaves fundamental questions unanswered: why three generations? what is the reason for the observed pattern of quark and lepton masses and mixings? And more importantly, there are several hints for physics beyond. Some are phenomenological and others more conceptual. Perhaps the most compelling is the muon magnetic dipole moment, whose very precise measurement is still challenging the SM prediction after many years of efforts both from the experimental and the theory side. There is also a bunch of flavor anomalies in $B$ physics that are gaining evidence. The neutrino sector is without doubt the Achilles heel of the model, that has already required an extension to accommodate neutrino masses and mixings in order to explain the flavor oscillations. The possibility that neutrinos are Majorana fermions, theoretically well motivated and under intense experimental exploration, would open the window to lepton number violation and, linked to this, would suggest the existence of extra neutrinos at a very heavy scale that might contribute to solve the baryon asymmetry problem\footnote{The SM violates the conservation of baryon number non-perturbatively, thanks to a global U(1) anomaly, but in an amount that is not enough to explain the matter-antimatter asymmetry of the universe.} \cite{Kuzmin:1985mm} through leptogenesis \cite{Davidson:2008bu}. Another problem is dark matter. If it is composed of hypothetical particles interacting with ordinary matter only through gravity \cite{Bertone:2004pz}, the SM does not provide any appropriate candidate, although there are interesting alternatives \cite{Clesse:2017bsw}. Nonetheless, it is very suggestive that the most popular solution to the strong CP problem (the Peccei-Quinn mechanism \cite{Peccei:2006as}) introduces a new global anomalous symmetry spontaneously broken at low energies giving rise to a pseudo-Goldstone boson, the axion, considered a viable candidate for dark matter.

Of course the SM cannot be the `theory of everything', since it does not include the gravitational interaction that governs the universe dynamics at large scales. But it has something to say about the value of the vacuum energy density, $\rho_{\text{vac}}$, that is related to the cosmological constant\footnote{The cosmological constant is the simplest form of dark energy \cite{Peebles:2002gy} so far indistinguishable from the more general quintessence.} by $\Omega_\Lambda=\rho_{\text{vac}}/\rho_c$ where $\rho_c = 3 H_0^2/(8\pi G_N)$ is the critical density of the universe. According to current cosmological measurements of the cosmic expansion acceleration \cite{Planck:2018vyg}, $\Omega_\Lambda\approx 0.7$, that implies $\rho_{\text{vac}}\approx (2\times10^{-3}\mbox{ eV})^4$. In the SM, as in any quantum field theory, the values of quantities like the masses, couplings or the cosmological constant cannot be predicted. They are fixed by the renormalization procedure: the bare parameters are chosen so that they cancel the divergent corrections and leave us with the desired renormalized quantity. The computation of the vacuum energy density yields a result that diverges quartically with the cutoff (physics scale up to which the theory is meaningful),\footnote{The divergent result comes from the zero-point energy we subtracted for convenience in the normal ordering prescription,
$$
\int \frac{\dd^3 p}{(2\pi)^3} E_{\vec{p}} \sim \int^{\Lambda_{\rm cut}} p^3 \dd p \sim \Lambda_{\rm cut}^4.
$$
 }
\eq{
\rho_{\text{vac}} \approx \rho_0(\Lambda_{\rm cut}) + c\,\Lambda_{\rm cut}^4 .
}
If we assume no new physics until the Planck scale ($\Lambda_{\rm cut} \sim M_P \sim 10^{19}$~GeV), where gravity becomes relevant, then $\rho_0(\Lambda_{\rm cut})$ has to be chosen so that a very fine-tuned cancelation with the correction of more than 120 digits will be required. Even if new physics were behind the corner, say $\Lambda_{\rm cut} \sim 1$~TeV, the fine-tuning would be of about 60 digits. Although $\rho_0(\Lambda_{\rm cut})$ has no physical meaning and can be chosen at will, this level of fine-tuning is considered very unnatural.

Another naturalness problem of the SM has to do with the renormalization of the mass of scalar fields. The corrections to the mass squared of a scalar field, like the Higgs, diverge quadratically with the cutoff,
\eq{
M_H^2 \approx (M^0_H)^2(\Lambda_{\rm cut}) + c\,\Lambda_{\rm cut}^2 .
}
This is in contrast to the masses of fermion or gauge fields whose corrections grow only logarithmically with the cutoff, because they are a protected by a symmetry (they would be massless if chiral or gauge symmetries were unbroken). If we take $\Lambda_{\rm cut} \sim M_P$ then a cancellation of 34 digits is needed to match the observed Higgs mass $M_H\simeq 125$~GeV. However, this hierarchy problem is different from the cosmological constant problem, because it could be solved if there were new physics not far above the electroweak scale (at $\Lambda_{\rm cut} \sim 1$~TeV for example) as in the case of supersymmetric extensions of the SM \cite{Martin:1997ns}, or if the `true' Planck scale is $M_P\sim 1$~TeV as in the case of models with extra dimensions \cite{Arkani-Hamed:1998jmv,Randall:1999ee}. Unfortunately there is no experimental clue of any of them.

In the absence of signals from a better fundamental theory that can tie up the loose ends of the SM, we can always consider the SM as a low-energy effective field theory \cite{Buchmuller:1985jz,Grzadkowski:2010es} (SMEFT) valid up to some energy scale, like the 4-Fermi model is a good effective theory for $E\ll M_W$. The effective Lagrangian can be  written as
\eq{
{\cal L} = {\cal L}_{\rm SM} + \sum_{i,n} \frac{c_i^{(n)}\,{\cal O}_i^{(n)}}{\Lambda_{\rm NP}^{n-4}},
}
where ${\cal L}_{\rm SM}$ is the renormalizable part of the Lagrangian, that we had so far identified with the SM. The new physics is parametrized by a set of higher dimensional (Lorentz and gauge invariant) operators ${\cal O}_i^{(n)}$ made of SM fields, where $n>4$ is the canonical dimension. $\Lambda_{\rm NP}$ is the new physics scale, such as the mass of a new particle. Their effects are suppressed by $(E/\Lambda_{\rm NP})^{n-4}$ with respect to the SM operators where $E$ is any low energy scale or mass, so the higher the dimension of the operator the smaller its contribution at low energies. Therefore, given a finite experimental precision we only need operators up to certain dimension and, since there are a finite number of these, their coefficients can be renormalized. The lack of predictivity on the (remaining) coefficients above some order is irrelevant. This is why the SMEFT, though `non-renormalizable', is perfectly acceptable to describe physics below $\Lambda_{\rm NP}$ and is used as a very powerful framework \cite{Ellis:2020unq}.

\acknowledgments

We thank the organizers of the COST CA18108 First Training School at the Corfu Summer Institute for the invitation and for creating such a nice and stimulating atmosphere. 
Part of these lectures had been given earlier online at the XIX Mexican School of Particles and Fields. The students of both events are also gratefully acknowledged for their very pertinent questions.
We are indebted to Manuel Masip, Manuel P\'erez-Victoria and Jos\'e Santiago for useful comments and suggestions that helped to improve the manuscript.
The work of JII is supported in part by the Spanish Ministry of Science, Innovation and Universities (FPA2016-78220-C3, PID2019–107844GB-C21/AEI/10.13039/501100011033), and by Junta de Andaluc{\'\i}a (FQM 101, SOMM17/6104/UGR, P18-FR-1962, P18-FR-5057). AJC is supported by the European Regional Development Fund through the Center of Excellence TK133 ``The Dark Side of the Universe" and by the Mobilitas Pluss postdoctoral grant MOBJD1035.


\end{document}